\documentclass[12pt]{article}

\usepackage{fullpage}
\usepackage[letterpaper, margin=1in]{geometry}
\usepackage{amsmath,amsthm,amsfonts,amssymb,amscd}
\usepackage{lastpage}
\usepackage{enumerate}
\usepackage{algorithm}
\usepackage{algpseudocode}
\usepackage{fancyhdr}
\usepackage{mathrsfs}
\usepackage{xcolor}
\usepackage{graphicx}
\usepackage{listings}
\usepackage[colorlinks]{hyperref}
\hypersetup{
    colorlinks=true,
    linkcolor=blue,
    citecolor=blue,
    filecolor=magenta,      
    urlcolor=cyan
    }
\usepackage{multirow}
\usepackage{verbatim}
\usepackage{natbib}
\usepackage{booktabs}

\usepackage[utf8]{inputenc}
\usepackage{url}
\usepackage{pdflscape,booktabs}
\usepackage{amsmath,color,amsfonts,rotating,amsthm,amssymb,mathtools,setspace,fullpage,natbib,mathabx,bbm,multirow,threeparttable,graphicx,caption,lipsum,subcaption,multicol,float,tabularx}
\usepackage{appendix}
\usepackage{tikz}
\usepackage{threeparttable}
\usetikzlibrary{matrix}
\DeclareMathOperator*{\argmax}{argmax}
\usepackage{setspace}
\graphicspath{ {./Figures/} }
\newcommand{\calB}{\mathcal B}

\newcommand{\calS}{\mathcal S}

\newcommand{\calL}{\mathcal L}

\newcommand{\calD}{\mathcal D}

\newcommand{\ba}{ {\boldsymbol a} }
\newcommand{\bA}{ {\boldsymbol A} }
\newcommand{\bb}{ {\boldsymbol b} }
\newcommand{\bB}{ {\boldsymbol B} }

\newcommand{\bC}{ {\boldsymbol C} }
\newcommand{\bd}{ {\boldsymbol d} }
\newcommand{\bD}{ {\boldsymbol D} }

\newcommand{\bF}{ {\boldsymbol F} }

\newcommand{\bG}{ {\boldsymbol G} }

\newcommand{\bH}{ {\boldsymbol H} }

\newcommand{\bI}{ {\boldsymbol I} }

\newcommand{\bJ}{ {\boldsymbol J} }

\newcommand{\bl}{ {\boldsymbol l} }
\newcommand{\bL}{ {\boldsymbol L} }

\newcommand{\bs}{ {\boldsymbol s} }

\newcommand{\bu}{ {\boldsymbol u} }

\newcommand{\bV}{ {\boldsymbol V} }
\newcommand{\bw}{ {\boldsymbol w} }

\newcommand{\bx}{ {\boldsymbol x} }
\newcommand{\bX}{ {\boldsymbol X} }
\newcommand{\by}{ {\boldsymbol y} }

\newcommand{\bz}{ {\boldsymbol z} }

\newcommand{\eps}{\epsilon}
\newcommand{\sigs}{\sigma^2}

\newcommand{\bone}{ {\bf 1} }
\newcommand{\bzero}{ {\bf 0} }

\newcommand{\given}{\,|\,}

\newcommand{\boeta}{ {\boldsymbol \eta} }

\newcommand{\bbeta}{ {\boldsymbol \beta} }

\newcommand{\bgamma}{ {\boldsymbol \gamma} }

\newcommand{\bpsi}{ {\boldsymbol \psi} }
\newcommand{\bxi}{ {\boldsymbol \xi} }

\newcommand{\blam}{ {\boldsymbol \lambda} }
\newcommand{\blambda}{ {\boldsymbol \lambda} }

\newcommand{\bmu}{ {\boldsymbol \mu} }

\newcommand{\bSigma}{ {\boldsymbol \Sigma} }
\newcommand{\btheta}{ {\boldsymbol \theta} }

\newcommand{\iid}{\overset{\mbox{iid}} \sim}

\title{Fast Variational Bayes for Large Spatial Data}
\author{
  Jiafang Song, Abhirup Datta\thanks{abhidatta@jhu.edu} \\
  Department of Biostatistics, Johns Hopkins University
}
\date{}

\begin{document}

\maketitle

\begin{abstract}
\noindent
Recent variational Bayes methods for geospatial regression, proposed as an alternative to computationally expensive Markov chain Monte Carlo (MCMC) sampling, have leveraged Nearest Neighbor Gaussian processes (NNGP) to achieve scalability. Yet, these variational methods remain inferior in accuracy and speed compared to spNNGP, the state-of-the-art MCMC-based software for NNGP. We introduce spVarBayes, a suite of fast variational Bayesian approaches for large-scale geospatial data analysis using NNGP. Our contributions are primarily computational. We replace auto-differentiation with a combination of calculus of variations, closed-form gradient updates, and linear response corrections for improved variance estimation. We also accommodate covariates (fixed effects) in the model and offer inference on the variance parameters. Simulation experiments demonstrate that we achieve comparable accuracy to spNNGP but with reduced computational costs, and considerably outperform existing variational inference methods in terms of both accuracy and speed. Analysis of a large forest canopy height dataset illustrates the practical implementation of proposed methods and shows that the inference results are consistent with those obtained from the MCMC approach. The proposed methods are implemented in publicly available R-package \texttt{ \href{https://github.com/jfsong100/spVarBayes}{spVarBayes}}. 
\end{abstract}


\section{Introduction}
Geospatial data are commonly analyzed using mixed effect models, specifying the outcome in terms of a linear regression (fixed effects) on observed covariates and location-specific random effects modeled as a Gaussian process. Hierarchical Bayesian models have often been a preferred choice for such geostatistical analysis \citep{banerjee2014hierarchical,cressie2015statistics}, offering inferences on all unknown parameters, including the spatial random effects and the parameters dictating its process properties, from Markov Chain Monte Carlo (MCMC) sampling methods. However, for large spatial datasets, MCMC algorithms can lead to significant computational costs. It requires sequentially sampling a large number random effects, resulting in a high-dimensional MCMC. Also, each update necessitates the expensive inversion operation of the dense $n \times n$ spatial covariance matrix requiring $O(n^3)$ floating point operations (FLOPs) and $O(n^2)$ storage,  $n$ being the sample size. 

Methods to reduce the computational costs associated with MCMC sampling for Gaussian process models include low rank predictive processes \citep{banerjee2008gaussian, finley2009improving}, Nearest Neighbor Gaussian Processes 
\citep[NNGP,][]{datta2016hierarchical,finley2019efficient}, multiresolutional approximations \citep{katzfuss2017multi},  and mesh-based approaches \citep{peruzzi2022highly}. Detailed reviews and comparisons of these approaches can be found in \cite{banerjee2017high} and \cite{heaton2019case}. NNGP is motivated by earlier work on likelihood approximations using nearest neighbor-based conditioning sets \citep{vecchia1988estimation,stein2004approximating}. \cite{datta2016hierarchical} showed that such nearest neighbor-based likelihood approximations correspond to a valid Gaussian process representation, admitting a sparse Cholesky factorization of the precision matrix over any finite set of realizations. As we will discuss subsequently, these are properties that make NNGP attractive for use in variational inference. The process representation enables the use of NNGP as a prior for spatial random effects in any spatial hierarchical model. The software spNNGP \citep{finley2020spnngp} implements hierarchical Bayesian spatial models through MCMC using the NNGP prior for the spatial random effects, offering massive computational improvements over other MCMC approaches for Bayesian spatial models. Even with the NNGP prior, the sequential sampling from a high-dimensional MCMC with $n$ latent random effects can still be slow for large datasets. \cite{finley2019efficient} proposed  marginalizing out the random effects, and use NNGP for the response. However, this {\em `response NNGP'} model loses the ability to infer on the spatial random effects which is often central to scientific insights on spatial variations of the outcome process that are not explained by the covariates. 

As an important alternative to MCMC, variational inference (VI) approximates the posterior distribution through optimization that requires reduced computation compared with MCMC sampling \citep{blei2017variational}. Variational inference aims to find a distribution within a specified {\em variational family} of distributions that minimizes the Kullback–Leibler divergence from the variational distribution to the posterior distribution. 

To our knowledge, \cite{ren2011variational} presented one of the first variational inference algorithms for geospatial mixed effects models. However, their algorithm does not scale to large datasets on account of using a full GP prior for the spatial random effects and also a Gaussian variation family with unstructured covariance matrix, both involving dense $n \times n$ matrices. Subsequent developments, including ours, have thus attempted to reduce the computational burden from both the prior and the variational distribution for the spatial random effects. \cite{wu2022variational} proposed the Variational Nearest Neighbor Gaussian Process (VNNGP) method that adopts NNGP prior but uses a {\em mean-field approximation} (MFA) for the variational distribution. MFA, while gaining speed, loses the ability to model posterior correlations among latent random variables, potentially leading to variance underestimation \citep{wang2005inadequacy, tan2018gaussian}. \cite{cao2023variational} developed a double-Kullback-Leibler-optimal Gaussian-process approximation (DKLGP), a method that utilizes both an NNGP prior and an NNGP-type variational family for the random effects, that parsimoniously model posterior correlations via a sparse Cholesky factor of the precision matrix, with the sparsity also determined by nearest neighbors. \cite{lee2024variational} proposed an Integrated Nonfactorized Variational Bayes (INFVB) for spatial generalized linear mixed models, modeling the spatial random effects using a low-rank GP specified via basis function expansions. These methods are reviewed in more details in Section \ref{sec:vireview}.  

We found that the evaluation of these existing fast spatial variational inference approaches has been inadequate. Comparisons to MCMC, if provided, only consider the na\"ive MCMC implementation for the full rank GP model, which is known to be prohibitively slow for large-scale Bayesian spatial analysis. There was no benchmarking against faster MCMC-based methods implemented in publicly available software, e.g., spNNGP for NNGP  \citep{finley2020spnngp} and spBayes for predictive processes \citep{finley2013spbayes}.
Our evaluations, detailed later, revealed that the default implementations of existing spatial variational inference methods fail to yield accurate results in posterior inference as obtained from spNNGP, while more optimized implementations were often slower than spNNGP. This is undesirable, as computational gains are the primary reason to opt for variational inference over MCMC. 

This article aims to develop better computational algorithms for variational inference on large spatial data. We present three algorithms of increasing complexity: spVB-MFA, spVB-MFA-LR, and spVB-NNGP. All three use NNGP priors for the spatial random effects, but differ in the choice of the variational families. spVB-NNGP uses a correlated Gaussian variational family with a nearest neighbor-based sparse Cholesky factor of the precision matrix.  spVB-MFA and spVB-MFA-LR use a mean-field approximation variational family. Additionally, spVB-MFA-LR applies a one-step {\em linear response correction} \citep{giordano2015linear} to spVB-MFA for improved estimation of posterior  covariance matrix, mitigating the well-known variance underestimation issue of mean-field approximation.  

Our proposed spVB-MFA and spVB-NNGP can be seen as respective analogs of the VNNGP \citep{wu2022variational} and DKLGP  \citep{cao2023variational}, deploying similar core methodological ideas. However, there are notable differences both in the scope and implementation. Our models allow covariates (i.e., fixed effects), enabling inference on the association of these with the outcome via the variational distribution of the regression coefficients. We also estimate variational distributions for the spatial variance and random error variance, as in a fully Bayesian practice. In comparison, current implementations of DKLGP and VNNGP do not allow covariates and provide only point estimates of the variance parameters. In terms of implementation, we deploy several computational innovations. First, we exploit the celebrated {\em reparameterization trick} and Monte Carlo simulation, resulting in closed-form gradients and speed up from a novel vanishing approximation for gradients. This stochastic method of constructing the gradient estimator accelerates computations beyond those of DKLGP, which relies on auto-differentiation. Additionally, by deploying the {\em calculus of variations}, we obtain conjugate and closed form variational distributions for the regression and variance parameters. These innovations result in significantly faster and more accurate algorithms compared to existing methods. Finally, for spVB-MFA, we show how the use of NNGP prior is naturally amenable to a linear response correction, resulting in the scalable spVB-MFA-LR algorithm with considerably improved inferential properties. 

Simulation studies demonstrate that spVB-NNGP and spVB-MFA-LR outperform the existing variational inference methods in terms of both accuracy and computational efficiency. spVB-MFA closely replicates the accuracy achieved by VNNGP, while offering substantial computational gains. Furthermore, both spVB-NNGP and spVB-MFA-LR produce inference on the fixed and random effects that align closely with those obtained from the MCMC-based spNNGP, but at reduced computational cost. We implement the proposed methods in a publicly available R-package \texttt{\href{https://github.com/jfsong100/spVarBayes}{spVarBayes}}. An accompanying \href{https://jfsong100.github.io/spVarBayes/spVarBayes-tutorial.html}{vignette} demonstrates its usage.

The remainder of this paper is organized as follows: In Section \ref{sec:rev}, we review related work. Section \ref{sec:method} presents our methodology and algorithm in detail. Section \ref{sec:sim} details the simulation experiments, comparing variational inference and MCMC-based  methods on accuracy and running time. Section \ref{sec:rwe} analyzes a large forest canopy height dataset using the proposed methods.  Section \ref{sec:discussion}  concludes the paper with a discussion.

\section{Review of Related Work}
\label{sec:rev}
\subsection{Bayesian Framework for Spatial Modeling}\label{sec:bayes}

Consider data collected at a set of locations $\calS = \{\bs_1, \bs_2, ..., \bs_n\}$ in a study region of interest $\mathcal{D}$. Let $y(\bs)$ and $\bX(\bs)$ respectively denote the response and the $p$-dimensional vector of covariates at location $\bs$.  A common way to model the response as a stochastic process over the entire domain is to use the spatial linear mixed effects model,
\begin{equation}\label{eq:splmprocess}
    y(\bs) = \bX(\bs)\bbeta + w(\bs) + \eps(\bs),
\end{equation}
where $w(\bs)$ is the spatial random effect and $\eps(\bs) \iid N(0,\tau^2)$ are the random errors. 
The conventional process specification for $w(\bs)$ is a {\em Gaussian process} (GP), such that $\bw = (w(\bs_1),...,w(\bs_n))^{T} \sim N(\boldsymbol{0},\bC)$, where $\bC:=\bC(\boldsymbol{\theta})=(C(\bs_i,\bs_j \given \btheta))$ is the $n \times n$ covariance matrix. Here $C(\cdot,\cdot \given \btheta)$ is the covariance function indexed by a set of spatial covariance parameters $\boldsymbol{\theta}$. The methods proposed in this article can be applied with any valid stationary covariance function $C$. However, for simplicity of presentation,  throughout, we use the popular exponential covariance function, expressed as $C(\bs_i,\bs_j|\sigma^2,\phi)= \sigma^2 \exp(-\phi ||\bs_i - \bs_j||)$, where $||\bs_i - \bs_j||$ is the Euclidean distance between locations $\bs_i$ and $\bs_j$. Thus the spatial covariance parameters are $\boldsymbol{\theta} = (\sigma^2, \phi)$. $\phi$ measures the rate of decay in spatial correlation with increasing distance, and $\sigma^2$ is the spatial marginal variance. 

For the full Bayesian specification, we also need to define prior distributions for $\bbeta$, $\tau^2$, and $\boldsymbol{\theta}$. Prior specifications tend to exploit conjugacy in the MCMC updates where possible. Typical choices include a flat prior for $\bbeta$, represented as $N(\boldsymbol{0},\boldsymbol{\Sigma_{\beta}})$, where $\boldsymbol{\Sigma_{\beta}} \to \infty$. Inverse Gamma priors for the variance parameters $\tau^2 \sim IG(a_{\tau}, b_{\tau})$ and $\sigs \sim IG(a_{\sigma}, b_{\sigma})$. For the remaining spatial parameters in $\btheta$, conjugacy is usually not attainable. 
We consider a Uniform prior for $\phi$, represented as $\text{Unif}(\alpha_{\phi}, \beta_{\phi})$. Let $y_i:=y(\bs_i)$, $\by :=(y_i, y_2,\ldots,y_n)^T$, and define $w_i$, $\bw$ and $x_i$, $\bX$ similarly. 
Given these priors for the parameters, and the GP prior for $\bw$, the posterior is: 
\begin{equation}\label{eq:hier}
\begin{split}
p(\bbeta, \boldsymbol{\theta}, \bw|\by, \bX)\, &\propto\, p(\by|\bbeta, \boldsymbol{\theta}, \bw, \bX) p(\bw) p(\bbeta) p(\boldsymbol{\theta}) \\
&\propto\, (\frac{1}{\tau^2})^{\frac{n}{2}} \exp\Big(-\frac{1}{2\tau^2}\sum_{i=1}^n (y_i - \bx_i^T\bbeta - w_i)^2\Big) \frac 1{\text{det}(\bC)^{\frac 12}}\exp(-\frac{1}{2}\bw^T\bC^{-1}\bw) \\
&\quad \times(\frac{1}{\tau^2})^{a_{\tau}+1}\exp(-\frac{b_{\tau}}{\tau^2})  (\frac{1}{\sigma^2})^{a_{\sigma}+1}\exp(-\frac{b_{\sigma}}{\sigma^2}) \mathbb{I}(\phi \in [\alpha_{\phi},\beta_{\phi}])\frac{1}{\beta_{\phi}-\alpha_{\phi}},
\end{split}
\end{equation}
where $\text{det}(\cdot)$ is the determinant and $\mathbb{I}(\cdot)$ is the indicator function.
Calculating the inverse and determinant of the prior covariance matrix $\bC$ requires $O(n^3)$ floating-point operations (flops) and $O(n^2)$ memory, rendering this model prohibitively slow for large sample sizes. 
The Nearest Neighbor Gaussian Process (NNGP) prior \citep{datta2016hierarchical} introduces sparsity in the prior precision matrix of $\bw$ to address the high computational costs associated with using a traditional GP prior. 
For every $\bs_i \in \calS$, define the neighbor set $N[\bs_1]=\{\}$, an empty set, and for $i \geq 2$, $N[\bs_i]$ as the $m$ nearest neighbors of $\bs_i$ in $\{\bs_1,\bs_2,...,\bs_{i-1}\}$, and $\bw_{N[\bs_i]}$ is a vector formed by the realizations of $w(\bs)$ over $N[\bs_i]$. Also, for any two finite sets $U,V \subseteq \calD$, let $C_{U,V}$ denote the scalar/vector/matrix with entries $C(u,v)$ for every pair of locations $(u,v) \in U \times V$ (for simplicity, we denote $C_{U \times U}$ by $C_U$). The NNGP prior is specified as
\begin{equation}\label{eq:NNGP_prior}
\begin{split}
    \tilde{p}(\bw) &= \prod_{i=1}^n p(w_i|\bw_{N[\bs_i]})=\prod_{i=1}^n N(w_i|\bb_{N[\bs_i]}\bw_{N[\bs_i]},\sigma^2 F_{\bs_i}),\\
    \bb_{N[\bs_i]} &= C_{\bs_i,N[\bs_i]}C^{-1}_{N[\bs_i]} , \,
    \sigma^2 F_{\bs_i} = C(\bs_i,\bs_i) - C_{\bs_i,N[\bs_i]}C^{-1}_{N[\bs_i]}C_{N[\bs_i],\bs_i}.
    \end{split}
\end{equation} 
Both $\bb_{N[\bs_i]}$ and $F_{\bs_i}$ are just functions of $\phi$. In an NNGP, $\bw$ follows a multivariate Gaussian distribution $\bw \sim N(\boldsymbol{0},\tilde{\bC})$, where $\tilde{\bC} = (\bI-\bB)^{-1}\bF(\bI-\bB)^{-T}$, $\bF = \sigma^2\mathrm{diag}(F_{\bs_1},F_{\bs_2}...,F_{\bs_n})$ and $\bB=(\bb_{ij})$ is a strictly lower-triangular matrix, with only $m$ non-zero entries per row, equaling the entries of $\bb_{N[\bs_i]}$. Consequently, the NNGP prior can be evaluated using only $O(n m^3)$ flops. We refer the readers to the review articles \citep{datta2016nearest,datta2022nearest} for more information on NNGP. For simplicity of notations, in the following sections, we will often use $i$ to refer to location $\bs_i$, e.g., $N[i]:=N[\bs_i]$, $\bb_{N[i]}:=\bb_{N[\bs_i]}$, $F_i:=F_{\bs_i}$. Also, note that while $\bb_{N[i]}$ and $F_i$ are functions of $\phi$, we suppress this notation, unless needed. 

\subsection{Review of Spatial Variational Inference}\label{sec:vireview}
We define $\boldsymbol{\psi}$ as the set of all parameters and latent variables: $\boldsymbol{\psi} = (\bbeta^T, \tau^2, \sigma^2, \phi, \bw^T)^T$, and  let $p(\by|\boldsymbol{\psi})$ denote the likelihood for the observed data $\by$. Under the Bayesian framework, let $p(\boldsymbol{\psi})$ denote the prior distribution and $p(\boldsymbol{\psi}|\by)$ denote the posterior distribution. The idea of variational inference is to seek a {\em variational distribution} or density $q^*(\boldsymbol{\psi})$ within a specified distribution family $Q$, that minimizes the Kullback-Leibler (KL) divergence from $q(\boldsymbol{\psi})$ to $p(\boldsymbol{\psi}|\by)$, denoted as $\text{KL}(q(\boldsymbol{\psi}) || p(\boldsymbol{\psi}|\by))$. Given the observed data $\by$, minimizing the KL divergence is equivalent to maximizing {\em evidence lower bound (ELBO)} 
\begin{equation}\label{eq:vi_elbo}
    q^{*}(\boldsymbol{\psi}) = \argmax_{q(\boldsymbol{\psi})\in Q} \mathcal{L}(q),\, \mbox{ where } \mathcal{L}(q)=E_{q}[\log p(\by \given \boldsymbol{\psi}) + \log p(\boldsymbol{\psi})-\log q(\boldsymbol{\psi})].
    \end{equation}
Here $E_{q}(\cdot)$ denotes the expectation with respect to $q(\boldsymbol{\psi})$. 

\cite{ren2011variational} presented one of the first implementations of variational inference for the hierarchical spatial linear mixed effects model of (\ref{eq:hier}). For the variational density $q(\boldsymbol{\psi})$, they considered a block independence structure 
$q(\boldsymbol{\psi}) = q(\bw) q(\bbeta)q(\tau^2)q(\sigma^2)q(\phi)$, thereby offering variational posterior distribution estimates for all parameters. The variational family $q(\bw)$ was chosen to be a multivariate Gaussian distribution with an unstructured covariance. However, the use of a full GP prior $p(\bw)=N(\bw \given \bzero, \bC)$ and the unstructured full rank Gaussian variational family for $\bw$, both involving dense $n \times n$ matrices, restrict the use of their methodology for large datasets, as both the log-prior $p(\bpsi)$ and the log-variational density $q(\bpsi)$ feature in the ELBO, and need to be evaluated during the variational optimization. \cite{wu2022variational} proposed replacing the full GP prior $p(\bw)$ with the  NNGP prior $\tilde p(\bw)$ in (\ref{eq:NNGP_prior}) which can be evaluated using linear time and memory. To simplify the computations involving the variational distribution, they consider a mean-field approximation (MFA), i.e., $q(\bw) = \prod_i q(w_i)$, with each $w_i$ modeled as an independent univariate Gaussian density. Although they discuss a richer variational family using {\em inducing points} leading to low-rank covariances, the available implementation of VNNGP is for the MFA version. mean-field approximations in variational inference are known to underestimate posterior variances \citep{wang2005inadequacy,tan2018gaussian}. Random effects for two nearby locations are likely to have spatial correlation in the posterior. \cite{cao2023variational} modeled this spatial correlation in the variational distribution $q(\bw)$ by using an NNGP type precision matrix for the variational distribution of $q(\bw)$, specified via a sparse Cholesky factor, with sparsity pattern dictated by nearest neighbors. Thus, both their prior and variational family for $\bw$ exploit NNGP-type approximations to mitigate computational issues, and their method is referred to as Double Kullback-Leibler GP (DKLGP). 

The methods of both \cite{wu2022variational} and \cite{cao2023variational} do not incorporate covariates (i.e., fixed effects), and are limited to providing variational distributions only for the spatial random effects. For the variance parameters only the point estimate was offered. Thus, the outputs of these approaches are limited in scope compared to \cite{ren2011variational} and typical Bayesian implementations of spatial mixed models. Additionally, they heavily rely on auto-differentiation for finding the variational parameters. Our experiments revealed that these algorithms can be suboptimal both in terms of accuracy and speed compared to Bayesian MCMC methods based on NNGP. \citet{lee2024variational} proposed an Integrated Nonfactorized Variational Bayes (INFVB) approach tailored specifically for spatial generalized linear mixed models. For Gaussian responses they use a full GP prior. For non-Gaussian responses, their method leverages a low-rank Gaussian Process representation of the spatial random effects via fixed basis function expansions, providing  scalability to large scale analysis. Empirically it has been observed that low-rank Gaussian process approximations can tend to oversmooth compared to NNGP \citep{datta2016hierarchical}.   

\section{Methodology}\label{sec:method}

Our proposed algorithm combines the strengths of the aforementioned contributions to spatial variational inference. It operates in the same general setup as \cite{ren2011variational}, accommodating covariates and offering variational posterior distributions for all variance parameters, while exploiting the numerous benefits of NNGP as priors and variational distributions, as in \cite{wu2022variational,cao2023variational}. Crucially, we deploy computational techniques that result in algorithms that are considerably faster and more accurate. The ELBO for the spatial mixed model using an NNGP prior can be expressed as 
\begin{align}
    \mathcal{L}(q)
     =& \mathbb{E}_q\Big[\log p(\by|\bX,\boldsymbol{\psi}) + \log p(\boldsymbol{\psi})\Big] - \mathbb{E}_q\Big[\log q(\boldsymbol{\psi})\Big] \nonumber\\
     =& \mathbb{E}_q\Big[\log p(\by|\bX,\bw, \bbeta,\tau^2,\sigma^2,\phi) + \log (\tilde{p}(\bw)p( \bbeta)p(\tau^2)p(\sigma^2)p(\phi))\Big] - \mathbb{E}_q\Big[\log q(\bw, \bbeta,\tau^2,\sigma^2,\phi)\Big]\nonumber\\
    =&\sum_{i=1}^n \mathbb{E}_q\Big[\frac{1}{2}\log \frac{1}{2\pi \tau^2} - \frac{1}{2\tau^2}(y_i-\bx_i^T\bbeta - w_i)^2\Big] +\sum_{i=1}^n \mathbb{E}_q\Big[\frac{1}{2}\log \frac{1}{2\pi \sigma^2 F_i} - \frac{1}{2\sigma^2F_i}(w_i - \bb_{N[i]}\bw_{N[i]})^2\Big] \nonumber\\
    &+\mathbb{E}_q\Big[(a_{\tau}+1)\log \frac{1}{\tau^2} - \frac{b_{\tau}}{\tau^2} +(a_{\sigma}+1)\log \frac{1}{\sigma^2} - \frac{b_{\sigma}}{\sigma^2} + 
    \log(\frac{1}{\beta_{\phi}-\alpha_{\phi}})\mathbb{I}(\phi \in (\alpha_{\phi},\beta_{\phi}))\Big] \nonumber\\
    & -\mathbb{E}_q\Big[\log(q(\bw, \bbeta,\tau^2,\sigma^2,\phi))\Big] + \text{constant} \label{ini_ELBO}. 
\end{align}
The following sections describe the key computational pieces and the algorithms. 

\subsection{Calculus of Variations}
\label{sec:others}

We consider a block independence structure $q(\bw, \bbeta,\tau^2,\sigma^2,\phi) = q(\bw)q( \bbeta)q(\tau^2)q(\sigma^2)q(\phi)$ for the variational distributions, as used in \cite{ren2011variational}. We use the calculus of variations \citep{gelfand1963calculus} to obtain variational solutions for regression coefficients $\bbeta$, spatial variance parameter $\sigma^2$, and the random error variance $\tau^2$ without having to make parametric assumptions about their variational families. We first briefly review the main idea. Let $\boldsymbol{\psi} =(\psi_1,\psi_2,...,\psi_k)$ to denote the entire collection of parameters with blocks $\psi_i$ such that the variational family $q = \prod_i q_i(\psi_i)$ is factorized over the blocks (i.e., block independent). By employing the calculus of variations, the ELBO is maximized when each variational factor is proportional to the exponential of the expected complete log joint density.  Formally, by solving Euler-Lagrange equations, the optimal variational density, $q^{*}_i(\psi_i)$, is $$q^{*}_i(\psi_i) \propto\exp\Big\{E_{-q(\psi_i)}\Big [\log p(\by,\boldsymbol{\psi}|\bX)\Big ]\Big\},$$ where $E_{-q(\psi_i)} \Big [\log p(\by,\boldsymbol{\psi}|\bX)\Big ]$ is the expectation of $\log p(\by,\boldsymbol{\psi} \given \bX)$ over other parameters leaving out $i^{\text{th}}$ block of the $\boldsymbol{\psi}$. 

Using the equation above, we can iteratively obtain  variational solutions $q^{*}(\bbeta)$,  $q^{*}({\tau^2})$, $q^{*}({\sigma^2})$, resulting in a conjugate variational distribution. Specifically, using the priors from (\ref{ini_ELBO}), if $q$ denotes the variational solution at the current iteration, then we obtain the following conjugate variational solutions $q^*$ for the next iteration:
\begin{equation*}
\begin{aligned}
    q^*(\bbeta) &= N\Big(({\bX^T\bX})^{-1}\bX^T(\by-\mathbb{E}_q[\bw]),\mathbb{E}_q[\frac{1}{\tau^2}]^{-1}({\bX^T\bX})^{-1})\Big),\\
    q^*({\tau^2}) &= IG\Big(a_{\tau}+\frac{n}{2}, b_{\tau}+\frac{1}{2}\Big [Tr(\bV_w)+p\mathbb{E}_q[\frac{1}{\tau^2}]^{-1} + (\by - \mathbb{E}_q[\bw])^T(\bI-\bH)(\by - \mathbb{E}_q[\bw])\Big ]\Big),\\
    q^*({\sigma^2}) &= IG\Big(a_{\sigma}+\frac{n}{2},b_{\sigma}+\frac{1}{2}\sum_{i=1}^n \mathbb{E}_q\Big [\frac{1}{F_i}(w_i - {\bb_{N[i]}\bw_{N[i]}})^2\Big ]\Big),
    \end{aligned}
\end{equation*}
where $\bH = \bX(\bX^T\bX)^{-1}\bX^T$, $\mathbb{E}_q[\bw]$ and $\bV_w$ are respectively the expectation and the covariance matrix of $\bw$ under $q$, and with $Tr$ denoting trace. The detailed derivations of each variational distribution can be found in Appendix \ref{appendix:calculus}. Thus, using calculus of variations, we get optimal variational distributions for these parameters without placing any restrictions on the variational distributions. \cite{ren2011variational} also adopted the calculus of variations for the spatial linear model. However, their use of full GP prior led to quadratic forms of dense $n \times n$ precision matrices in their conjugate updates. This requires $O(n^3)$ computation and is infeasible for large data. Our algorithm combines calculus of variations with the use of NNGP prior and variational distributions to facilitate computational scalability. 
The quadratic forms in our conjugate updates do not involve any $n \times n$ matrix.  

With our conjugate variational solutions for $\boldsymbol{\beta}$, $\tau^2$ and $\sigma^2$, we compute the terms needed for updating parameters in the next iteration: $\mathbb{E}_q[\boldsymbol{\beta}] = ({\bX^T\bX})^{-1}\bX^T(\by-\mathbb{E}_q[\bw])$, $\mathbb{E}_q[\frac{1}{\tau^2}] = \frac{a_{\tau}^{*}}{b_{\tau}^{*}}$, $\mathbb{E}_q[\frac{1}{\sigma^2}] = \frac{a_{\sigma}^{*}}{b_{\sigma}^{*}}$, where $a_{\tau}^{*}$, $a_{\sigma}^{*}$ are the updated shape parameters, $b_{\tau}^{*}$, $b_{\sigma}^{*}$ are the updated scale parameters of respectively $q^*({\tau^2})$ and $q^*({\sigma^2})$. Efficient computation of $Tr(\bV_w)$ and $\mathbb{E}_q\Big [\frac{1}{F_i}(w_i - {\bb_{N[i]}\bw_{N[i]}})^2\Big ]$ via Monte Carlo simulation is introduced later in (\ref{trace_cal}) and (\ref{eq:quadratic}).

For the spatial decay parameter $\phi$,  the solution from calculus of variations is not conjugate or analytically tractable and needs computationally expensive importance sampling approximations \citep{ren2011variational}.  
Instead, akin to VNNGP and DKLGP, we use a point-mass variational distribution for $\phi$ and optimize it via gradient ascent to maximize the ELBO:  $$\mathcal{L}(\phi) = \frac{1}{2}\sum_{i=1}^n \Bigl(\log \mathbb{E}_q[\frac{1}{\sigma^2}]
\frac{1}{F_i(\phi)} - \frac{1}{F_i(\phi)} \mathbb{E}_q[\frac{1}{\sigma^2}] \mathbb{E}_{q(\bw)}\Big[(w_i - {\bb_{N[i]}(\phi)\bw_{N[i]}})^2 \Big] \Bigl).$$ 
The update for $\phi$ is given by $\phi^{*} = \underset{\phi \in [\phi_{\min}, \phi_{\max}]}{\arg\max} \ \mathcal{L}(\phi)$. Subsequently, we determine instances of  $\bb_{N[i]}(\phi)$ and $F_i(\phi)$ used in updates of the other parameters and evaluate these by plugging in the latest $\phi^*$. 

\subsection{Random Effect Updates}
\label{sec:w}
Rather than assuming a full-rank variational covariance structure for $q(\bw)$ as in \cite{ren2011variational}, we incorporate nearest neighbor sparsity into the variational precision matrix, employing a technique similar to that used in the NNGP prior. Formally we consider the following family of variational density for $q$ which can be fully factorized into the product of a series of conditional distributions, i.e., 
\begin{align}
    q(\bw) &= \prod_{i=1}^n q(w_i|\bw_{N_q[i]}) =\prod_{i=1}^n N(w_i|\mu_{i}+{\ba_{N_q[i]}}{\bw_{N_q[i]}},d_i), \nonumber 
\end{align} where ${N_q[i]} \subseteq \{\bs_1,...,\bs_{i-1}\}$ is the directed neighbor set for $i$ in the variational distribution ($N_q[1]=\{\}$). We allow the size of the neighbor sets in the variational distribution to be different from that in the prior. This is because our experiments revealed that much smaller neighbor sets suffice for the variational distribution as we are also conditioning on the observed data. The term ${\ba_{N_q[i]}}$ captures how much the location $i$ is affected by its neighbors. With a simple parameterization, the joint distribution is given as
\begin{equation}\label{eq:var_w}
q(\bw) = N\Bigl((\bI-\bA)^{-1}\boldsymbol{\mu},(\bI-\bA)^{-1}\bD(\bI-\bA)^{-T}\Bigl) = N\Bigl(\boldsymbol{\eta},(\bI-\bA)^{-1}\bD(\bI-\bA)^{-T}\Bigl),
\end{equation}
with mean vector $\boldsymbol{\eta} = (\bI-\bA)^{-1}\boldsymbol{\mu}$, where $\boldsymbol{\mu} = (\mu_{1},...,\mu_{n})^T$. The matrix  $\bD$ is diagonal with entries $\bd = (d_1, d_2, ..., d_n)$ and ${\ba_{N_q[i]}}$ are the non-zero entries in the $i^{th}$ row of the strictly lower triangular matrix $\bA$. The precision matrix in the variational distribution enjoy sparsity akin to the NNGP prior distribution, and the sparse Cholesky factor is obtained as $(\bI-\bA)^{-1}\bD^{1/2}$. We denote the method with this choice of NNGP variational distribution as {\em spVB-NNGP} (spatial variational Bayes with NNGP). 

The parameters $\boeta$, $\bA$ and $\bD$ are all unknown that will be estimated as part of the variational optimization. 
From (\ref{ini_ELBO}), the ELBO for these parameters is 
\begin{align}
    \mathcal{L}(\{\ba_{N_q}\},\bd,\boldsymbol{\eta}) = \sum_{i=1}^n \mathbb{E}_q\Big[&
    -\frac{1}{2\tau^2}(y_i-{\bx_i}^T\bbeta - w_i)^2
    -\frac{1}{2\sigma^2F_i}(w_i - {\bb_{N[i]}
    \bw_{N[i]}})^2\Big] 
    + \frac{1}{2}\sum_{i=1}^n\log d_i + \frac{n}{2}. \nonumber
\end{align}

Writing $\bw = \boeta + \bu$, where $q(\bu)=N(\bu \given \bzero, (\bI-\bA)^{-1}\bD(\bI-\bA)^{-T})$, we can write 
\begin{equation}\label{eq:elbow}
\begin{split}
    \mathcal{L}(\{\ba_{N_q}\},\bd,\boldsymbol{\eta}) = \sum_{i=1}^n \mathbb{E}_q &
   \Big[ -\frac{1}{2\tau^2} (y_i-{\bx_i}^T\bbeta - \eta_i)^2 
    -\frac{1}{2\tau^2} u_i^2 
    -\frac{1}{2\sigma^2F_i}(\eta_{i} - {\bb_{N[i]}
    \boeta_{N[i]}})^2 \\
    & - \frac{1}{2\sigma^2F_i}(u_i - {\bb_{N[i]}
    \bu_{N[i]}})^2 \Big] 
    + \frac{1}{2} \sum_{i=1}^n \log d_i + \frac{n}{2}.
\end{split}
\end{equation}

Note that in above, the cross-terms, being linear in $\bu$, vanish as $\bu$ has zero mean under $q$. Details for deriving (\ref{eq:elbow}) are provided in the Appendix \ref{appendix:spVB-NNGP_w}.  This form of the ELBO facilitates closed-form gradient expressions for all the parameters of $q(\bw)$ either directly or via the reparametrization trick (introduced later in this section). 

\noindent \textbf{Closed-form gradient for $\boeta$.} Gradient ascent is used to update the parameters and maximize the ELBO. For the mean  $\boldsymbol{\eta}$, the expression in (\ref{eq:elbow}) and the block independence structure of $q$ enable us to directly compute the gradients of $\mathcal{L}(\{\ba_{N_q}\},\bd,\boldsymbol{\eta})$ with respect to $\boldsymbol{\eta}$ as 
\begin{align}
    \frac{\partial  \mathcal{L}(\{\ba_{N_q}\},\bd,\boldsymbol{\eta})}{\partial {\eta}_i} =&\mathbb{E}_q[\frac{1}{\tau^2}](y_i-\bx_i^T\mathbb{E}_q[\bbeta]-{\eta}_i)-\mathbb{E}_q[\frac{1}{\sigma^2}]\mathbb{E}_q[\frac{1}{F_i}]({\eta}_i - \mathbb{E}_q[\bb_{N[i]}]{\boldsymbol{\eta}_{N[i]}})\nonumber \\
    & + \mathbb{E}_q[\frac{1}{\sigma^2}]\sum_{l:i \in N[l]}\mathbb{E}_q [\frac{1}{F_l}]\Big(\mathbb{E}_q [b_{N[l],i}] {\eta}_l - \mathbb{E}_q [b_{N[l],i}\bb_{N[l]}]{\boldsymbol \eta}_{N[l]}\Big), \label{eq:NNGPVIupdate_mu}
\end{align}
where $b_{N[l],i}$ denotes the entry of $b_{N[l]}$ corresponding to the neighbor $i$ of location $l$. Calculation of the the expectations terms  $\mathbb{E}_q[\frac{1}{\tau^2}]$ and  $\mathbb{E}_q[\frac{1}{\sigma^2}]$ is already available using calculus of variations (see Section \ref{sec:others}). Similarly, the expectations $\mathbb{E}_q[\frac{1}{F_i}]$, $\mathbb{E}_q [b_{N[i]}]$ and $\mathbb{E}_q [b_{n[l],i}b_{N[i]}]$ are obtained by simply plugging in $\phi^*$ in these expressions. 

For the other variational parameters of $\bw$, i.e., $\{\ba_{N_q}\}$ and $\bd$, directly calculating the gradient of $\mathcal{L}(\{\ba_{N_q}\},\bd,\boldsymbol{\eta})$ can be complicated since the expectation (with respect to $q$) of the quadratic forms of $\bu$ in (\ref{eq:elbow}) involves the covariance matrix of $\bu$ (or $\bw$), i.e., we need to do the matrix inversion to get elements in ${\bV_w} = (\bI-\bA)^{-1}\bD(\bI-\bA)^{-T}$. As we discussed before, large matrix inversion can cause a big computational burden. In the following, we first discuss the computation of closed-form gradients based on the reparameterization trick and Monte Carlo simulation. Then, we propose using the vanishing approximations for the gradients to reduce computational costs. 

\noindent \textbf{Reparameterization trick.} 
 The key to obtaining gradients efficiently in variational inference is the use of the celebrated \emph{reparameterization trick} to calculate expectations of expressions under the variational distribution $q$ \citep{kingma2013auto,ong2018gaussian,xu2019variance,tran2020bayesian}. Broadly, the reparameterization trick is to express a variational distribution $q(\bu)$, characterized by a set of parameters $\blambda$, using a deterministic function: $\bu = h(\boldsymbol{\lambda},\boldsymbol \xi)$, where $\boldsymbol \xi$ is a random vector with a fixed density function $f_\bxi$ that is independent of the variational parameters $\boldsymbol{\lambda}$. This enables approximating the ELBO using Monte Carlo samples from $f_\bxi$. In our case, we first transform the $d_i$'s to the log-scale to, defining $\gamma_i = \frac{1}{2}\log(d_i)$ to be supported on the real line. Then we  
 rewrite $\bu = h(\blambda,\bxi) := (\bI - \bA)^{-1} \boldsymbol D^{1/2} \boldsymbol \xi$, where $\boldsymbol \xi \sim N(\boldsymbol{0}, \bI_n)$, $\boldsymbol{\lambda} = (\{\ba_{N_q}\},\bgamma)$ and $\bgamma=(\gamma_1,\ldots,\gamma_n)$. 
 Ignoring the terms in (\ref{eq:elbow}) not involving $\blambda$, we have
 \begin{equation}\label{eq:elbou}
\begin{split}
    \mathcal{L}(\blambda) = \mathcal L_1(\blambda) 
    + \sum_{i=1}^n \gamma_i, \mbox{ where } \mathcal L_1(\blambda):=\sum_{i=1}^n \mathbb{E}_{f_\bxi} &
   \Big[ 
    -\frac{1}{2\tau^2} u_i^2 
    -\frac{1}{2\sigma^2F_i}(u_i - {\bb_{N[i]}
    \bu_{N[i]}})^2 \Big]. 
\end{split}
\end{equation}

Then a Monte Carlo estimate $\widehat{\mathcal L_1 (\lambda)}$ is obtained from generating $\bxi^{(j)} \iid N(\bzero,\bI_n)$ for $j =1 ,\ldots,N_{MC}$, and evaluating $\mathcal L_1 (\lambda)$ in (\ref{eq:elbou}) using $\bxi=\bxi^{(j)}$ and 
averaging over the $N_{MC}$ values. However, the computational challenge for this step is to solve the linear system $\bu = (\bI - \bA)^{-1} \boldsymbol D^{1/2} \boldsymbol \xi$, in order to express the quadratic forms of $\bu$ in $\mathcal L_1(\blambda)$ explicitly in terms of $\bxi$ and $\blam$. We use a sparse back-solve to avoid brute-force inversion of the matrix $(\bI - \bA)^{-1}$.  
Letting $\{\ba_{N_q}\} = ({\ba_{N_q[2]}},...,{\ba_{N_q[n]}})$, the back-solve can proceed as follows:
\begin{equation}\label{eq:trick}
\begin{split}
    u_1&=\exp(\gamma_1) \xi_1 \\
    u_2&=\exp(\gamma_2) \xi_2 + {\ba_{N_q[2]}}{\bu_{N_q[2]}}\\
    \vdots\\
    u_n&=\exp(\gamma_n) \xi_n + {\ba_{N_q[n]}}{\bu_{N_q[n]}}.
\end{split}
\end{equation}
 
As $(\bI - \bA)$ is the lower triangular matrix with at most $m_q$ non-zero entries for each row, the back-solve (\ref{eq:trick}) requires $O(nm_q)$ total flops. 
 
Denoting the solution as $\bu^{(j)} := h(\blambda,\bxi^{(j)})$, and using the chain rule we have 
\begin{align*}
    \widehat{\nabla_{\boldsymbol \lambda} \mathcal{L}_1(\boldsymbol{\lambda})} &= \frac{1}{N_{MC}}\sum_{j=1}^{N_{MC}} \nabla_{\boldsymbol  \lambda} h(\blambda,\bxi^{(j)}) \nabla_{\boldsymbol u^{(j)}}  \widehat{\mathcal{L}_1(\boldsymbol{\lambda})}.
\end{align*}

The $i^{\text{th}}$ element of $\nabla_{\boldsymbol u^{(j)}}  \widehat{\mathcal{L}_1(\boldsymbol{\lambda})}$ is obtained in closed-form as:

\begin{align*}
        \frac{\partial \widehat{\mathcal{L}_1(\boldsymbol{\lambda})}}{\partial u_i^{(j)}} = \frac{1}{N_{MC}}\sum_{j=1}^{N_{MC}}\Big[&-\mathbb{E}_q[\frac{1}{\tau^2}]  u_i^{(j)} -\mathbb{E}_q[\frac{1}{\sigma^2}]\mathbb{E}_q[\frac{1}{F_i}]\Big(u_i^{(j)} - \mathbb E_{q}[\bb_{N[i]}]\bu_{N[i]}^{(j)}\Big)\\
        &+\mathbb{E}_q[\frac{1}{\sigma^2}]\sum_{l:i \in N[l]} \mathbb E_q[\frac{1}{F_l}] \Big(  \mathbb E_q[b_{N[l],i}]u_l^{(j)} - \mathbb E_q [b_{N[l],i}\bb_{N[l]}]\boldsymbol{u}^{(j)}_{N[l]}\Big)\Big]. 
    \end{align*}

For the other term $\nabla_{\boldsymbol  \lambda} h(\blambda,\bxi^{(j)})$ in the gradient, we use the following strategy. 

\noindent \textbf{Vanishing gradients.} We use $\nabla_{\boldsymbol \gamma} h(\blambda,\bxi)$ as the example to illustrate the calculation of $\nabla_{\boldsymbol  \lambda} h(\blambda,\bxi^{(j)})$. Note that $\bu = h(\blambda,\bxi)$ is an $n \times 1$ vector. The gradient of $\bu$ with respect to $\boldsymbol{\gamma}$, another $n \times 1$ vector yields an $n \times n$ matrix $\nabla_{\boldsymbol \gamma} h(\blambda,\bxi)$ whose $(k,i)^{th}$ element is $(\frac{\partial u_k}{\partial \gamma_i})$. 

First, we can observe from (\ref{eq:trick}) that this is a lower triangular matrix since $u_k$ does not involve $\gamma_i$ when $k<i$. The first non-zero entry for $i^{\text{th}}$ column is shown in the diagonal, i.e., $\frac{\partial u_i}{\partial \gamma_i} = \exp(\gamma_i)\xi_i$. For $k > i$, we use chain rule to get $$\frac{\partial u_k}{\partial \gamma_i} = \frac{\partial u_{k}}{\partial u_{k-1}}\frac{\partial u_{k-1}}{\partial u_{k-2}}\hdots\frac{\partial u_{i}}{\partial \gamma_i} = \Big[\prod_{j=i+1}^k \frac{\partial u_{j}}{\partial u_{j-1}}\Big]\frac{\partial u_{i}}{\partial \gamma_i} = \Big[\prod_{j=i+1}^k \frac{\partial u_{j}}{\partial u_{j-1}}\Big]\exp(\gamma_i)\xi_i.$$ For $i = 1,2,...,n$ columns, we need $\frac{1}{2}(n^2+n)$ entries for storage and calculations involving this matrix. As $k$ is far away from $i$, the cumulative product can be small (vanishing gradient) or extremely large (exploding gradient) if $i$ is not a direct neighbor of $k$, i.e., $i \notin N_q[k]$. To avoid expensive calculations of such gradient terms, we further introduce a first-order "vanishing" gradient approximation where we only preserve the non-zero entries in the gradient matrix if $i$ is directly a neighbor of $k$. In that case, from (\ref{eq:trick}), using $u_k = \exp(\gamma_k)\xi_k + a_{k,i}u_i + \sum_{j\neq i, j \in N_q[k]} a_{k,j} u_j$ we get $\partial u_k/\partial \gamma_i = a_{k,i}\partial u_i/\partial \gamma_i + \sum_{j\neq i, j \in N_q[k]} a_{k,j} (a_{j,i} \partial u_i/\partial \gamma_i + \ldots)$. Here $a_{k,i}$ is the element in the ${\ba_{N_q[k]}}$ that corresponding to neighbor $i$ of location $k$. Ignoring the second- and higher-order terms, we thus have the following vanishing-gradient approximation. 

$$\frac{\partial h(\blambda,\bxi)_k}{\partial \gamma_i} = \frac{\partial {u}_k}{\partial \gamma_i} \approx \begin{cases}
0 \ \ &\text{if $k<i$},\\
\exp(\gamma_i)\xi_i \ \ &\text{if $k=i$},\\
\mathbb{I}(i \in N_q[k]) a_{k,i} \exp(\gamma_i)\xi_i \ \ &\text{if $k>i$}.
\end{cases}$$
Similarly, we get the first-order "vanishing" gradient approximations for $\{\ba_{N_q}\}$ as \[\frac{\partial h(\blambda,\bxi)_k}{\partial {\ba_{N_q[i]}}} = \frac{\partial {u}_k}{\partial {\ba_{N_q[i]}}} \approx \begin{cases}
{\bu_{N_q[i]}}  \ \ &\text{$i = k$},\\
0 \ \ &\text{otherwise}.
\end{cases}\]

By applying these "vanishing" approximated gradients, the gradients of $\widehat{\nabla \mathcal{L}(\boldsymbol{\lambda})}$ for $\boldsymbol \gamma$ and $\{\ba_{N_q}\}$ are as follows:
\begin{align}
    \widehat{\nabla_{\boldsymbol \gamma} \mathcal{L}(\boldsymbol{\lambda})} &= \frac{1}{N_{MC}}\sum_{j=1}^{N_{MC}} \nabla_{\boldsymbol  \gamma} h(\blambda,\bxi^{(j)}) \nabla_{\boldsymbol u^{(j)}} \widehat{\mathcal{L}_1(\boldsymbol{\lambda})}\label{eq:nngpgamma_grad} + \bone,\\
    \widehat{\nabla_{\{\ba_{N_q}\}} \mathcal{L}(\boldsymbol{\lambda})} &= \frac{1}{N_{MC}}\sum_{j=1}^{N_{MC}} \nabla_{\{\ba_{N_q}\}} h(\blambda,\bxi^{(j)})\nabla_{\boldsymbol u^{(j)}}  \widehat{\mathcal{L}_1(\boldsymbol{\lambda})}.\label{eq:nngpA_grad}
\end{align}
Here, the additional vector of ones appears in the gradient with respect to $\bgamma$ due to the $\sum \gamma_i$ term in $\calL(\blambda)$ in (\ref{eq:elbou}).

\noindent \textbf {Trace and quadratic terms calculation:} Another benefit of the reparameterization trick is that it reduces the computational complexity of estimating the trace term $Tr(\bV_w)$ which features in the conjugate update $q^*(\tau^2)$ in Section \ref{sec:others}. 
We have: 
\begin{equation}\label{trace_cal}
\begin{aligned}
    Tr(\bV_w) &= Tr(\mbox{Cov}_q(\bw)) =  Tr(\mbox{Cov}_q(\bu)) = Tr(\mathbb{E}[\bu\bu^T]) \\
    &\approx Tr(\frac{1}{N_{MC}}\sum_{k=1}^{N_{MC}}\bu^{(k)}\bu^{(k),T}) 
    =\frac{1}{N_{MC}}\sum_{k=1}^{N_{MC}} Tr(\bu^{(k),T}\bu^{(k)}) = \frac{1}{N_{MC}}\sum_{k=1}^{N_{MC}} ||\bu||^2,
    \end{aligned}
\end{equation}
where $\bu^{(1)}$, ..., $\bu^{(N_{MC})} \iid q(\bu)$ are the Monte Carlo samples. Since generating each sample $\bu$ requires $O(nm_q)$ operations, as discussed after (\ref{eq:trick}), the total cost for estimating the trace is $O(nm_qN_{MC})$, yielding considerable computational gain as in practice $N_{MC} \ll n$. 

We also estimate the quadratic forms of $\bu$ using the Monte Carlo samples, e.g., 
\begin{equation}
\begin{split}
    \mathbb{E}_q\Big [\frac{1}{F_i}(w_i - {\bb_{N[i]}\bw_{N[i]}})^2\Big ] & \approx  \mathbb{E}_q[\frac{1}{F_i}](\eta_{w_i} - \mathbb{E}_q[\bb_{N[i]}]\boldsymbol{\eta}_{w_{N[i]}})^2 + \\
    & \mathbb{E}_q[\frac{1}{F_i}] \frac{1}{N_{MC}}\sum_{j=1}^{N_{MC}} (u_i^{(j)} -  \mathbb{E}_q[\bb_{N[i]}] \bu_{N[i]}^{(j)})^2. \label{eq:quadratic}
    \end{split}
\end{equation}

The algorithm for spVB-NNGP with full batch size is shown in Algorithm \ref{alg:NNGP}. From the expression of the ELBO in (\ref{eq:elbow}), it is evident that our algorithm can be naturally extended to a minibatch version, resulting in a doubly stochastic variational inference procedure. Implementation details, including learning rate schedules, stopping criteria, and the minibatch extension, are provided in the Appendix \ref{appendix:al_implement}. We also investigate the choice of the number of nearest neighbors used in both the prior and the variational family. While the NNGP prior typically requires 15–20 neighbors to adequately capture the underlying spatial dependence, the posterior can often be well-approximated using fewer neighbors (e.g., 3–5). This is because the posterior, conditioning on the data, leverages both the prior and the information from the likelihood, allowing accurate inference even with a smaller set of nearest neighbors.

\begin{algorithm}[!t]
\footnotesize
\caption{spVB-NNGP: NNGP based Variational Approximation.}\label{alg:NNGP}
\begin{algorithmic}
\State Specify the value of the hyper-parameters in the prior distribution for $\tau^2 \sim IG(a_{\tau},b_{\tau})$, $\sigma^2 \sim IG(a_{\sigma},b_{\sigma})$.
\State Assign initial values to $\mathbb{E}_q[\phi]^{(0)}$, $\mathbb{E}_q^{(0)}[\frac{1}{\tau^2}]$, $\mathbb{E}_q^{(0)}[\frac{1}{\sigma^2}]$, $\boldsymbol{\eta}^{(0)}$, $\bd^{(0)}$ and  $\{\ba_{N_q}^{(0)}\}$. 
\State Assign values for $\phi_{min}$, $\phi_{max}$, $N_{MC}$, AdaDelta algorithm noise $\delta$ and input rate $r$.
\For{$t=1$ to $T$} 
\State \textbf{Step 1:} Update the distribution of $\bbeta \sim N(\boldsymbol{\mu_\beta}^{(t)},\bV_{\boldsymbol\beta}^{(t)})$ where 

$\bV_{\boldsymbol \beta}^{(t)} = \Big [\mathbb{E}_q^{(t-1)}(\frac{1}{\tau^2})\Big ]^{-1} (\bX^T\bX)^{-1}$, $\boldsymbol \mu_{\boldsymbol \beta}^{(t)} = (\bX^T\bX)^{-1} \bX^T(\by-\boldsymbol \eta^{(t-1)})$.

\State \textbf{Step 2:} Update the distribution of 
$\tau^2 \sim IG$ with $a_{\tau}^{*(t)} = a_\tau +\frac{n}{2}$ and

$b_{\tau}^{*(t)} = b_{\tau}+\frac{1}{2}\Big[Tr(\bV_w^{(t-1)})+p\{\mathbb{E}_q^{(t-1)}[\frac{1}{\tau^2}]\}^{-1}+(\by-\boldsymbol{\eta}^{(t-1)})^T(\bI-\bH)(\by-\boldsymbol{\eta}^{(t-1)})\Big]$
, 

where $\bH = \bX(\bX^T\bX)^{-1} \bX^T$ and the trace is approximated by (\ref{trace_cal}).

\State Calculate $\mathbb{E}_q^{(t)}[\frac{1}{\tau^2}] = \frac{a_{\tau}^{*(t)}}{b_{\tau}^{*(t)}}$.
\State \textbf{Step 3:} Update $\sigma^2 \sim IG$ with parameters $a_{\sigma}^{*(t)} = a_{\sigma} + \frac{n}{2}$ and

$b_{\sigma}^{*(t)} = b_{\sigma}+\frac{1}{2}\sum_{i=1}^n\mathbb{E}_q^{(t-1)}\Big[\frac{1}{F_i}(w_i - \bb_{N[i]}\bw_{N[i]})^2\Big]$, where  $\mathbb{E}^{(t-1)}_q[\frac{1}{F_i}(w_i - \bb_{N[i]}\bw_{N[i]})^2]$ is approximated by (\ref{eq:quadratic}).

\State Calculate $\mathbb{E}_q^{(t)}[\frac{1}{\sigma^2}] = \frac{a_{\sigma}^{*(t)}}{b_{\sigma}^{*(t)}}$.
\State \textbf{Step 4:} 
Update $\phi$ with a degenerate (single point-mass) distribution.

    \State Calculate numerical gradient for $\nabla_{\phi} \mathcal{L}(\phi)$, 
    
    where $\mathcal{L}(\phi) = \frac{1}{2}\sum_{i=1}^n \Bigl(\log \mathbb{E}^{(t)}_q[\frac{1}{\sigma^2}]
\frac{1}{F_i(\phi)} - \frac{1}{F_i(\phi)} \mathbb{E}^{(t)}_q[\frac{1}{\sigma^2}] \mathbb{E}_{q(\bw)}^{(t-1)}[(w_i - {\bb_{N[i]}(\phi)\bw_{N[i]}})^2 ] \Bigl).$ 

    \State Using AdaDelta optimizer in (\ref{eq:adadelta}) to adapt the learning rate and get $\Delta_{\phi}^{(t)}$.
 \State Update $  \phi^{(t)} = \phi^{(t-1)} + \Delta_{\phi}^{(t)}$.

\State Update $\mathbb{E}_q^{(t)}[\frac{1}{F_i}] = \frac{1}{F_i(\phi^{(t)})}$ and $\mathbb{E}_q^{(t)}[\bb_{N[i]}]=\bb_{N[i]}(\phi^{(t)})$ using (\ref{eq:NNGP_prior}). 

\State \textbf{Step 5:} Update the distribution of $\bw$ with parameters  $\boldsymbol{\eta}$, $\bd = (d_1,d_2,...,d_n)$, $\{\ba_{N_q}\} = (\ba_{N_q[2]},...,\ba_{N_q[n]})$.

 \State Compute gradient $\nabla_{  \boldsymbol{\eta}} \mathcal{L}(\{\ba_{N_q}\},\bd,\boldsymbol{\eta})$ using (\ref{eq:NNGPVIupdate_mu}).
 \State Using AdaDelta optimizer in (\ref{eq:adadelta}) to adapt the learning rate and get $\Delta_{  \boldsymbol{\eta}}^{(t)}$.
 \State Update ${  \boldsymbol{\eta}}^{(t)} = {  \boldsymbol{\eta}}^{(t-1)} + \Delta_{  \boldsymbol{\eta}}^{(t)}$.

 \State Compute vanishing gradient $\widehat{\nabla_{\boldsymbol{\gamma}} \mathcal{L}}(\{\ba_{N_q}\},\bd,\boldsymbol{\eta})$ use (\ref{eq:nngpgamma_grad}).
 \State Using AdaDelta optimizer in (\ref{eq:adadelta}) to adapt the learning rate and get $\Delta_{  \boldsymbol{\gamma}}^{(t)}$.
 \State Update $ \boldsymbol{\gamma}^{(t)} = \boldsymbol{\gamma}^{(t-1)} + \Delta_{ \boldsymbol{\gamma}}^{(t)}$.
 \State Update $  \bd^{(t)} = \exp(\boldsymbol{\gamma}^{(t)})^2$.

 \State Compute vanishing gradient $\widehat{\nabla_{\{\ba_{N_q}\}} \mathcal{L}}(\{\ba_{N_q}\},\bd,\boldsymbol{\eta})$ use (\ref{eq:nngpA_grad}).
 \State Using AdaDelta optimizer in (\ref{eq:adadelta}) to adapt the learning rate and get $\Delta_{  \{\ba_{N_q}\}}^{(t)}$.
 \State Update $ \{\ba_{N_q}\}^{(t)} = \{\ba_{N_q}\}^{(t-1)} + \Delta_{\{\ba_{N_q}\}}^{(t)}$.
\State \textbf{Early Stop regularization:} Terminate if ELBO does not improve for $T_{\text{patience}}$ consecutive iterations
\EndFor
\end{algorithmic}
\end{algorithm}

\subsection{Joint Variational Distribution for Fixed and Random Effects}\label{sec:joint}

Since spVB-NNGP assumes an independent block for regression coefficients $\bbeta$ and spatial random effects $\bw$, it may fail to capture the correlations between these parameters, potentially leading to underestimation of posterior variances. To address this, we offer an extension to a joint variational model 
$q(\bbeta,\bw) = N\Big ((\boldsymbol{\mu_\beta}^T, \boldsymbol{\eta}^T)^T
, (\bI-\bA^{*})^{-1}\bD^{*}(\bI-\bA^{*})^{-T}\Big )$. 

In the joint modeling approach, which we refer to as {\em spVB-NNGP-joint}, we extend the neighbor sets of each $w_i$ to include the regression coefficients $\boldsymbol{\beta}$, enabling the variational distribution $q(\bbeta,\bw)$ to capture both the covariance within each parameter and also the correlation between them. This joint structure retains sparsity pattern in the spatial component while introducing correlation between $\boldsymbol{\beta}$ and $\bw$, allowing efficient sampling and gradient computation. Details can be found in the Appendix \ref{appendix:spVB-NNGP-joint}.

\noindent \textbf{Detailed comaprision with DKLGP.} As mentioned earlier, the core methodological idea of spVB-NNGP is similar to that of DKLGP \citep{cao2023variational}. Both introduce nearest neighbor based models when specifying both the prior and the variational family for the latent spatial random effects $\bw$. Hence, we want to clarify the difference between the two methods in more details, as enumerated in Table \ref{tab:compareDKL}. From a model setting perspective, spVB-NNGP is more general in scope as it includes covariates $\bX$, a feature not currently present in DKLGP. Regarding the variational distribution, DKLGP provides only point estimations for the variance parameters (spatial variance $\sigma^2$ and random error variance $\tau^2$). In contrast, spVB-NNGP computes variational posterior distributions for them. In fact, spVB-NNGP requires no apriori restriction or parametrization of the variational distribution of the regression coefficients $\bbeta$ and variance parameters ($\sigma^2$, $\tau^2$), using calculus of variations to obtain the optimal solution. Also, DKLGP relies on auto-differentiation for gradient calculations, while spVB-NNGP utilizes the reparameterization trick, Monte Carlo simulation, and vanishing gradients to get closed-form gradient updates and trace calculations as reduced computational costs. DKLGP calculates the trace directly, which leads to a significant computational burden as the sample size increases.

\begin{table}[]
\caption{Comparison for spVB-NNGP and DKLGP methods}\label{tab:compareDKL}
\scriptsize
\begin{tabular}{p{0.29\linewidth} | p{0.31\linewidth} | p{0.31\linewidth}}
\hline
 & spVB-NNGP & DKLGP \\ \hline
Model setting & Include covariates (fixed effects) & Does not include covariates \\
Initial values for $\bw$ & Linear model residuals & Laplace Approximation \\
Prior for spatial random effects  $\bw$ & NNGP & NNGP \\
Update for $\bw$ & Gradient ascent (combines closed form terms, reparametrization trick, Monte Carlo sampling, and vanishing gradients) & Auto-differentiation \\
Calculating trace of $\mbox{Cov}_q(\bw)$ & Monte Carlo simulation (fast) & Exact trace (slow) \\
Estimation for regression  
and variance parameters & Posterior distributions allowing uncertainty quantification & Point estimations for variance parameters (no regression parameters) \\
Updates of regression and variance parameters & Closed-form distributions (using calculus of variations) & Auto-differentiation \\
Stopping Criteria & Fixed iterations or change in ELBO & Fixed iterations (default implementation) \\ \hline
\end{tabular}
\end{table}

\subsection{Mean-Field Approximation}
\label{sec:mfa}
We also design a considerably faster algorithm spVB-MFA, which employs an NNGP prior for $\bw$ but with a mean-field approximation (MFA) as the variational distribution for $\bw$, i.e., \begin{align*}
    q(\bw)=\prod_{i=1}^n N(w_i \given \mu_{w_i}, G_i).
\end{align*} 
The parameter update strategies are similar as for spVB-NNGP. 
The methodological idea of spVB-MFA is thus similar to the Variational NNGP (VNNGP) of \cite{wu2022variational}, with the key differences being that spVB-MFA accommodates covariates, providing posterior distributions of variance parameters, and being implemented via an algorithm with fundamentally different computational strategies (similar to the differences between spVB-NNGP and DKLGP). The detailed algorithm for spVB-MFA can be found in Appendix \ref{appendix:MFA} Algorithm \ref{alg:MFA}. As revealed in our experiments later, spVB-MFA stands out in terms of computation speed, and offers very accurate posterior mean estimates. However, posterior variance estimates are expectedly poor, on account of the well-known variance underestimation issue of mean-field approximations. 
spVB-MFA is thus an algorithm for fast posterior mean estimation for very large-scale spatial analysis.

\subsection{spVB-MFA with Linear Response Correction}

Linear response methods, originally developed in statistical physics, have been applied to variational inference \citep{giordano2015linear, giordano2018covariances} to address the known drawbacks of MFA in underestimating the uncertainty of parameters. 
We incorporate this idea into spVB-MFA to improve the estimation of posterior variance. We first briefly review the linear response correction method. The method assumes that variational distribution $q$ belongs to the exponential family with natural parameters $\boldsymbol{\eta}_{LR}$, sufficient statistics $\boldsymbol{\theta}$ and log partition function $\bA_{LR}$, i.e., 
    $q(\boldsymbol{\theta}) \propto \exp\{ \boldsymbol{\eta}_{LR}^T \boldsymbol{\theta} - A_{LR}(\boldsymbol{\eta}_{LR})\}$.

Let $\bV = \mbox{Cov}_q(\boldsymbol{\theta})$ denote the covariance matrix of the sufficient statistics under the variational distribution, and let $M:= \mathbb{E}_q[\boldsymbol{\theta}]$ denote the corresponding mean. The goal is to estimate the true posterior covariance matrix $\bSigma = \mbox{Cov}_{p(\boldsymbol{\theta}|\by,\bX)}(\boldsymbol{\theta})$. The linear response approach corrects the the variational covariance $\bV$ by accounting for the second-order behavior of the variational objective at its optimum:
\begin{align*}
    \widehat{\bSigma} = (\bI-\bV\bH)^{-1} \bV,
\end{align*}
where $\bH = \frac{\partial^2 \bL}{\partial M\partial M^T}$ is the Hessian matrix of the expected log posterior $\bL = \mathbb{E}_q\Big[\log p(\boldsymbol{\theta}|\by, \bX)\Big]$.

A key assumption for the linear response correction to be valid is accurate estimation of the posterior means of the parameters. In spatial settings, mean-field approximations typically yield biased estimates for random error variance ($\tau^2$) and spatial covariance parameters ($\sigma^2$,$\phi$), ruling out applicability of this correction for them. Hence, we apply the correction only to the regression coefficients $\bbeta$ and spatial random effects $\bw$, using  initial estimates of the spatial covariance parameters from the BRISC R-package \citep{saha2018brisc}. 

For our settings, under the mean-field approximation, the variational distribution factorizes as $q(\bbeta, \bw) = \prod_{j=1}^p q(\beta_j) \prod_{i=1}^n q(w_i)$, with $q(\beta_j) = N({\mu}_{\beta_j}, \sigma^2_{\beta_j})$ and $q(w_i) = N(\mu_{w_i}, G_i)$ for $j = 1, \dots, p$,  $ i = 1, \dots, n$.  
This is an exponential family with the sufficient statistics including linear and quadratic terms involving $\bbeta$ and $\bw$: $\btheta =(\{\beta_j\},\{\beta_j^2\},\{w_i\},\{w_i^2\})$. 

There are four main steps for linear response correction: (1) finding the optimal solution $q^*$; (2) computing $\bV = \mbox{Cov}_q^*(\boldsymbol{\theta})$; (3) computing the Hessian matrix of the expected log posterior; (4) computing the corrected covariance matrix. In our case, we focus on the covariance of subvector $\boldsymbol{\alpha} = \{\beta_1,...,\beta_p,w_1,...,w_n\} \subset \btheta$, and let $\bz$ denote the remaining terms. Using properties of the Gaussian distribution, we show that the covariance of $\boldsymbol{\alpha}$ can be computed using only the relevant block of the full matrices: $$\widehat{\bSigma}_{\boldsymbol{\alpha}} = (\bI_{\boldsymbol{\alpha}} - \bV_{\boldsymbol{\alpha}} \bH_{\boldsymbol{\alpha}})^{-1} \bV_{\boldsymbol{\alpha}}.$$ This simplification avoids inverting the full matrix and improves efficiency by limiting computation to the subset of interest. While detailed derivations are shown in the Appendix~\ref{appendix:LRVB}, we summarize below the key steps for scalability. 

The covariance matrix $\bV$ can be directly obtained after finding the optimal solution $q^*$, the top block structure of $\bV_{\boldsymbol{\alpha}}$ is
\begin{align*}
    \bV_{\boldsymbol{\alpha}} = \begin{bmatrix}
\mathrm{diag}(\sigma^2_{\beta_1}, \dots, \sigma^2_{\beta_p}) & \mathbf{0} \\
\mathbf{0} & \mathrm{diag}(G_1, \dots,G_n)
\end{bmatrix},
\end{align*}
where $\sigma^2_{\beta_1}, \dots, \sigma^2_{\beta_p}$ and the spatial random effect variances $G_1, \dots, G_n$ are obtained from the variational posterior. The expected log posterior under the variational distribution is given by 
    $\bL = \mathbb{E}_q[\log p(\boldsymbol{\theta} | \by, \bX)]$
which expands as:
\begin{align*}
\bL 
&= -\frac{1}{2\tau^2} \sum_{i=1}^n 
\Bigg[ 
      \sum_{j} x_{ij}^2 \, \mathbb{E}_q[\beta_j^2]
    + \sum_{\substack{j,k \\ j \ne k}} x_{ij} x_{ik} \, \mathbb{E}_q[\beta_j] \mathbb{E}_q[\beta_k] 
    - 2 y_i \sum_{j} x_{ij} \mathbb{E}_q[\beta_j]
    + 2 \mathbb{E}_q[w_i] \sum_{j} x_{ij} \mathbb{E}_q[\beta_j] \\
&\phantom{++++++}\quad 
    + y_i^2 
    - 2 y_i \mathbb{E}_q[w_i]
    + \mathbb{E}_q[w_i^2]
\Bigg] \\
&\quad - \frac{1}{2\sigma^2} \sum_{i=1}^n \frac{1}{F_i} 
\Bigg[ 
      \mathbb{E}_q[w_i^2]
    - 2 \sum_{j \in N[i]} b_{ij} \mathbb{E}_q[w_i] \mathbb{E}_q[w_j] 
    + \sum_{j \in N[i]} b_{ij}^2 \, \mathbb{E}_q[w_j^2] 
    + \sum_{\substack{j, k \in N[i] \\ j \ne k}} b_{ij} b_{ik} \, \mathbb{E}_q[w_j] \mathbb{E}_q[w_k]
\Bigg].
\end{align*}

Given the expected log posterior, the top block of the Hessian matrix $\bH_{\boldsymbol{\alpha}}$ is given as 
\begin{align*}
\bH_{\boldsymbol{\alpha}} = \begin{bmatrix}
\left[\frac{\partial^2\bL}
     {\partial \mathbb{E}_q[\beta_j]\,\partial \mathbb{E}_q[\beta_k]}\right]       & -\frac{1}{\tau^2}\bX       \\
-\frac{1}{\tau^2}\bX^\top 
         & \left[ \frac{\partial^2 \bL}{\partial \mathbb{E}_q[w_i] \partial \mathbb{E}_q[w_j]} \right]
\end{bmatrix},    
\end{align*}
where the upper and lower block involves second derivatives: 
\begin{align*}
\frac{\partial^2\bL}
     {\partial 
     \mathbb{E}_q[\beta_j]\,\partial \mathbb{E}_q[\beta_k]}
&=
-\frac{\mathbb I(j \neq k)}{\tau^2}\sum_{i=1}^n x_{ij}x_{ik},
\\
\frac{\partial^2\bL}
     {\partial  \mathbb{E}_q[w_i]\,\partial  \mathbb{E}_q[w_j]}
&=
\dfrac{\mathbb I(i \neq j)}{\sigma^2}\left(
\dfrac{b_{ij}}{F_i}\,\mathbb{I}_{j \in N[i]}
+ \dfrac{b_{ji}}{F_j}\,\mathbb{I}_{i \in N[j]}
- \displaystyle\sum_{k : i,j \in N[k]} \dfrac{b_{ki}b_{kj}}{F_k}
\right).
\end{align*}

Thus  $\bH_{\alpha}$ is sparse, with the bottom-right block having the same off-diagonal sparsity pattern as the NNGP precision matrix. As $\bV_{\alpha}$ is diagonal due to the mean-field factorization, the final corrected covariance matrix $\widehat{\bSigma}_{\boldsymbol{\alpha}}$ is computed via solving the sparse linear systems
$$(\bI_{\boldsymbol{\alpha}} - \bV_{\boldsymbol{\alpha}} \bH_{\boldsymbol{\alpha}})c_j = v_j,$$ where $v_j$ is the $j$th column of $\bV_\alpha$ and $c_j$ is the corresponding column of $\widehat \bSigma_{\boldsymbol{\alpha}}$. These systems are solved efficiently via sparse LU decomposition and parallelized across columns using \texttt{RcppParallel}, allowing linear response correction to scale to large spatial datasets. Also, note that computation of $\widehat \bSigma_{\boldsymbol{\alpha}}$ is a one time cost, at the end of the spVB-MFA estimation. 

\subsection{Predictions}\label{sec:pred}

Subsequent to estimating the variational posteriors from any of our three methods,  posterior distributions of the response at a new location can be obtained using simple Monte Carlo sampling. Details are in Appendix \ref{sec:predapp}. 

\section{Simulation}
\label{sec:sim}

\begin{table}[!b]
\caption{Methods used in the simulation studies} \label{tab:competing}
\footnotesize
\resizebox{\textwidth}{!}{
\begin{tabular}{ll}
\hline
Methods & Description \\ \hline
spVB-MFA & NNGP prior with MFA variational distribution. \\
spVB-MFA-LR & spVB-MFA with one-step linear response correction for the covariance matrix. \\
spVB-NNGP & NNGP for both prior and variational distributions. \\
spVB-NNGP-joint & NNGP for prior, NNGP-type variational distribution jointly for $(\bbeta,\bw)$ (Section \ref{sec:joint}). \\
VNNGP & Variational nearest neighbor GP \citep{wu2022variational}. \\
DKLGP-default & DKLGP with 35 (default) running epochs \citep{cao2023variational}. \\
DKLGP & DKLGP with running epochs until convergence \citep{cao2023variational}. \\
spNNGP & MCMC-based methods \citep{finley2020spnngp}. \\ 
\hline
\end{tabular}
}
\end{table}

We conduct simulation studies to assess the model performance. The competing methods are shown in Table \ref{tab:competing}, and includes our proposed methods, VNNGP \citep{wu2022variational}, DKLGP \citep{cao2023variational}, and the MCMC-based spNNGP \citep{finley2019efficient,finley2020spnngp} as the benchmark. 
We do not include INFVB \citep{lee2024variational} in our comparison as the current implementation of the method for Gaussian data is based on full GP prior which cannot be used for even moderately large datasets. Also, this method marginalizes out the spatial random effects in the Gaussian model, preventing direct comparisons with other methods. The details of implementation of each method is given in Appendix \ref{sec:impl}. 

We generate data for $n = 1000, 5000, 10000, 50000, 100000$ on a fixed domain from a $c\times c$ square, where $c = 10$. The data generation model is $\by = \bX\bbeta + \bw + \boldsymbol{\epsilon}$, where the coefficients are fixed at $\bbeta = (2,5)^T$ and both two covariates are generated from $N(0,1)$. $\bw$ follows a Nearest Neighbor Gaussian Process with mean $\boldsymbol{0}$ and covariance matrix $\tilde{\bC}$ with the number of nearest neighbors = 15 as described in Section \ref{sec:bayes}, and $\boldsymbol{\epsilon} \sim N(\boldsymbol{0},\tau^2\bI)$. The true values of parameters are as follows: $\tau^2 = 0.5$, $\sigma^2 = 10$, and $\phi = 1$. 

We evaluate the methods based on a suite of metrics. Using the MCMC-based method (spNNGP) as the reference posterior, we  assess the accuracy of the  the VI-based methods using posterior means and variances. For the small or moderate sample sizes ($n \leq 10000$), we also compare the KL divergence between the posterior distribution of $\bw$ to the analytically available posterior distribution $p(\bw|\by,\bX, \bbeta,\tau^2, \sigma^2,\phi) \sim  N\Big(\frac{1}{\tau^2} \Big[\frac{1}{\tau^2}\bI_n + \tilde{\bC}^{-1}\Big]^{-1} (\by-\bX\bbeta), \Big[\frac{1}{\tau^2}\bI_n + \tilde{\bC}^{-1}\Big]^{-1}\Big)$. Additionally, we compare the continuous ranked probability score (CRPS), 95\% weighted Interval Score \citep{gneiting2007strictly} and 95\% coverage for spatial random effects. The CRPS evaluates the accuracy of the predictive distribution by comparing it to the observed value. The 95\% weighted interval score assesses the accuracy of prediction while taking the quantiles of the predicted samples into account. For both metrics, lower scores indicate better performance. We also compare the average running time on simulated data for each method. 

For prediction, we generate $n = 1100, 5500, 11000$ observations and randomly select $100, 500, 1000$ observations as the test dataset. After using the training data to fit the model, we predict the missing data (test data). We use a 95\% weighted interval score, CRPS, mean squared error (MSE) and 95\% coverage to assess the prediction for both $\by$  and $\bw$. 

When comparing accuracy, we simulate 100 replicates for the small sample size setting ($n \leq 10000$) and 10 replicates for the large sample size setting and summarize the results. To determine the average running time, we run 10 replicates for each setting. spVB-NNGP, spVB-NNGP-joint, spVB-MFA, spVB-MFA-LR, and spNNGP are implemented in R 4.4.0, while VNNGP and DKLGP are implemented in Python 3.10. 

\normalsize

\subsection{Inference on Spatial Random Effects}
\label{sec:results}
\begin{figure}[t]
    \centering
    \begin{subfigure}[b]{\textwidth}
        \centering
        \includegraphics[width=\textwidth]{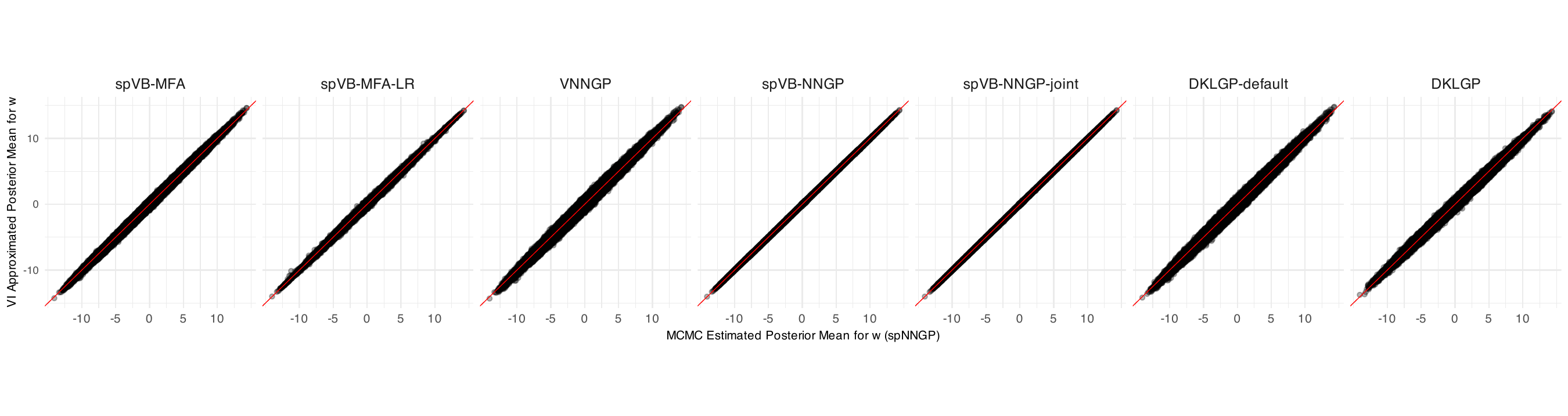}
        \caption{VI approximated posterior mean vs MCMC estimated posterior mean (spNNGP).}
        \label{fig:mean}
    \end{subfigure}
    \vspace{0.5cm}
    \begin{subfigure}[b]{\textwidth}
        \centering
        \includegraphics[width=\textwidth]{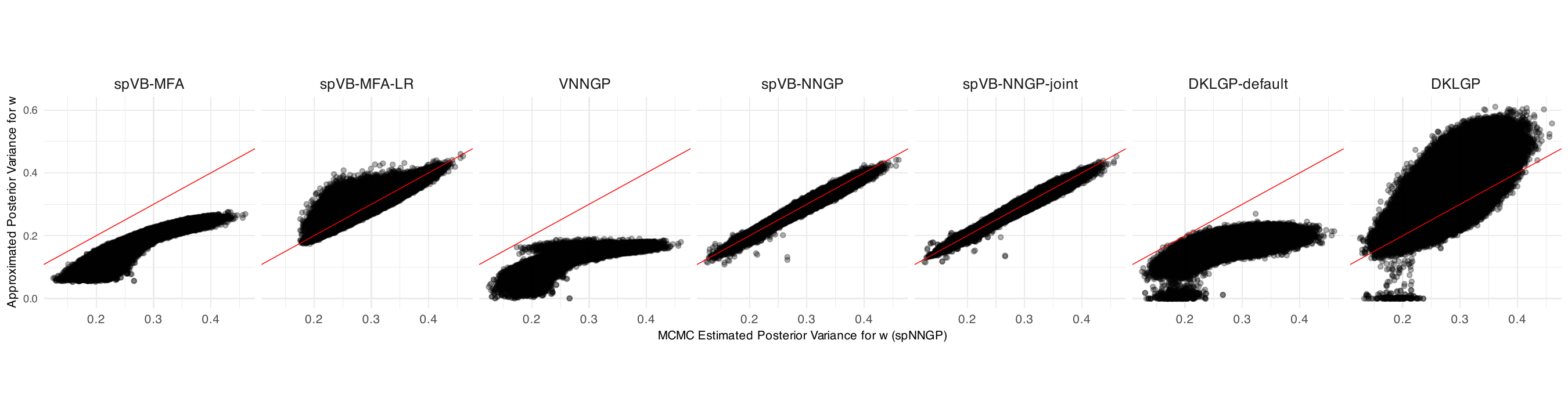}
        \caption{VI approximated posterior variance vs MCMC estimated posterior variance (spNNGP).}
        \label{fig:var}
    \end{subfigure}
    \caption{VI approximated posterior mean and variance vs MCMC (spNNGP) estimated posterior mean and variance for $\bw$ under the setting $n = 10000$ with $100$ simulated replicates. The columns correspond to spVB-MFA, spVB-MFA-LR, VNNGP, spVB-NNGP, spVB-NNGP-joint, DKLGP-default and DKLGP (optimized).
    }
    \label{fig:combined_posterior}
\end{figure}

Figure \ref{fig:mean} and Figure \ref{fig:var} respectively compare the posterior means and variances for $\bw$ of each VI-based method to those from the MCMC-based method, for sample size $n=10000$. All methods recover the posterior mean well. Among them, spVB-NNGP, spVB-NNGP-joint, and spVB-MFA-LR exhibit tighter alignment with the diagonal, indicating lower variability and better consistency compared to spVB-MFA, VNNGP, and DKLGP. In Figure \ref{fig:var}, spVB-MFA and VNNGP tend to underestimate the variance across all settings. In contrast, spVB-NNGP, spVB-NNGP-joint and spVB-MFA-LR again align well with the MCMC estimated variance. The improvement in the variance estimate from spVB-MFA-LR compared to spVB-MFA shows the ability of the linear response correction to mitigate variance underestimation from mean-field approximation. The DKLGP-default with 35 epochs fails to update the variance effectively, resulting in highly variable estimates mostly lying considerably below the diagonal. Running DKLGP till convergence produces higher variance estimates. There are still clusters of points that are distantly below the diagonal line. These clusters might be removed by running more epochs, 
but this would require more time. 

Complete results for all sample sizes can be found in the Appendix \ref{appendix:sim_w} which shows similar trends. DKLGP occasionally returns NA values during optimization. For large sample sizes, we include results from the first 10 seeds where DKLGP completed 35 epochs. We limit comparisons to the default DKLGP setting for large-scale experiments because (i) it already requires substantially more time than other VI-based methods, and (ii) longer training often leads to more frequent NA failures and fewer usable seeds. For $n=100000$, DKLGP-default fails to produce output within 7 days under a 500 GB memory limit. As a result, we also exclude DKLGP-default from those comparisons. 

In summary, spVB-NNGP, spVB-NNGP-joint and spVB-MFA-LR provide the most accurate and stable posterior mean and variance approximations for spatial random effects across all settings, while DKLGP-default, DKLGP, spVB-MFA, and VNNGP struggle with variance estimation.

\begin{figure}[H]
\centering
\includegraphics[width=0.9\textwidth]{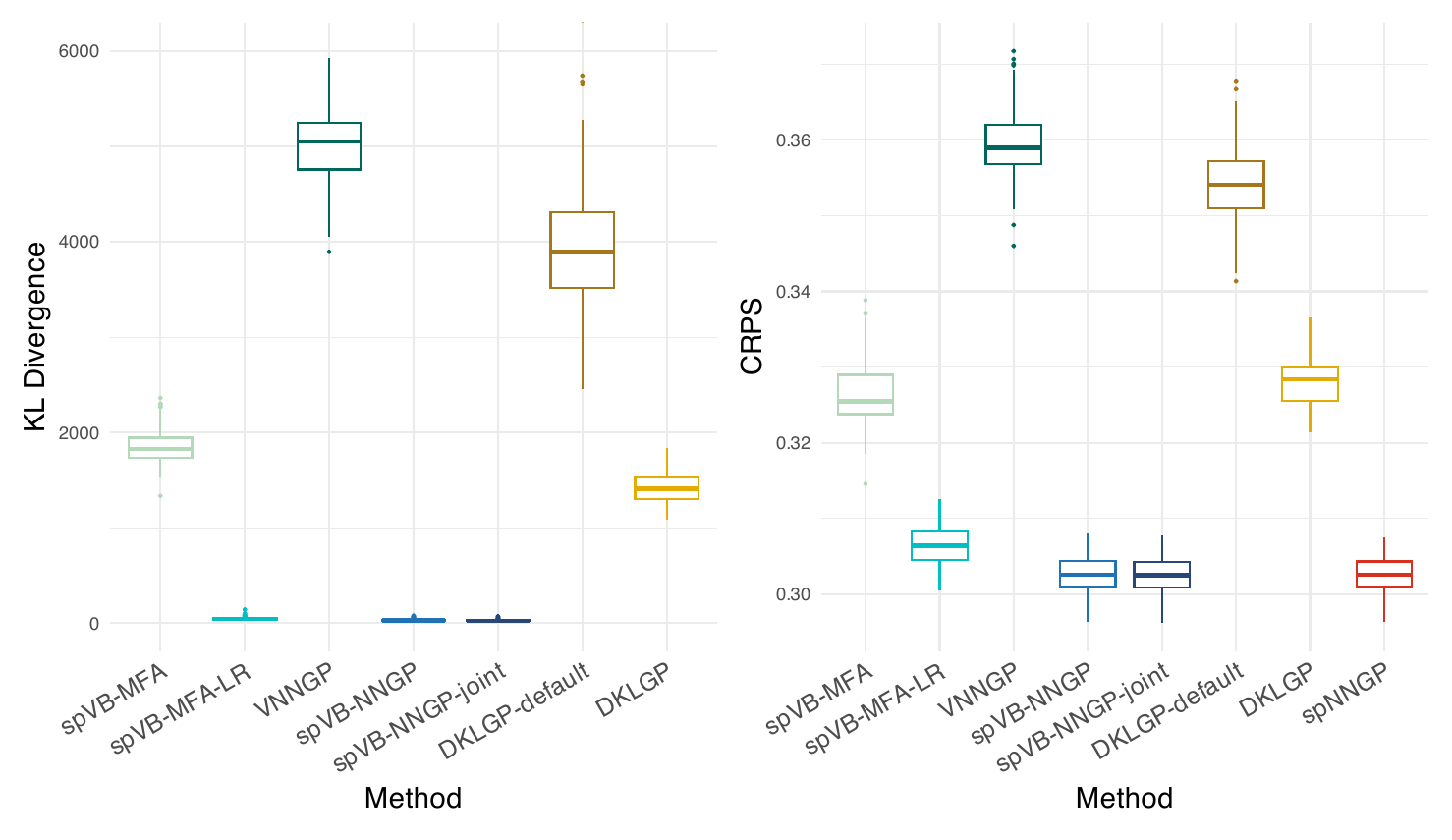}
\caption{Kullback–Leibler (KL) divergence from approximated posterior distribution to the pseudo-reference posterior distribution (left) and continuous ranked probability score (CRPS) for $\bw$ (right) under the setting $n = 10000$ with $100$ simulated replicates.}\label{fig:KL}
\end{figure}

Figure \ref{fig:KL} shows the boxplots of the Kullback–Leibler (KL) divergence from approximated posterior $q(\bw)$ to the pseudo-reference posterior distribution, across different VI-based methods. spVB-NNGP and spVB-NNGP-joint consistently achieve the lowest KL divergence, indicating a close approximation to the pseudo-reference posterior. DKLGP-default fails to converge, resulting in significantly higher KL divergence, while the overestimation of variance when DKLGP is run till convergence results in somewhat high KL. spVB-MFA also has large KL divergence values due to the underestimation of posterior variances. This is mitigated to a large extent by the linear response correction as seen by the lower KL values of spVB-MFA-LR. The right panel of Figure \ref{fig:KL} displays the CRPS comparison. Consistent with the KL divergence results, spVB-NNGP, spVB-NNGP-joint and spNNGP achieve the lowest CRPS across all sample sizes, suggesting more accurate approximate distributions. In contrast, spVB-MFA, VNNGP and DKLGPs generally yield higher CRPS values, indicating less accurate approximations and  estimation of uncertainty.

The KL divergence results demonstrate that our proposed spVB-NNGPs outperforms existing variational inference approaches in approximating the posterior distribution. Furthermore, the CRPS comparison confirms that our method not only improves upon variational methods but also closely matches the performance of the MCMC-based (spNNGP), offering a reliable and scalable alternative for large-scale spatial inference. Results for all sample sizes and across all metrics (KL divergence, CRPS, 95\% weighted interval score, 95\% coverage) are provided in the Appendix \ref{appendix:sim_w}.

\subsection{Inference on Regression Coefficients}

We also compare inference on the regression coefficients $\bbeta$, as summarized in Table \ref{tab:beta}. DKLGPs and VNNGP do not accommodate fixed effects, so we only compare the other methods.  spNNGP achieves roughly 95\% coverage. spVB-MFA and spVB-NNGP exhibit lower coverage rates, largely due to the block independence assumptions in their variational distributions. By addressing these limitations, the corresponding enhanced versions, spVB-MFA-LR and spVB-NNGP-joint, show markedly improved coverage. These results highlight the effectiveness of incorporating covariance adjustments and joint modeling in variational inference.  

\begin{table}[ht]

\centering
\caption{Coverage probabilities for regression coefficients across methods and sample sizes}
\begin{tabular}{llccc}
\toprule
$n$ & Method & $\beta_1$ & $\beta_2$ \\
\midrule
\multirow{5}{*}{1000} 
  & spVB-MFA            & 0.551 & 0.541 \\
  & spVB-MFA-LR         & 0.918 & 0.949 \\
  & spVB-NNGP           & 0.663 & 0.633 \\
  & spVB-NNGP-joint     & 0.786 & 0.827 \\
  & spNNGP              & 0.939 & 0.939 \\
\midrule
\multirow{5}{*}{5000} 
  & spVB-MFA            & 0.645 & 0.570 \\
  & spVB-MFA-LR         & 0.957 & 0.957 \\
  & spVB-NNGP           & 0.763 & 0.731 \\
  & spVB-NNGP-joint     & 0.935 & 0.925 \\
  & spNNGP              & 0.946 & 0.957 \\
\midrule
\multirow{5}{*}{10000} 
  & spVB-MFA            & 0.731 & 0.615 \\
  & spVB-MFA-LR         & 0.971 & 0.904 \\
  & spVB-NNGP           & 0.856 & 0.788 \\
  & spVB-NNGP-joint     & 0.981 & 0.942 \\
  & spNNGP              & 0.971 & 0.942 \\
\bottomrule
\end{tabular}
\label{tab:beta}
\end{table}

A comparison of point estimates for the random error variance $\tau^2$ and the spatial variance parameter $\sigma^2$ is provided in the Appendix \ref{appendix:other_var}. Among the methods, spVB-NNGPs, spVB-MFA-LR, and spNNGP yield point estimates that are closest to the true values.

\subsection{Computational Comparison}

\begin{figure}
\centering
\includegraphics[width=\textwidth]{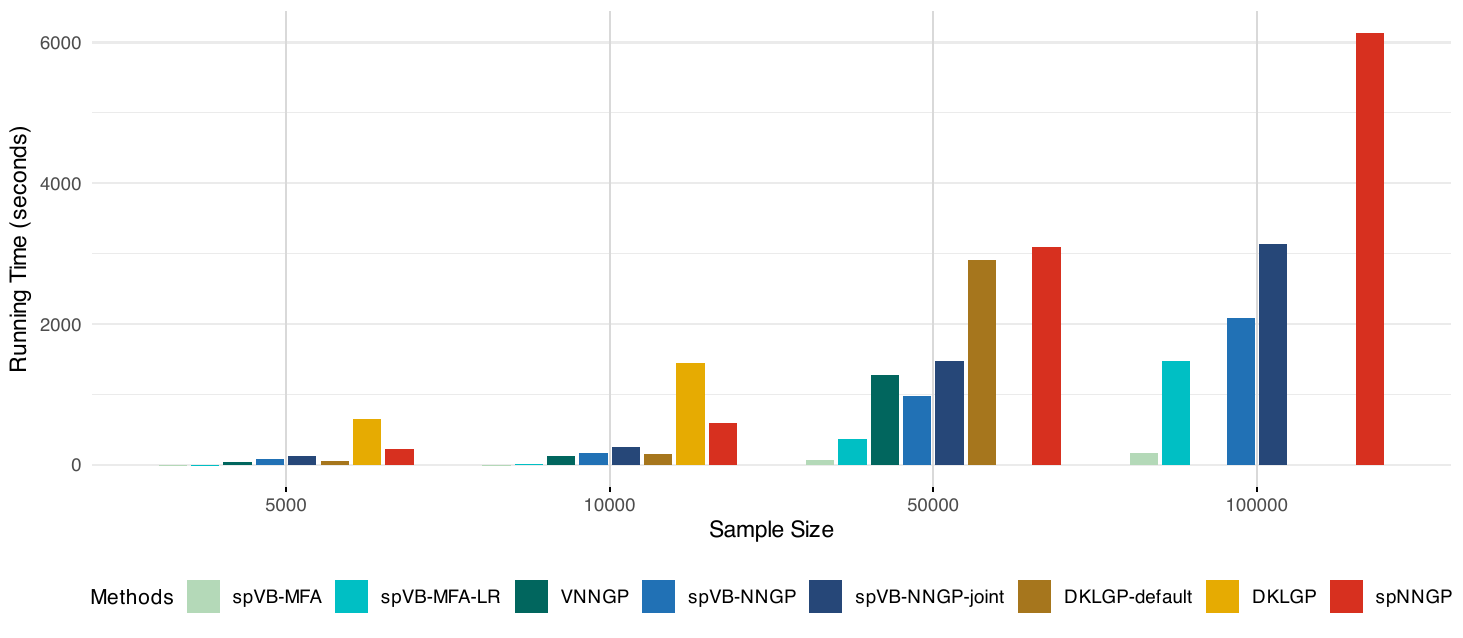}
\caption{Average running time (in seconds) for each method across sample sizes. VNNGP timing is omitted at $n = 100000$ due to exceeding the 500 GB memory limit on the local machine. DKLGP is excluded at large $n$ due to failures and long running time. DKLGP-default is excluded at $n = 100000$ for the same memory reason. Computational experiments are conducted on the local machine.}\label{fig:time}
\end{figure}

Figure \ref{fig:time} shows the average running time (in seconds) for each method. DKLGP requires substantially more computation time compared with other methods. spVB-MFA and spVB-MFA-LR are significantly faster than both spVB-NNGPs and spNNGP. Also, spVB-MFA considerably outperforms VNNGP in terms of speed, with the computational advantage becoming more significant as sample size increases. spVB-NNGP achieves approximately one-third the running time of spNNGP, primarily due to requiring fewer epochs to reach convergence than the number of MCMC iterations. The spVB-MFA-LR exhibits comparable running time to spVB-MFA for small and moderate sample sizes, and remains much faster than the remaining methods for large sample sizes. Additionally, spVB-MFA, spVB-NNGPs, and spNNGP all exhibit computational scaling that is approximately linear in the sample size, i.e., $O(n)$. For $n=100000$, VNNGP exceeds 500 GB of memory usage and cannot be executed on a local machine. Instead, we rely on high-performance computing resources for running this method, and its timing is therefore not reported. 

Finally, assessment of the predictive performances of the methods are detailed in Appendix \ref{appendix:results_pred}. In summary, we see that spVB-NNGPs and spVB-MFA-LR not only improve upon existing variational inference approaches such as VNNGP and DKLGP in predictive accuracy but are also comparable to the MCMC-based spNNGP. 

\section{Real World Data Example}
\label{sec:rwe}
We examine the performance of the models using the large Bonanza Creek Experimental Forest (BCEF) dataset, originally provided in the \texttt{spNNGP} R-package. It comprises forest canopy height (FCH) estimates from NASA Goddard’s LiDAR instruments, encompassing $188,717$ locations across Alaska. In this experiment, we focus on a block-random subsample of $101,620$ spatial locations for training, using the remaining locations for testing. A primary covariate included in the analysis is percent tree cover (PTC), derived from Landsat imagery, which provides additional ecological context for the spatial modeling. The detailed information for this data can be found in \cite{finley2020spnngp}. 

\begin{figure}[H]
\centering
\begin{subfigure}[t]{0.48\textwidth}
    \centering
    \includegraphics[width=\textwidth]{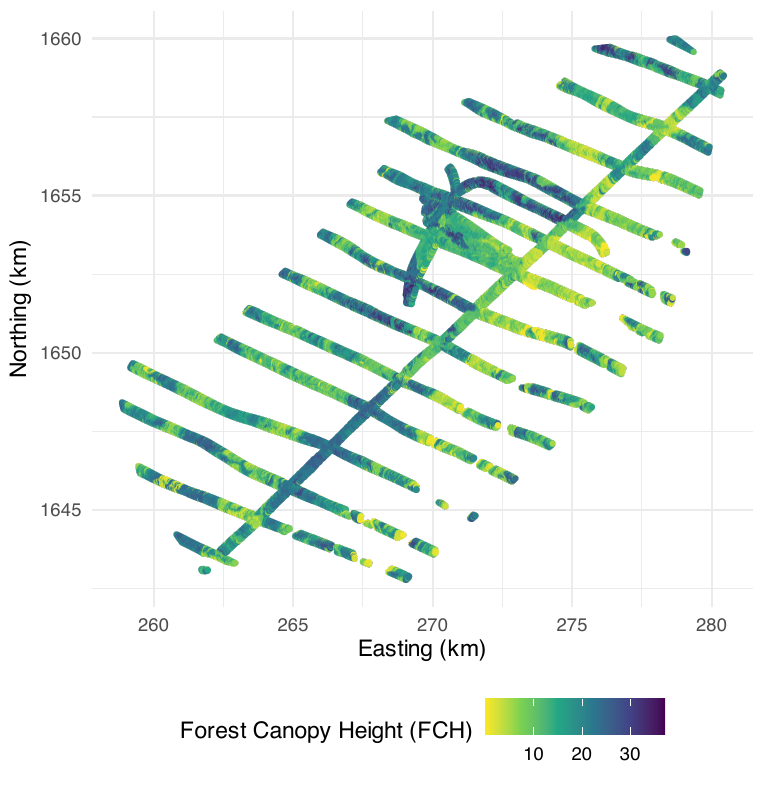}
    \caption{Forest canopy height (FCH) from G-LiHT LiDAR.}
    \label{fig:FCH}
\end{subfigure}
\hfill
\begin{subfigure}[t]{0.48\textwidth}
    \centering
    \includegraphics[width=\textwidth]{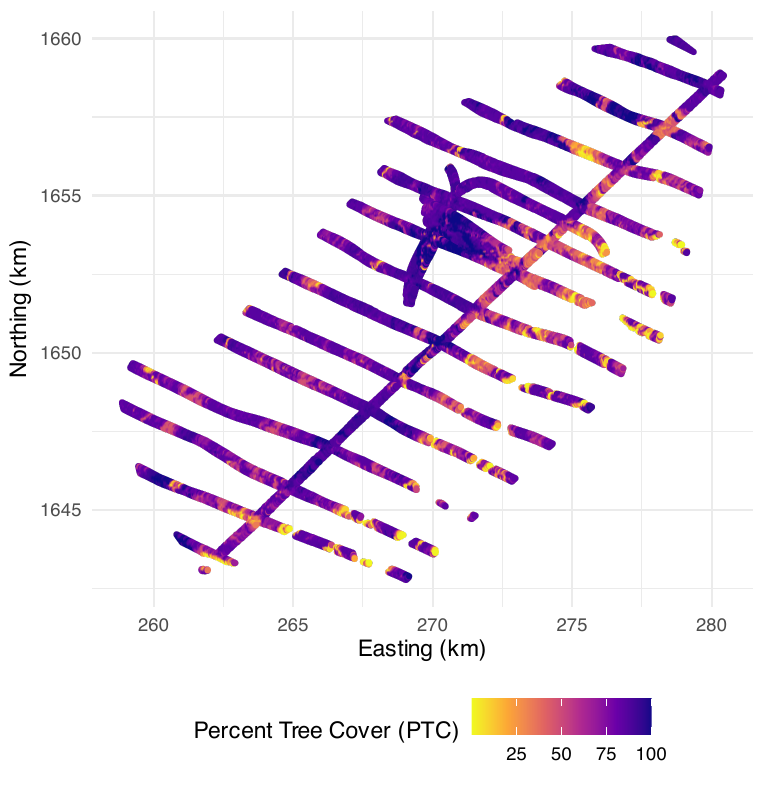}
    \caption{Landsat-derived percent tree cover estimates (PTC).}
    \label{fig:PTC}
\end{subfigure}
\caption{Distribution of forest canopy height (FCH) estimates from G-LiHT LiDAR and percent tree cover (PTC) estimates derived from Landsat in the Bonanza Creek Experimental Forest (BCEF) dataset.}
\label{fig:real}
\end{figure}

An overview of the FCH and PTC distribution in the analyzed data is shown in Figure \ref{fig:real}.  The spatial linear mixed effects model to analyze the data is as follows: $$\by = \bx\beta_{PTC} + \bw + \boldsymbol{\epsilon},$$ where $\by$ is the vector of observed (centered) FCH estimates, $\beta_{PTC}$ is the slope coefficient associated with the centered PTC predictor variable denoted as $\bx$. We consider the following methods as introduced in the simulation study: spVB-MFA, spVB-MFA-LR, spVB-NNGP, spVB-NNGP-joint, VNNGP and DKLGP-default. The corresponding running epochs are 4000, 4500, 3500, 4500, 500, 40 (the default running epochs of DKLGP for real data is set to be 40). For spNNGP, we run 150000 iterations with 5000 as burn-in. As VNNGP and DKLGP, cannot incorporate covariates, to enable a fair comparison, we include the covariate as an additional spatial dimension for fitting these two methods. The spatial coordinates are rescaled to match the range of the covariates, ensuring a stable lengthscale. Due to memory constraints, VNNGP and DKLGP-default could not be run on a local machine. Hence, all methods are executed on a high-performance computing (HPC) cluster using the same node. 

\begin{figure}[h]
    \centering
    \begin{subfigure}[b]{\textwidth}
        \centering
        \includegraphics[width=\textwidth]{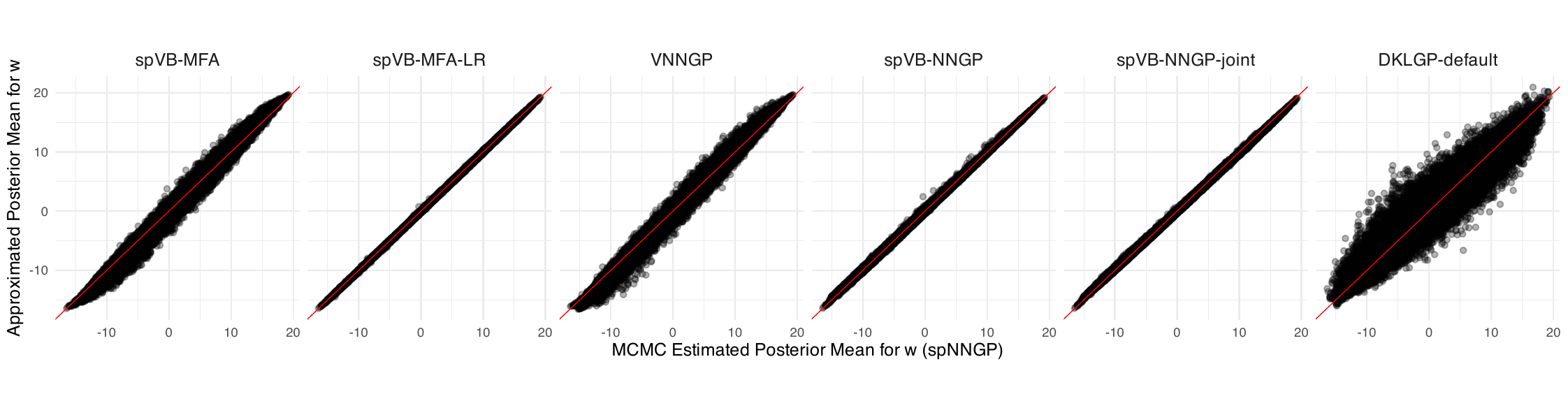}
        \caption{VI approximated posterior mean vs MCMC estimated mean.}
        \label{fig:real_mean}
    \end{subfigure}
    \vspace{0.5cm}
    \begin{subfigure}[b]{\textwidth}
        \centering
        \includegraphics[width=\textwidth]{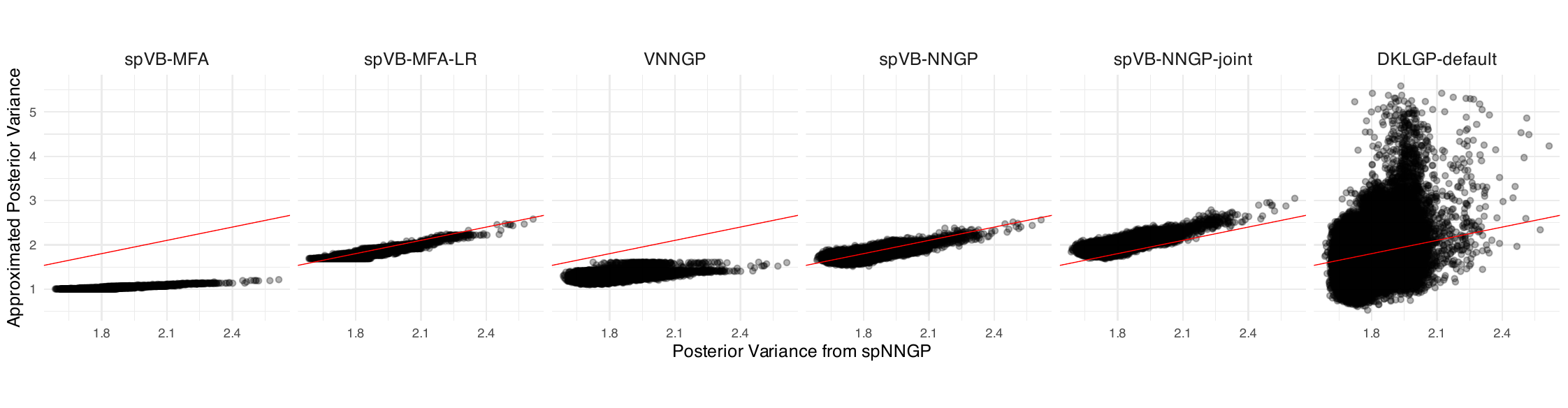}
        \caption{VI approximated posterior variance vs MCMC estimated variance.}
        \label{fig:real_var}
    \end{subfigure}
    \caption{Comparison of VI and MCMC estimation for posterior mean (top) and variance (bottom). For fair comparison, VNNGP and DKLGP-default are also fitted with covariates as additional input dimensions, with spatial coordinates rescaled to ensure stable lengthscales.
    }
    \label{fig:combined_posterior}
\end{figure}

We evaluate the variational approximations of the posterior distribution $q(\bw)$ using the training data, with MCMC-based spNNGP results serving as the reference. Figure \ref{fig:real_mean} and Figure \ref{fig:real_var} display comparisons of the posterior mean and variance, respectively. For posterior mean, spVB-NNGPs and spVB-MFA-LR perform similarly to spNNGP. In terms of posterior variance, spVB-NNGPs and spVB-MFA-LR align well with the diagonal reference line, indicating accurate uncertainty quantification, while spVB-MFA expectedly underestimates the variance. For VNNGP and DKLGP-default, the posterior mean is distributed substantially dispersed around the reference, while the posterior variances are inaccurate with VNNGP underestimating and DKLGP-default both under- and over-estimating considerably. 

Table \ref{tab:real} summarizes the posterior means and 95\% credible intervals of the other parameters. The spVB-NNGPs give similar point estimations for $\sigma^2$ and $\tau^2$ compared with spNNGP, but with narrower 95\% credible intervals. The spVB-MFA method tends to overestimate $\sigma^2$ and underestimate the random error variance $\tau^2$. VNNGP gives a similar estimation for spatial variance, but underestimates the random error variance as well. DKLGP-default fails to capture the true values for those variance parameters. 

\begin{table}[h]
\centering
\caption{Summary of estimation and predictive performance for analyzing BCEF data}
\resizebox{\textwidth}{!}{
\begin{tabular}{lcccccccc}
\hline
 & $\beta_{\text{PTC}}$ & $\sigma^2$ & $\tau^2$ & CRPS & Interval Score & MSE & 95\% Coverage (\%) \\
\hline
spVB-MFA 
    & 0.0157 (0.0153, 0.0160) 
    & 74.4 (73.8, 75.1) 
    & 1.26 (1.25, 1.27) 
    & 3.47 & 0.770 & 38.1 & 98.3 \\
spVB-MFA-LR 
    & 0.0204 (0.0166, 0.0243) 
    & 53.9 (53.4, 54.4) 
    & 2.82 (2.78, 2.83) 
    & 3.42 & 0.710 & 37.7 & 96.8 \\
VNNGP 
    & -- 
    & 56.6$^{*}$ 
    & 1.66$^{*}$ 
    & 4.28 & 0.825 & 55.9 & 95.6 \\
spVB-NNGP 
    & 0.0288 (0.0281, 0.0293) 
    & 55.0 (54.4, 55.4) 
    & 2.82 (2.80, 2.85) 
    & 3.39 & 0.712 & 37.0 & 97.0 \\
spVB-NNGP-joint 
    & 0.0273 (0.0219, 0.0330) 
    & 50.8 (50.4, 51.3) 
    & 3.32 (3.30, 3.35) 
    & 3.39 & 0.703 & 37.2 & 96.6 \\
DKLGP-default 
    & -- 
    & 21.9$^{*}$ 
    & 7.28$^{*}$ 
    & 3.98 & 0.917 & 46.5 & 85.0 \\
spNNGP 
    & 0.0202 (0.0166, 0.0236) 
    & 57.8 (52.4, 60.7) 
    & 2.95 (2.85, 3.05) 
    & 3.38 & 0.719 & 37.0 & 97.1 \\
\hline
\end{tabular}
} 
\\
\raggedright\footnotesize $^{*}$ Credible intervals not available for VNNGP and DKLGP-default.
\label{tab:real}
\end{table}

\begin{figure}[H]
\centering
\includegraphics[width=\textwidth]{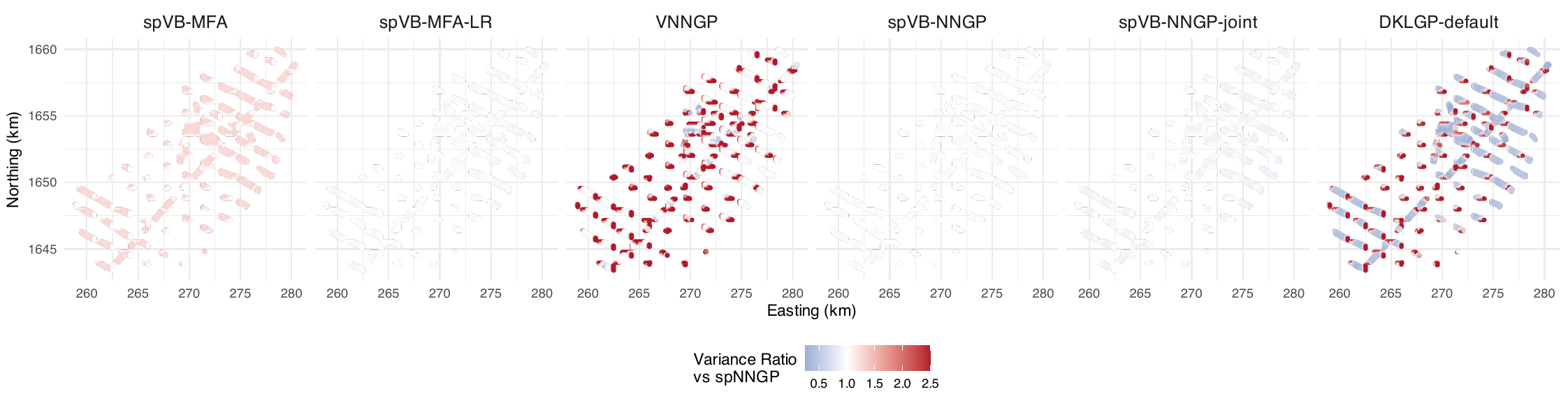}
\caption{Posterior predictive variance ratios compared to the MCMC-based (spNNGP) on the test set. Each ratio is computed as the predicted posterior variance divided by the corresponding spNNGP variance. VNNGP and DKLGP-default show substantial deviations from the MCMC benchmark, either overestimating or underestimating the variance. In contrast, spVB-MFA-LR and spVB-NNGPs yield ratios closer to 1, indicating closer alignment with the spNNGP predicted variance.}\label{fig:real_pred_var}
\end{figure}

We evaluate predictive performance using average CRPS, 95\% weighted interval score, mean squared error (MSE), and 95\% coverage over the test data. spVB-NNGP and spVB-NNGP-joint achieve predictive accuracy and uncertainty quantification comparable to spNNGP, while VNNGP and DKLGP-default yield the highest CRPS and interval scores, implying suboptimal predictive performance. This is also confirmed from the spatial prediction variance comparison plotted in Figure \ref{fig:real_pred_var}. 

\begin{figure}[H]
\centering
\includegraphics[width=0.6\textwidth]{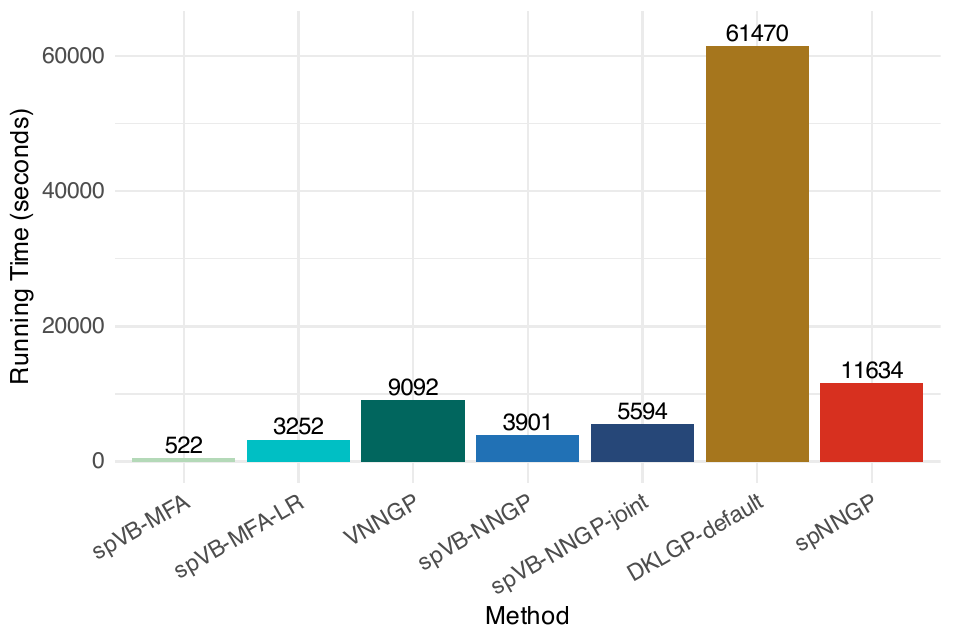}
\caption{Running time (seconds) for each method. Experiments are performed on the high-performance computing (HPC) cluster using the same node, as VNNGP and DKLGP-default exceed the memory limits of a local machine. Exact values are shown above each bar.}\label{fig:real_time}
\end{figure}

Finally, Figure \ref{fig:real_time} presents the computation times across methods, showing that spVB-NNGPs and spVB-LR offer substantial computational savings compared to spNNGP with comparable approximation accuracy. 

\section{Discussion}
\label{sec:discussion}
We review existing variational inference methods and introduce spVarBayes, a suite of scalable spatial variational Bayes methods incorporating Nearest Neighbor Gaussian Processes (NNGP) in both the prior and variational distributions. We propose both independent and joint formulations of the method with the latter capturing the posterior correlation between fixed and random effects. To enable efficient optimization, we derive closed-form gradients using the reparametrization trick and further accelerate computation through a vanishing gradient approximation. Using calculus of variations, we provide assumption-free closed-form variational distributions for the regression coefficients and variance parameters, thereby enabling full uncertainty quantification in parameter estimation. We also develop a faster mean-field approximation based method and a novel linear response correction tailored specifically to the spatial mixed effects model, improving uncertainty estimates from MFA.

Extensive simulation studies demonstrate that spVB-NNGP and spVB-NNGP-joint achieve posterior and predictive accuracy comparable to that of MCMC-based approaches, while offering reduced computation time. Compared to existing variational methods, spVB-NNGPs and spVB-MFA-LR consistently show superior performance in both posterior approximation and predictive metrics. The spVB-MFA offers comparable performance to VNNGP, a similar method, but with greater computational efficiency, making it more suitable for large-scale applications. Finally, analysis of the BCEF dataset demonstrates the practical utility of spVB-NNGPs and spVB-MFA-LR, with inference results closely aligned with those obtained via MCMC.

One limitation of the spVB-NNGP framework is the use of an independent block structure in the variational distribution, where spatial parameters are assumed to be independent of other model parameters. As noted in \citet{ren2011variational}, such independence assumptions can lead to overly narrow credible intervals, indicating caution when interpreting uncertainty estimates. One possible remedy involves integrated variational methods, similar to those used in INFVB. However, these approaches typically require repeated evaluations or integration over subsets of the parameter space, which can become computationally infeasible for large datasets and may not outperform MCMC-based methods such as spNNGP in practice.

To further improve the accuracy and flexibility of variational inference in spatial settings, several promising directions remain open. First, richer variational families, such as semi-implicit variational inference \citep{yin2018semi}, can be employed to better capture complex posterior dependencies. Second, normalizing flows can be used to transform simple base distributions into more expressive variational posteriors through a sequence of invertible mappings, enabling the approximation to capture complex correlations \citep{rezende2015variational, papamakarios2021normalizing}. 

We also plan to generalize our current framework by incorporating additional covariance functions from the Matérn family and extending the variational inference approach to accommodate non-Gaussian outcomes, including Poisson and Bernoulli data. The developed methods are implemented in a publicly available R package \texttt{\href{https://github.com/jfsong100/spVarBayes}{spVarBayes}}, currently accessible 
on GitHub, and 
will be made available on CRAN.

\section*{Acknowledgments} This work was supported by National Institute of Environmental Health Sciences (NIEHS) grant
R01 ES033739. 

\section*{Data Availability Statement} The Bonanza Creek Experimental
Forest (BCEF) dataset used in this manuscript is publicly available in the spNNGP R-package.

\bibliographystyle{chicago}
\bibliography{references}
\clearpage 

\appendix
\counterwithin{figure}{section}

\section{Algorithm for spVarBayes}
\subsection{Calculus of variations}
\label{appendix:calculus}

\paragraph{For $\bbeta$}

From the ELBO in (\ref{ini_ELBO}), and after omitting terms irrelevant to $q(\bbeta)$, the objective function $\mathcal{L}(q(\bbeta))$ is as follows:
$$\mathcal{L}(q(\bbeta)) = \sum_{i=1}^n \mathbb{E}_q\Big[\frac{1}{2}\log \frac{1}{2\pi \tau^2} - \frac{1}{2\tau^2}(y_i- \bx_i^T\bbeta- w_i)^2\Big]- \mathbb{E}_q\Big[\log q(\bbeta)\Big]+\text{constant not depends on $q(\bbeta)$}.$$

We want to find $q(\bbeta)$ that maximizes $\mathcal{L}(q(\bbeta))$.
\begin{align*}
    \frac{\partial}{\partial q(\bbeta)}\mathcal{L}(q(\bbeta)) &\propto \frac{\partial}{\partial q(\bbeta)} \Big\{\sum_{i=1}^n \mathbb{E}_q\Big[- \frac{1}{2\tau^2}(y_i- \bx_i^T\bbeta -w_i)^2\Big]- \mathbb{E}_q\Big[\log q(\bbeta)\Big]\Big\}\\
     &\propto \frac{\partial}{\partial q(\bbeta)}  \Big\{\sum_{i=1}^n\int \Big[- \frac{1}{2\tau^2}(y_i- \bx_i^T\bbeta -w_i)^2\Big]q(\bbeta)d \boldsymbol{\psi}_{-\bbeta}- \int\log q(\bbeta)q(\bbeta)d\boldsymbol{\psi}_{-\bbeta}\Big\}\\
     &\propto\Big\{\sum_{i=1}^n[- \frac{1}{2}\mathbb{E}_q[\frac{1}{\tau^2}](y_i- \bx_i^T\bbeta-\mathbb{E}_{q}[w_i])^2] - \log q(\bbeta) - 1\Big\},
\end{align*}
where $\boldsymbol{\psi}_{-\bbeta}$ means all the parameters except $\bbeta$. Set $ \frac{\partial}{\partial q(\bbeta)}\mathcal{L}(q(\bbeta))  = 0$:
\begin{align*}
    q(\bbeta) & \propto \exp\Big\{\sum_{i=1}^n [- \frac{1}{2}\mathbb{E}_q[\frac{1}{\tau^2}](y_i- \bx_i^T\bbeta -\mathbb{E}_q[w_i])^2]\Big\}\\
    & \propto \exp\Big\{ - \frac{1}{2}\mathbb{E}_q[\frac{1}{\tau^2}](\by- \bX\bbeta -\mathbb{E}_{q}[\bw])^T(\by- \bX\bbeta -\mathbb{E}_{q}[\bw])\Big\}.\\
    q(\bbeta) &= N\Big((\bX^T\bX)^{-1}\bX^T(\by-\mathbb{E}_{q}[\bw]),\mathbb{E}_q[\frac{1}{\tau^2}]^{-1}(\bX^T\bX)^{-1}\Big).
\end{align*}

We can compute $\mathbb{E}_q[\bbeta] = (\bX^T\bX)^{-1}\bX^T(\by-\mathbb{E}_{q}[\bw])$ and $\bV_{\bbeta} = \mathbb{E}_q[\frac{1}{\tau^2}]^{-1}(\bX^T\bX)^{-1}$.

\paragraph{For $\tau^2$} From the ELBO in (\ref{ini_ELBO}), and after omitting terms irrelevant to $q({\tau^2})$, the objective function $\mathcal{L}(q({\tau^2}))$ is as follows:
\begin{align*}
    \mathcal{L}(q({\tau^2})) = &\sum_{i=1}^n \mathbb{E}_q\Big[\frac{1}{2}\log \frac{1}{2\pi \tau^2} - \frac{1}{2\tau^2}(y_i-\bx_i^T\bbeta - w_i)^2\Big] \\
    &+ \mathbb{E}_q\Big[(a_{\tau}+1)\log \frac{1}{\tau^2} - \frac{b_{\tau}}{\tau^2}\Big] - \mathbb{E}_q\Big[\log q(\tau^2)\Big]+\text{constant not depends on $q(\tau^2)$}.
\end{align*}

We want to find $q(\tau^2)$ that maximizes $\mathcal{L}(q(\tau^2))$.
\begin{align*}
    \frac{\partial}{\partial q(\tau^2)}\mathcal{L}(q({\tau^2})) &\propto \frac{\partial}{\partial q(\tau^2)} \Big\{\sum_{i=1}^n \mathbb{E}_q\Big[\frac{1}{2}\log \frac{1}{2\pi\tau^2} - \frac{1}{2\tau^2}(y_i-\bx_i^T\bbeta - w_i)^2\Big] \\
    &\phantom{+++++}+ \mathbb{E}_q\Big[(a_{\tau}+1)\log \frac{1}{\tau^2}  - \frac{b_{\tau}}{\tau^2}\Big] - \mathbb{E}_q\Big[\log q(\tau^2)\Big]\Big\}\\
    &\propto\frac{\partial}{\partial q(\tau^2)} \Big\{\sum_{i=1}^n \int \Big[\frac{1}{2}\log \frac{1}{\tau^2}  - \frac{1}{2\tau^2}(y_i-\bx_i^T\bbeta-w_i)^2\Big]q(\tau^2)d\boldsymbol{\psi}_{-\tau^2} \\
    &\quad \quad \quad \quad+ \int\Big[(a_{\tau}+1)\log \frac{1}{\tau^2}  - \frac{b_{\tau}}{\tau^2}\Big]q(\tau^2)d\boldsymbol{\psi}_{-\tau^2} - \int[\log q(\tau^2)]q(\tau^2)d\boldsymbol{\psi}_{-\tau^2}\Big\}\\
     \propto&\Big\{\frac{n}{2}\log \frac{1}{\tau^2}  - \frac{1}{2\tau^2}\Big[\sum_{i=1}^n (y_i-\bx_i^T\mathbb{E}_q[\bbeta]-\mathbb{E}_q[w_i])^2 + Tr(\sum_{i=1}^n \bx_i\bx_i^T \bV_{\bbeta})+Tr(\bV_{w}) \Big]\\
    &\quad (a_{\tau}+1)\log \frac{1}{\tau^2}  - \frac{b_{\tau}}{\tau^2} - \log q(\tau^2) - 1\Big\}\\
    \propto&\Big\{\frac{n}{2}\log \frac{1}{\tau^2}  - \frac{1}{2\tau^2}\Big[(\by- \mathbb{E}_{q}[\bw])^T(\bI - \bH)(\by- \mathbb{E}_{q}[\bw])+ p \mathbb{E}_q[\frac{1}{\tau^2}]^{-1} +Tr(\bV_{w}) \Big]\\
    &\quad (a_{\tau}+1)\log \frac{1}{\tau^2}  - \frac{b_{\tau}}{\tau^2} - \log q(\tau^2) - 1\Big\},
\end{align*}
where $\boldsymbol{\psi}_{-\tau^2}$ means all the parameters except $\tau^2$, $\mathbb{E}_q[{\bbeta}]$ and $\bV_{\bbeta}$ is the mean and variance in the $q(\bbeta)$. Since we have computed $\mathbb{E}_q[{\bbeta}]$ and $\bV_{\bbeta}$ with the closed form as shown in the above, we can plug in the equation to simplify the formula. $\bV_{w}$ is the covariance of $\bw$ in the variational distribution $q(\bw)$. Set $ \frac{\partial}{\partial q(\tau^2)}\mathcal{L}(q(\tau^2))  = 0$:
\begin{align*}
    q(\tau^2) \propto & \exp\Big\{\frac{n}{2}\log \frac{1}{\tau^2}  - \frac{1}{2\tau^2}\Big[ (\by- \mathbb{E}_{q}[\bw])^T(\bI - \bH)(\by- \mathbb{E}_{q}[\bw]) + p \mathbb{E}_q[\frac{1}{\tau^2}]^{-1} +Tr(\bV_{w})\Big] \\
    &\phantom{++}+ (a_{\tau}+1)\log \frac{1}{\tau^2}  - \frac{b_{\tau}}{\tau^2}\Big\}\\
    \propto & (\frac{1}{\tau^2}) ^{\frac{n}{2}+a_{\tau}+1} \exp\Big\{-\frac{1}{\tau^2}\Big(b_{\tau}+\frac{1}{2}\Big[ (\by- \mathbb{E}_{q}[\bw])^T(\bI - \bH)(\by- \mathbb{E}_{q}[\bw] )+ p \mathbb{E}_q[\frac{1}{\tau^2}]^{-1} + Tr(\bV_{w})\Big]\Big)\Big\}.\\
    q(\tau^2) =  & IG(\frac{n}{2}+a_{\tau},b_{\tau}+\frac{1}{2}\Big[ (\by- \mathbb{E}_{q}[\bw])^T(\bI - \bH)(\by- \mathbb{E}_{q}[\bw])+ p \mathbb{E}_q[\frac{1}{\tau^2}]^{-1}+Tr(\bV_{w})\Big]).
\end{align*}
The trace is approximated using (\ref{trace_cal}) for spVB-NNGP. For spVB-MFA, the trace can be directly calculated as the sum of the posterior variance.

We can compute $\mathbb{E}_q[\frac{1}{\tau^2}] = \frac{a_\tau^{'}}{b_{\tau}^{'}}$, where $a_\tau^{'}$ and $b_{\tau}^{'}$ is the updated shape and scale parameters for $q(\tau^2)$.

\paragraph{For $\sigma^2$} The objective function $\mathcal{L}(q({\sigma^2}))$ is as follows:
\begin{align*}
    \mathcal{L}(q({\sigma^2})) =& \sum_{i=1}^n \mathbb{E}_q\Big[\frac{1}{2}\log\frac{1}{2\pi\sigma^2}-\frac{1}{2\sigma^2}\frac{1}{F_i}(w_i - \bb_{N[i]}\bw_{N[i]})^2\Big] \\
    &+ \mathbb{E}_q\Big[(a_{\sigma}+1)\log\frac{1}{\sigma^2} - \frac{b_{\sigma}}{\sigma^2}\Big] -\mathbb{E}_q\Big[\log q(\sigma^2)\Big]+\text{constant not depends on $q(\sigma^2)$}.
\end{align*}

We want to find $q(\sigma^2)$ that maximizes $\mathcal{L}(q(\sigma^2))$.
\begin{align*}
    \frac{\partial}{\partial q(\sigma^2)}\mathcal{L}(q({\sigma^2})) &\propto\frac{\partial}{\partial q(\sigma^2)} \Big\{\sum_{i=1}^n \mathbb{E}_q\Big[\frac{1}{2}\log\frac{1}{\sigma^2}-\frac{1}{2\sigma^2}\frac{1}{F_i}(w_i - \bb_{N[i]}\bw_{N[i]})^2\Big] \\
    &\quad \quad \quad \quad + \mathbb{E}_q\Big[(a_{\sigma}+1)\log\frac{1}{\sigma^2} - \frac{b_{\sigma}}{\sigma^2}\Big]-\mathbb{E}_q\Big[\log q(\sigma^2)\Big] \Big\}\\
    &\propto\frac{\partial}{\partial q(\sigma^2)} \Big\{\sum_{i=1}^n \int\Big[\frac{1}{2}\log\frac{1}{\sigma^2}-\frac{1}{2\sigma^2}\frac{1}{F_i}(w_i - \bb_{N[i]}\bw_{N[i]})^2\Big]q(\sigma^2)d\boldsymbol{\psi}_{-\sigma^2}\\
    & \quad \quad \quad \quad + \int\Big[(a_{\sigma}+1)\log\frac{1}{\sigma^2} - \frac{b_{\sigma}}{\sigma^2}\Big]q(\sigma^2)d\boldsymbol{\psi}_{-\sigma^2} - \int\log q(\sigma^2)q(\sigma^2)d\boldsymbol{\psi}_{-\sigma^2} \Big\}\\
    &\propto \sum_{i=1}^n\Big[\frac{1}{2}\log\frac{1}{\sigma^2}-\frac{1}{2\sigma^2}\mathbb{E}_q\Big[\frac{1}{F_i}(w_i - \bb_{N[i]}\bw_{N[i]})^2\Big]\Big]+(a_{\sigma}+1)\log\frac{1}{\sigma^2} - \frac{b_{\sigma}}{\sigma^2}-\log q(\sigma^2)-1,
\end{align*}
where $\boldsymbol{\psi}_{-\sigma^2}$ means all the parameters except $\sigma^2$. Set $ \frac{\partial}{\partial q(\sigma^2)}\mathcal{L}(q(\sigma^2))  = 0$:
\begin{align*}
    q(\sigma^2) \propto &\exp\Big\{ \sum_{i=1}^n\Big[\frac{1}{2}\log\frac{1}{\sigma^2}-\frac{1}{2\sigma^2}\mathbb{E}_q\Big[\frac{1}{F_i}(w_i - \bb_{N[i]}\bw_{N[i]})^2\Big]\Big]+(a_{\sigma}+1)\log\frac{1}{\sigma^2} - \frac{b_{\sigma}}{\sigma^2}\Big\}\\
    \propto & (\frac{1}{\sigma^2})^{\frac{n}{2}+a_{\sigma}+1} \exp\Big\{-\frac{1}{\sigma^2}\Big(b_{\sigma}+\frac{1}{2}\sum_{i=1}^n\mathbb{E}_q\Big[\frac{1}{F_i}(w_i - \bb_{N[i]}\bw_{N[i]})^2\Big]\Big\}.\\
    q(\sigma^2) = & IG(\frac{n}{2}+a_{\sigma}, b_{\sigma}+\frac{1}{2}\sum_{i=1}^n\mathbb{E}_q\Big[\frac{1}{F_i}(w_i - \bb_{N[i]}\bw_{N[i]})^2\Big]).
\end{align*}

For spVB-MFA, the expectation of the quadratic form can be directly calculated as 
\begin{align*}
    &\sum_{i=1}^n\mathbb{E}_q\Big[\frac{1}{F_i}(w_i - \bb_{N[i]}\bw_{N[i]})^2\Big]\\
    =&\sum_{i=1}^n\Big[\mathbb{E}_q[\frac{1}{F_i}]({\mu_w}_i - \mathbb{E}_q[\bb_{N[i]}]\boldsymbol{\mu}_{wN[i]})^2+\mathbb{E}_q[\frac{1}{F_i}]\exp(J_i) +\mathbb{E}_q[\frac{1}{F_i}] \mathbb{E}_q[\bb_{N[i]}]\odot \mathbb{E}_q[\bb_{N[i]}] \odot \exp(\bJ_{N[i]})\Big],
\end{align*}
where $\odot$ is component-wise multiplication, $\boldsymbol{\mu}_{w}$ and $\bJ$ are computed in Appendix \ref{appendix:MFA}.

For spVB-NNGP, the expectation of the quadratic form can be estimated using the reparameterize trick and Monte Carlo simulation (See (\ref{eq:quadratic})).

We can compute $\mathbb{E}_q[\frac{1}{\sigma^2}] = \frac{a_\sigma^{'}}{b_{\sigma}^{'}}$, where $a_\sigma^{'}$ and $b_{\sigma}^{'}$ is the updated shape and scale parameters for $q(\sigma^2)$.

\paragraph{For $\phi$} Degenerated distribution.

The optimal point estimation of $\phi$ yields, i.e., $\phi^{*}$ the maximum value of $\mathcal{L}(\phi)$:
$$\mathcal{L}(\phi) = \frac{1}{2}\sum_{i=1}^n \Bigl(\log \mathbb{E}_q[\frac{1}{\sigma^2}]\frac{1}{F_i(\phi)} - \frac{1}{F_i(\phi)} \mathbb{E}_q[\frac{1}{\sigma^2}] \mathbb{E}_{q(\bw)}[(w_i - {\bb_{N[i]}(\phi)\bw_{N[i]}})^2 ] \Bigl).$$

Subsequently, we determine instances of  $\bb_{N[i]}(\phi)$ and $F_i(\phi)$ used in updates of the other parameters and evaluate these by 
plugging in the latest $\phi^*$.
\clearpage
\subsection{Computing the ELBO for spVB-NNGP} 
\label{appendix:spVB-NNGP_w}
The objective function $\mathcal{L}(\{\ba_{N_q}\},\bd,\boldsymbol{\eta})$ is as follows:
\begin{align*}
\mathcal{L}(\{\ba_{N_q}\},\bd,\boldsymbol{\eta}) & = \mathbb{E}_q[\log p(\by|\bX,\bw,\bbeta,\sigma^2,\tau^2)\tilde{p}(\bw)p(\bbeta)p(\sigma^2)p(\tau^2)] - \mathbb{E}_q[\log {q}(\bw)q(\bbeta,\sigma^2,\tau^2)]\\
& =\mathbb{E}_q[\log p(\by|\bX,\bw,\bbeta,\sigma^2,\tau^2)\tilde{p}(\bw)] - \mathbb{E}_q[\log {q}(\bw)] + \text{constant}\\
  &= \sum_{i=1}^n \mathbb{E}_q\Big[\frac{1}{2}\log \frac{1}{2\pi \tau^2} -\frac{1}{2\tau^2}(y_i-\bx_i^T\bbeta-w_i)^2\Big]\\
  &+ \sum_{i=1}^n\mathbb{E}_q\Big[\frac{1}{2}\log \frac{1}{2\pi \sigma^2 F_i} -\frac{1}{2\sigma^2F_i}(w_i - {\bb_{N[i]}
    \bw_{N[i]}})^2\Big] \\
    &-  \mathbb{E}_q\Big[-\frac{1}{2}(\bw - \boldsymbol{\eta}) ^T (\bI-\bA)^T\bD^{-1}(\bI-\bA)(\bw - \boldsymbol{\eta}) -\frac{1}{2}\log \text{det}\Big((\bI-\bA)^{-1}\bD(\bI-\bA)^{-T}\Big)\Big] \\
    & = \sum_{i=1}^n \mathbb{E}_q\Big[\frac{1}{2}\log \frac{1}{2\pi \tau^2} -\frac{1}{2\tau^2}(y_i-\bx_i^T\bbeta-w_i)^2\Big]\\
    &+ \sum_{i=1}^n\mathbb{E}_q\Big[\frac{1}{2}\log \frac{1}{2\pi \sigma^2 F_i} -\frac{1}{2\sigma^2F_i}(w_i - {\bb_{N[i]}
    \bw_{N[i]}})^2\Big]\\
& +\frac{1}{2}\sum_{i=1}^n \log{d_i} + \frac{n}{2}.
\end{align*}

Writing $\bw = \boeta + \bu$, where $q(\bu)=N(\bu \given \bzero, (\bI-\bA)^{-1}\bD(\bI-\bA)^{-T})$. Since, $\mathbb{E}_q[\bu] = 0$, we simplify the expected likelihood term 
\begin{align*}
    \sum_{i=1}^n \mathbb{E}_q\Big[ -\frac{1}{2\tau^2}(y_i-\bx_i^T\bbeta-w_i)^2\Big] &= \sum_{i=1}^n \mathbb{E}_q\Big[ -\frac{1}{2\tau^2}(y_i-\bx_i^T\bbeta-\eta_i - u_i)^2\Big]\\
    &= \sum_{i=1}^n \mathbb{E}_q\Big[ -\frac{1}{2\tau^2}(y_i-\bx_i^T\bbeta-\eta_i)^2 -\frac{1}{2\tau^2} u_i^2 + \frac{1}{\tau^2}(y_i-\bx_i^T\bbeta-\eta_i)u_i\Big]\\
    &= \sum_{i=1}^n \mathbb{E}_q\Big[ -\frac{1}{2\tau^2}(y_i-\bx_i^T\bbeta-\eta_i)^2 -\frac{1}{2\tau^2} u_i^2\Big].
\end{align*}

Similarly, we simplify the expected prior as 
\begin{align*}
    \sum_{i=1}^n \mathbb{E}_q\Big[ -\frac{1}{2\sigma^2F_i}(w_i - {\bb_{N[i]}
    \bw_{N[i]}})^2\Big] &= \sum_{i=1}^n \mathbb{E}_q\Big[ -\frac{1}{2\sigma^2F_i}(\eta_i + u_i - {\bb_{N[i]}
    (\boldsymbol{\eta}_{N[i]} + \bu_{N[i]}}))^2\Big]\\
    &= \sum_{i=1}^n \mathbb{E}_q\Big[ -\frac{1}{2\sigma^2F_i}(\eta_i - {\bb_{N[i]}
    \boldsymbol{\eta}_{N[i]}})^2-\frac{1}{2\sigma^2F_i}(u_i - {\bb_{N[i]}
    \bu_{N[i]}})^2\\
    & - \frac{1}{\sigma^2F_i}(\eta_i - {\bb_{N[i]}
    \boldsymbol{\eta}_{N[i]}})(u_i - {\bb_{N[i]}
    \bu_{N[i]}}) \Big]\\
    &= \sum_{i=1}^n \mathbb{E}_q\Big[ -\frac{1}{2\sigma^2F_i}(\eta_i - {\bb_{N[i]}
    \boldsymbol{\eta}_{N[i]}})^2-\frac{1}{2\sigma^2F_i}(u_i - {\bb_{N[i]}
    \bu_{N[i]}})^2\Big].
\end{align*}

The objective function is given as 
\begin{equation}
\begin{split}
    \mathcal{L}(\{\ba_{N_q}\},\bd,\boldsymbol{\eta}) = \sum_{i=1}^n \mathbb{E}_q &
   \Big[ -\frac{1}{2\tau^2} (y_i-{\bx_i}^T\bbeta - \eta_i)^2 
    -\frac{1}{2\tau^2} u_i^2 
    -\frac{1}{2\sigma^2F_i}(\eta_{i} - {\bb_{N[i]}
    \boeta_{N[i]}})^2 \\
    & - \frac{1}{2\sigma^2F_i}(u_i - {\bb_{N[i]}
    \bu_{N[i]}})^2 \Big] 
    + \frac{1}{2} \sum_{i=1}^n \log d_i + \frac{n}{2}.
\end{split}\nonumber
\end{equation}

\clearpage

\subsection{Model structure for spVB-NNGP-joint}
\label{appendix:spVB-NNGP-joint}
We offer a joint variational model for $\bbeta$ and $\bw$, i.e., $q(\bbeta,\bw) = N\Big (\begin{pmatrix}
\boldsymbol{\mu_\beta} \\
\boldsymbol{\eta}
\end{pmatrix}, (\bI-\bA^{*})^{-1}\bD^{*}(\bI-\bA^{*})^{-T}\Big )$. 
Similarly, we use the reparameterization trick for $\begin{pmatrix}
\bbeta \\
\bw
\end{pmatrix} = \begin{pmatrix}
\boldsymbol{\mu_\beta} \\
\boldsymbol{\eta}
\end{pmatrix} + \begin{pmatrix}
\bu_{\bbeta} \\
\bu
\end{pmatrix}$, where $\begin{pmatrix}
\bu_{\bbeta} \\
\bu
\end{pmatrix} = \begin{pmatrix}
\bI-\bA^{*}
\end{pmatrix}^{-1}\bD^{*1/2}\boldsymbol{\xi}^{*}$ and $\boldsymbol{\xi}^{*} \sim N(0,\bI_{n+p})$. We can proceed $\begin{pmatrix}
\bu_{\bbeta} \\
\bu
\end{pmatrix} $ using back-solve:
\begin{equation}
\begin{split}
    u_{\bbeta_1} &= \exp(\gamma_{\bbeta_1}) \xi_{\bbeta_1} \\
    u_{\bbeta_2} &= \exp(\gamma_{\bbeta_2}) \xi_{\bbeta_2} + \bl_{\bbeta21}u_{\bbeta_1} \\
    \vdots\\
    u_{\bbeta_p} &= \exp(\gamma_{\bbeta_p}) \xi_{\bbeta_p} + \sum_{j=1}^{p-1} \bl_{\bbeta_pj}u_{\bbeta_j} \\    
    u_1&=\exp(\gamma_1) \xi_1 + \sum_{j=1}^{p} \ba_{\bbeta_1j}u_{\bbeta_j} \\
    u_2&=\exp(\gamma_2) \xi_2 + {\ba_{N_q[2]}}{\bu_{N_q[2]}} + \sum_{j=1}^{p} \ba_{\bbeta_2j}u_{\bbeta_j}\\
    \vdots\\
    u_n&=\exp(\gamma_n) \xi_n + {\ba_{N_q[n]}}{\bu_{N_q[n]}}  + \sum_{j=1}^{p} \ba_{\bbeta_nj}u_{\bbeta_j}.
\end{split}
\end{equation}

In the joint modeling approach, which we refer to as {\em spVB-NNGP-joint}, the variational distribution $q(\bbeta,\bw)$ captures both the covariance within each parameter and also the correlation between them. Specifically, the parameters $\bl_{\bbeta}$ quantify the correlation structure within the regression coefficients $\bbeta$, forming the dense top $p\times p$ block of matrix $\bA^{*}$, which resembles the Cholesky factor of the regression coefficient covariance. While the parameters $\ba_{\bbeta}$ characterize the dependencies between the regression coefficients $\bbeta$ and $\bw$, and correspond to the $n\times p$ cross-block in $\bA^{*}$. The lower $n\times n$ triangular block retains the same sparsity pattern as in the independent model (spVB-NNGP). $\bD^{*}$ is a diagonal matrix whose diagonal elements are given by $(\exp(\gamma_{\bbeta_1})^2,...,\exp(\gamma_{\bbeta_p})^2,\exp(\gamma_n)^2,...,\exp(\gamma_n)^2)$. Leveraging this sparsity, sampling from the joint variational distribution $q(\bbeta,\bw)$ remains computationally efficient, allowing the same reparameterization and sampling strategies previously discussed for spVB-NNGP. Moreover, the hierarchical formulation enables closed-form gradient computation with respect to the variational parameters $\gamma_{\bbeta}$, $\bl_{\bbeta}$ and $\ba_{\bbeta}$.

\clearpage

\section{Mean-field variational approximation: spVB-MFA}
\label{appendix:MFA}
Mean-field variational approximation (MFA) assumes $q(\bw)$ to follow a multivariate normal distribution characterized by a diagonal covariance matrix. This assumption allows the joint distribution of the parameters to be factorized into a product of independent normal distributions: $$q(\bw) = \prod_{i=1}^n q(w_i) = \prod_{i=1}^n N(w_i|\mu_{w_i}, G_i) = N(\bw| \boldsymbol{\mu}_{w}, \mathrm{diag}(G_1, G_2,..., G_n)).$$

To ensure the non-negativity of the variance term in the gradient optimization process that follows, we implement a log transformation for $G_i$ as $J_i = \log(G_i)$. The variational distribution $q(\bw)$ is characterized by parameters $\boldsymbol{\mu}_{w}$ and $\bJ = (J_1,J_2,...,J_n)$. Our objective is to identify the values of these parameters that maximize the ELBO: 
\begin{align*}
    \mathcal{L}(\bw;\bJ,\boldsymbol{\mu}_{w}) &= \sum_{i=1}^n \mathbb{E}_q\Big[\frac{1}{2}\log \frac{1}{2\pi\tau^2}-\frac{1}{2\tau^2}(y_i-\bx_i^T\bbeta - w_i)^2 +\frac{1}{2}\log \frac{1}{2\pi\sigma^2F_i}- \frac{1}{2\sigma^2}\frac{1}{F_i}(w_i - \bb_{N[i]}\bw_{N[i]})^2\Big] \\
    &- \sum_{i=1}^n \mathbb{E}_q[-\frac{1}{2}J_i-\frac{1}{2}\frac{1}{\exp(J_i)}(w_i - {\mu_w}_i)^2 ] +\text{constant} \\
    & = -\frac{1}{2}\mathbb{E}[\frac{1}{\tau^2}]\sum_{i=1}^n\Big[(y_i-\bx_i^T\mathbb{E}_q[\bbeta]-{\mu_w}_i)^2 + \exp(J_i)\Big] \\
    &+\sum_{i=1}^n\Big[\frac{1}{2}\log\frac{1}{2\pi} \mathbb{E}_q[\frac{1}{\sigma^2}]\mathbb{E}_q[\frac{1}{F_i}]-\frac{1}{2}\mathbb{E}_q[\frac{1}{\sigma^2}]\mathbb{E}_q[\frac{1}{F_i}]({\mu_w}_i - \mathbb{E}_q[\bb_{N[i]}]\boldsymbol{\mu}_{wN[i]})^2 \\
    &- \frac{1}{2}\mathbb{E}[\frac{1}{\sigma^2}]\mathbb{E}_q[\frac{1}{F_i}]\exp(J_i) -\frac{1}{2}\mathbb{E}[\frac{1}{\sigma^2}]\mathbb{E}_q[\frac{1}{F_i}] \mathbb{E}_q[\bb_{N[i]}]\odot \mathbb{E}_q[\bb_{N[i]}] \odot \exp(\bJ_{N[i]})\Big] \\
    &+\frac{n}{2} + \frac{1}{2}\sum_{i=1}^n J_i,
\end{align*}
where $\odot$ is component-wise multiplication.
We compute the gradients of $\mathcal{L}(\bw;\bJ,\boldsymbol{\mu}_{w})$ with respect to $\boldsymbol{\mu}_{w}$ and $\bJ$ to get $\Delta_{\boldsymbol{\mu}_{w}} \mathcal{L}(\bw;\bJ,\boldsymbol{\mu}_{w})$ and $\Delta_{{\bJ}} \mathcal{L}(\bw;\bJ,\boldsymbol{\mu}_{w})$. The $i^{\text{th}}$ elements of the n-length vector $\Delta_{\boldsymbol{\mu}_{w}} \mathcal{L}(\bw;\bJ,\boldsymbol{\mu}_{w})$ and $\Delta_{{\bJ}} \mathcal{L}(\bw;\bJ,\boldsymbol{\mu}_{w})$ are as follows: 
\begin{align}
    \frac{\partial \mathcal{L}(\bw;\bJ,\boldsymbol{\mu}_{w}) }{\partial {\mu_{wi}}}=&\mathbb{E}_q[\frac{1}{\tau^2}](y_i-\bx_i^T\mathbb{E}_q[\bbeta]-{\mu_{wi}})-\mathbb{E}_q[\frac{1}{\sigma^2}]\mathbb{E}_q[\frac{1}{F_i}]({\mu_{wi}} - \mathbb{E}_q[\bb_{N[i]}]\boldsymbol{\mu}_{wN[i]})\nonumber \\
    +&\mathbb{E}_q[\frac{1}{\sigma^2}]\sum_{l:i \in N[l]}\mathbb{E}_q[\frac{1}{F_l}]({\mu_{wl}} - \mathbb{E}_q[\bb_{N[l]}]\boldsymbol{\mu}_{w{N[l]}})\mathbb{E}_q[\bb_{N[l],i}], \label{eq:MFAupdate_mu}\\
    \frac{\partial \mathcal{L}(\bw;\bJ,\boldsymbol{\mu}_{w})}{\partial J_i} =& \Big[-\frac{1}{2}\mathbb{E}_q[\frac{1}{\tau^2}]-\frac{1}{2}\mathbb{E}_q[\frac{1}{\sigma^2}]\mathbb{E}_q[\frac{1}{F_i}] - \frac{1}{2}\mathbb{E}_q[\frac{1}{\sigma^2}]\sum_{l: i\in N_{[l]}}\Big(\mathbb{E}_q[\frac{1}{F_l}]\mathbb{E}_q[b^2_{N[l],t}]\Big)+\frac{1}{2}\frac{1}{\exp(J_i)}\Big]\exp(J_i).\label{eq:MFAupdate_J}
\end{align} We then apply $\Delta_{\boldsymbol{\mu}_{w}} \mathcal{L}(\bw;\bJ,\boldsymbol{\mu}_{w})$ and $\Delta_{{\bJ}} \mathcal{L}(\bw;\bJ,\boldsymbol{\mu}_{w})$ to optimize the parameters $\boldsymbol{\mu}_{w}$ and $\bJ$, through gradient ascent algorithm,  thereby achieving the maximization of the ELBO. The approach outlined in this section, which incorporates the NNGP prior and utilizes the MFA as the variational distribution, closely resembles the methodology presented in the Variational Nearest Neighbor Gaussian Process (VNNGP) framework \citep{wu2022variational}. The variational distributions for spatial covariance parameters, regression coefficients, and random error variance are the same as spVB-NNGP, as computed in Appendix \ref{appendix:calculus}. The algorithm for implementing spVB-MFA can be found in Algorithm \ref{alg:MFA}
.
 \begin{algorithm}[!ht]
\caption{spVB-MFA: Mean-Field Approximation Variational Inference}\label{alg:MFA}
\footnotesize
\begin{algorithmic}
\State Specify the value of parameters in the prior distribution for $\tau^2 \sim IG(a_{\tau},b_{\tau})$, $\sigma^2 \sim IG(a_{\sigma},b_{\sigma})$.
\State Give initial values to $\mathbb{E}_q[\frac{1}{\tau^2}]^{(0)}$,$\mathbb{E}_q[\frac{1}{\sigma^2}]^{(0)}$, $\mathbb{E}_q[\phi]^{(0)}$, $\boldsymbol{\mu}_{w}^{(0)}$ and  $\bG^{(0)}=(G_1^{(0)},G_2^{(0)},...,G_n^{(0)})$.
\State Give values for $\phi_{min}$, $\phi_{max}$, AdaDelta algorithm noise $\delta$ and input rate $r$.
\For{$t=1$ to $T$}
\State \textbf{Step 1:} Update the distribution of $\bbeta \sim N(\boldsymbol{\mu_\beta}^{(t)},\bV_{\boldsymbol\beta}^{(t)})$ where 

$\mathbf{V_{\boldsymbol \beta}}^{(t)} = \Big [\mathbb{E}_q^{(t-1)}(\frac{1}{\tau^2})\Big ]^{-1} (\bX^T\bX)^{-1}$, $\boldsymbol \mu_{\boldsymbol \beta}^{(t)} = (\bX^T\bX)^{-1} \bX^T(\by-\boldsymbol \mu_{w}^{(t-1)})$.

\State \textbf{Step 2:} Update the distribution of 
$\tau^2 \sim IG$ with $a_{\tau}^{*(t)} = a_\tau +\frac{n}{2}$ and

$b_{\tau}^{*(t)} = b_{\tau}+\frac{1}{2}\Big[\sum_{i=1}^n G_i^{(t-1)}+p\{\mathbb{E}_q^{(t-1)}[\frac{1}{\tau^2}]\}^{-1}+(\by-\boldsymbol{\mu}_w^{(t-1)})^T(\bI-\bH)(\by-\boldsymbol{\mu}_w^{(t-1)})\Big]$, 

where $\bH = \bX(\bX^T\bX)^{-1} \bX^T$.
\State Calculate $E^{(t)}(\frac{1}{\tau^2}) = \frac{a_{\tau}^{*(t)}}{b_{\tau}^{*(t)}}$.
\State \textbf{Step 3:} Update $\sigma^2 \sim IG$ with parameters $a_{\sigma}^{*(t)} = a_{\sigma} + \frac{n}{2}$ and 

$b_{\sigma}^{*(t)} = b_{\sigma}+\frac{1}{2}\sum_{i=1}^n\mathbb{E}_q^{(t-1)}[\frac{1}{F_i}(w_i - \bb_{N[i]}\bw_{N[i]})^2]$, where

$\sum_{i=1}^n\mathbb{E}_q^{(t-1)}\Big[\frac{1}{F_i}(w_i - \bb_{N[i]}\bw_{N[i]})^2\Big]
    =\sum_{i=1}^n\Big[\mathbb{E}_q^{(t-1)}[\frac{1}{F_i}]({\mu_w}^{(t)}_i - \mathbb{E}_q^{(t-1)}[\bb_{N[i]}]\boldsymbol{\mu}^{(t)}_{wN[i]})^2+\mathbb{E}_q^{(t-1)}[\frac{1}{F_i}]\exp(J_i^{(t-1)}) +\mathbb{E}_q^{(t-1)}[\frac{1}{F_i}] \mathbb{E}_q^{(t-1)}[\bb_{N[i]}]\odot \mathbb{E}_q^{(t-1)}[\bb_{N[i]}] \odot \exp(\bJ^{(t-1)}_{N[i]})\Big].$

\State Calculate $E^{(t)}(\frac{1}{\sigma^2}) = \frac{a_{\sigma}^{*(t)}}{b_{\sigma}^{*(t)}}$.

\State \textbf{Step 4:} 
Update $\phi$ with a degenerate (single point-mass) distribution.

    \State Calculate numerical gradient for $\nabla_{\phi} \mathcal{L}(\phi)$, 
    
    where $\mathcal{L}(\phi) = \frac{1}{2}\sum_{i=1}^n \Bigl(\log \mathbb{E}^{(t)}_q[\frac{1}{\sigma^2}]
\frac{1}{F_i(\phi)} - \frac{1}{F_i(\phi)} \mathbb{E}^{(t)}_q[\frac{1}{\sigma^2}] \mathbb{E}_{q(\bw)}[(w_i - {\bb_{N[i]}(\phi)\bw_{N[i]}})^2 ] \Bigl).$

    \State Using AdaDelta optimizer in (\ref{eq:adadelta}) to adapt the learning rate and get $\Delta_{\phi}^{(t)}$.
 \State Update $  \phi^{(t)} = \phi^{(t-1)} + \Delta_{\phi}^{(t)}$.

\State Update $\mathbb{E}_q^{(t)}[\frac{1}{F_i}] = \frac{1}{F_i(\phi^{(t)})}$ and $\mathbb{E}_q^{(t)}[\bb_{N[i]}]=\bb_{N[i]}(\phi^{(t)})$ using (\ref{eq:NNGP_prior}). 

\State \textbf{Step 5:} Update the distribution of $\bw$ with parameter  $\boldsymbol{\mu}_{w}$ and $\bG=(G_1,G_2,...,G_n)$.

 \State Compute gradient using $\nabla_{  \boldsymbol{\mu}_{w}} \mathcal{L}(\bw;\boldsymbol{\mu}_{w},\bJ)$ using (\ref{eq:MFAupdate_mu}).
 \State Using AdaDelta optimizer in (\ref{eq:adadelta}) to adapt the learning rate and get $\Delta_{  \boldsymbol{\mu}_{w}}^{(t)}$.
 \State Update ${  \boldsymbol{\mu}_{w}}^{(t)} = {  \boldsymbol{\mu}_{w}}^{(t-1)} + \Delta_{  \boldsymbol{\mu}_{w}}^{(t)}$.
 
 \State Compute gradient $\nabla_{\bJ} \mathcal{L}(\bw;\boldsymbol{\mu}_{w},\bJ)$ using (\ref{eq:MFAupdate_J}).
 \State Using AdaDelta optimizer in (\ref{eq:adadelta}) to adapt the learning rate and get $\Delta_{  \bJ}^{(t)}$.
 \State Update $  \bJ^{(t)} = \bJ^{(t-1)} + \Delta_{  \bJ}^{(t)}$.
 \State Update $  \bG^{(t)} = \exp(\bJ^{(t)})$.
 \State \textbf{Early Stop regularization:} Terminate if ELBO does not improve for $T_{\text{patience}}$ consecutive iterations
\EndFor
\end{algorithmic}
\end{algorithm}
\clearpage
\section{Linear response method for spatial data}
\label{appendix:LRVB}
In this section, we provide a detailed description of conducting linear response methods for spatial data. We follow the four steps for calculating linear response corrected covariance.

\paragraph{Find the spVB-MFA optimum $q^*$.}

For the mean-field approximation model (spVB-MFA), where $\bX \in \mathbb{R}^{n \times p}$, regression coefficients $\bbeta \in \mathbb{R}^p$, spatial random effects $\bw \in \mathbb{R}^n$, and outcome vector $\by \in \mathbb{R}^n$. As mentioned before, we employ the BRISC estimator for $\sigma^2$, $\tau^2$ and $\phi$ as linear response methods relying on the accuracy of estimating the posterior mean of parameters. Assuming a fully independent spVB-MFA model, the variational distribution is given as:
$$
q(\bbeta, \bw) =\prod_{j=1}^pq(\beta_j) \prod_{i=1}^n q(w_i),
$$
where
$$
q(\beta_j) = N({\mu}_{\beta_j}, \sigma^2_{\beta_j}), \quad
q(w_i) = N(\mu_{w_i}, G_i), \quad j = 1,..., p, \quad i = 1, \dots, n.
$$

For the optimal solution to $q(\bbeta) = N(\bmu_{\bbeta},\mathrm{diag}(\sigma^2_{\beta_j}))$, we first get the Evidence Lower Bound (ELBO) involving $\bmu_{\bbeta}, \mathrm{diag}(\sigma^2_{\beta_j})$ as:
\begin{align*}
\mathcal{L}(q(\bbeta)) 
&= - \frac{1}{2\tau^2} \mathbb{E}_q[ (\by - \bw - \bX \bbeta)^T(\by - \bw - \bX \bbeta)] - \mathbb{E}_q[\log(q(\bbeta))] + \text{constant}\\
& = - \frac{1}{2\tau^2} \mathbb{E}_q[ \bbeta^T\bX^T\bX\bbeta - (\by-\bw)^T\bX\bbeta - \bbeta^T\bX^T(\by-\bw)] + \frac{1}{2}\sum_{j=1}^p\log(\sigma^2_{\beta_j})\\
& = - \frac{1}{2\tau^2} [ \bmu_{\bbeta}^T\bX^T\bX\bmu_{\bbeta} +Tr(\bX^T\bX \mathrm{diag}(\sigma^2_{\beta_j})) - (\by-\bmu_\bw)^T\bX\bmu_\bbeta - \bmu_\bbeta^T\bX^T(\by-\bmu_\bw)] \\
&+ \frac{1}{2}\sum_{j=1}^p\log(\sigma^2_{\beta_j}).
\end{align*}

Computing the gradient with respect to $\bmu_\bbeta$ and setting the gradient to zero gives:

$$
\frac{\partial \mathcal{L}}{\partial \bmu_\bbeta} 
= \frac{1}{\tau^2} \bX^T (\by - \bmu_\bw - \bX \bmu_\bbeta) 
= -\frac{1}{\tau^2} \bX^T \bX \bmu_\bbeta + \frac{1}{\tau^2} \bX^T (\by - \bmu_\bw).
$$

$$
\bX^T \bX \bmu_\bbeta = \bX^T (\by - \bmu_\bw) \quad \Rightarrow \quad
 \bmu_\bbeta = (\bX^T \bX )^{-1} \bX^T (\by - \bmu_\bw).
$$

The ELBO terms involving \( \sigma^2_{\beta_j} \) are:
$$
\mathcal{L}(\sigma^2_{\beta_j}) = -\frac{1}{2\tau^2} \sigma^2_{\beta_j} \|x_j\|^2 + \frac{1}{2} \log(\sigma^2_{\beta_j}),
$$
so the gradient is:
$$
\frac{\partial \mathcal{L}}{\partial \sigma^2_{\beta_j}} 
= -\frac{1}{2\tau^2} \|x_j\|^2 + \frac{1}{2\sigma^2_{\beta_j}}.
$$
Setting the derivative to zero yields:
$$ \sigma^2_{\beta_j} = \frac{\tau^2}{\|x_j\|^2}. $$

The gradient of $\mu_{w_i}$ and $G_i$ is the same as in (\ref{eq:MFAupdate_mu}) and (\ref{eq:MFAupdate_J}).

The optimal solution for $q(\bbeta)$ and $q(\bw)$ is obtained by optimizing the variational parameters $\bmu_{\bbeta}$, $\sigma^2_{\beta_j}$'s, $\mu_{w_i}$ and $G_i$ using an iterative gradient ascent procedure until convergence, following an approach analogous to spVB-MFA.

\paragraph{Computing the covariance $\bV$ of $q^*$.}

Define the sufficient statistics as $\boldsymbol{\theta} = (\{\beta_j\}, \{w_i\}, \{\beta_j^2\}, \{w_i^2\})$ and let ${M}:= \mathbb{E}_q[\boldsymbol{\theta}]$ denote the mean parameterization. $\boldsymbol{\theta}$ can be partition into subvector $\boldsymbol{\alpha} = (\{\beta_j\}, \{w_i\})$ and the remaining sufficient statistics $\bz = (\{\beta_j^2\}, \{w_i^2\})$. Thus, $M = \mathbb{E}_q[\boldsymbol{\theta}] = \mathbb{E}_q[(\boldsymbol{\alpha},\bz)]$. 

The structure of the covariance under $q$  is:
$$
\bV:=\mbox{Cov}_q(\boldsymbol{\theta})=\begin{bmatrix}
\mbox{Cov}(\boldsymbol{\alpha}) & \mbox{Cov}(\boldsymbol{\alpha},\bz)\\
\mbox{Cov}(\bz,\boldsymbol{\alpha})      & \mbox{Cov}(\bz)
\end{bmatrix} = \begin{bmatrix}
\bV_{\boldsymbol{\alpha}} & \bV_{\alpha \bz}\\
\bV_{\bz\alpha}      & \bV_{\bz}
\end{bmatrix},
\qquad
\bV_{\boldsymbol{\alpha}}=
\begin{bmatrix}
\mathrm{diag}(\sigma^2_{\beta_j}) & 0\\
0       & \mathrm{diag}(G_i)
\end{bmatrix}.
$$

\paragraph{Computing the Hessian of the expected log-posterior.}
The linear response objective is defined as the expected log-posterior under the variational distribution:
\begin{align*}
    \bL &= \mathbb{E}_q[\log p(\boldsymbol{\theta} | \by, \bX)]\\
     &= \mathbb{E}_q[\log p(\by|\boldsymbol{\theta},\bX) + \log p(\boldsymbol{\theta})] + \text{constant}\\
     &= \sum_{i=1}^n \mathbb{E}_q\left[
        -\frac{1}{2\tau^2}(y_i - \bx_i^\top \bbeta - w_i)^2 
        - \frac{1}{2\sigma^2 F_i}(w_i - \bb_{N[i]}^\top \bw_{N[i]})^2
    \right].
\end{align*}

Given fixed values of $\tau^2$, $\phi$ and $\sigma^2$ and assuming that the variational distributions are independent across $\beta_j$ and across $w_i$, we expand the squared terms accordingly.
\begin{align*}
\bL 
&= -\frac{1}{2\tau^2} \sum_{i=1}^n 
\Bigg[ 
      \sum_{j} x_{ij}^2 \, \mathbb{E}_q[\beta_j^2]
    + \sum_{\substack{j,k \\ j \ne k}} x_{ij} x_{ik} \, \mathbb{E}_q[\beta_j] \mathbb{E}_q[\beta_k] 
    - 2 y_i \sum_{j} x_{ij} \mathbb{E}_q[\beta_j]
    + 2 \mathbb{E}_q[w_i] \sum_{j} x_{ij} \mathbb{E}_q[\beta_j] \\
&\phantom{++++++}\quad 
    + y_i^2 
    - 2 y_i \mathbb{E}_q[w_i]
    + \mathbb{E}_q[w_i^2]
\Bigg] \\
&\quad - \frac{1}{2\sigma^2} \sum_{i=1}^n \frac{1}{F_i} 
\Bigg[ 
      \mathbb{E}_q[w_i^2]
    - 2 \sum_{j \in N[i]} b_{ij} \mathbb{E}_q[w_i] \mathbb{E}_q[w_j] 
    + \sum_{j \in N[i]} b_{ij}^2 \, \mathbb{E}_q[w_j^2] 
    + \sum_{\substack{j, k \in N[i] \\ j \ne k}} b_{ij} b_{ik} \, \mathbb{E}_q[w_j] \mathbb{E}_q[w_k]
\Bigg].
\end{align*}

The first-order derivatives of $\bL$ with respect to the mean parameters ${M}$ are:
\begin{align*}
\frac{\partial\bL}{\partial \mathbb{E}_q[\beta_j]}
&= -\frac{1}{\tau^2} \sum_{i=1}^n x_{ij} \left( \mathbb{E}_q[w_i] - y_i + x_{ik} \mathbb{E}_q[\beta_k] \right),\\
\frac{\partial\bL}{\partial \mathbb{E}_q[w_i]}
&= -\frac{1}{\tau^2} \left(\sum_{j} x_{ij} \mathbb{E}_q[\beta_j] - y_i \right) - \frac{1}{\sigma^2}  \left( 
        - \frac{1}{F_i}\sum_{j \in N[i]} b_{ij} \mathbb{E}_q[w_j]
        + \sum_{k: i \in N[k]}\frac{b_{ki}}{F_k} \sum_{j\in N[k]} b_{kj}\mathbb{E}_q[w_j] \right),\\
\frac{\partial\bL}{\partial \mathbb{E}_q[\beta_j^2]}
&= -\frac{1}{2\tau^2} \sum_{i=1}^n x_{ij}^2,\\
\frac{\partial\bL}{\partial \mathbb{E}_q[w_i^2]}
&= -\frac{1}{2\tau^2}-\frac{1}{2\sigma^2}(\frac{1}{F_i} + \sum_{k:i\in N[k]}\frac{b_{ki}^2}{F_k}).
\end{align*}

The second derivatives are:
\begin{align*}
\frac{\partial^2\bL}
     {\partial 
     \mathbb{E}_q[\beta_j]\,\partial \mathbb{E}_q[\beta_k]}
&=
-\frac{\mathbb I(j \neq k)}{\tau^2}\sum_{i=1}^n x_{ij}x_{ik},
\\
\frac{\partial^2\bL}
     {\partial \mathbb{E}_q[\beta_j]\,\partial \mathbb{E}_q[w_i]}
&=-\frac1{\tau^2}x_{ij},\\
\frac{\partial^2\bL}
     {\partial  \mathbb{E}_q[w_i]\,\partial  \mathbb{E}_q[w_j]}
&=
\dfrac{\mathbb I(i \neq j)}{\sigma^2}\left(
\dfrac{b_{ij}}{F_i}\,\mathbb{I}_{j \in N[i]}
+ \dfrac{b_{ji}}{F_j}\,\mathbb{I}_{i \in N[j]}
- \displaystyle\sum_{k : i,j \in N[k]} \dfrac{b_{ki}b_{kj}}{F_k}
\right).
\end{align*}

Higher-order moment terms, i.e., $\mathbb{E}_q[\bz] =(\{\mathbb{E}_q[\beta_j^2]\}, \{\mathbb{E}_q[w_i^2]\})$ do not interact with each other or with the first-order terms $\mathbb{E}_q[\boldsymbol{\alpha}] =(\{\mathbb{E}_q[\beta_j]\}, \{\mathbb{E}_q[w_i]\})$ in the expansion of $\bL$. As a result, all second partial derivatives involving $\mathbb{E}_q[\bz]$ are zero.

Given the above expression, the full Hessian with respect to the expectation parameters $M = \mathbb{E}_q[\boldsymbol{\theta}] = \mathbb{E}_q[(\boldsymbol{\alpha},\bz)] $ can be written as:
$$
\bH = \begin{bmatrix}
\frac{\partial^2 \bL}{\partial \mathbb{E}_q[\boldsymbol{\alpha}]\partial \mathbb{E}[\boldsymbol{\alpha}]^T} & \frac{\partial^2 \bL}{\partial\mathbb{E}_q[\boldsymbol{\alpha}]\partial \mathbb{E}[\bz]^T}\\
\frac{\partial^2 \bL}{\partial \mathbb{E}[\bz]\partial\mathbb{E}_q[\boldsymbol{\alpha}]^T}                & \frac{\partial^2 \bL}{\partial \mathbb{E}[\bz]\partial \mathbb{E}[\bz]^T}
\end{bmatrix}=\begin{bmatrix}
\bH_{\alpha} & 0\\
0                & 0
\end{bmatrix},\quad
\bH_{\boldsymbol{\alpha}} = 
\begin{bmatrix}
\left[\frac{\partial^2\bL}
     {\partial \mathbb{E}_q[\beta_j]\,\partial \mathbb{E}_q[\beta_k]}\right]       & -\frac{1}{\tau^2}\bX       \\
-\frac{1}{\tau^2}\bX^\top 
         & \left[ \frac{\partial^2\bL}{\partial \mathbb{E}_q[w_i] \partial \mathbb{E}_q[w_j]} \right]
\end{bmatrix}.
$$

\paragraph{Computing the matrix inverse to obtain the corrected covariance.}

As only the covariance of the subvector $\boldsymbol{\alpha}$ is of interest, we show that the top $(\boldsymbol{\alpha}, \boldsymbol{\alpha})$  block of $\bSigma=(\bI - \bV \bH)^{-1} \bV$ is equal to 
$$
(\bI_{\boldsymbol{\alpha}} - \bV_{\boldsymbol{\alpha}} \bH_{\boldsymbol{\alpha}})^{-1} \bV_{\boldsymbol{\alpha}}.
$$

We first expand:
\begin{align*}
\bI - \bV \bH 
= \begin{bmatrix}
\bI_{\boldsymbol{\alpha}} - \bV_{\boldsymbol{\alpha}} \bH_{\boldsymbol{\alpha}} & \mathbf{0} \\
- \bV_{\bz\boldsymbol{\alpha}} \bH_{\boldsymbol{\alpha}} & \bI_{\bz}
\end{bmatrix},
\end{align*}
and apply the block matrix inverse identity:
\begin{align*}
(\bI - \bV \bH)^{-1} 
= \begin{bmatrix}
(\bI_{\boldsymbol{\alpha}} - \bV_{\boldsymbol{\alpha}} \bH_{\boldsymbol{\alpha}})^{-1} & \mathbf{0} \\
\bV_{\bz\boldsymbol{\alpha}} \bH_{\boldsymbol{\alpha}} (\bI_{\boldsymbol{\alpha}} - \bV_{\boldsymbol{\alpha}} \bH_{\boldsymbol{\alpha}})^{-1} & \bI_\bz
\end{bmatrix}.
\end{align*}

Multiplying by $\bV$, we obtain:
\begin{align*}
\widehat{\bSigma} 
= (\bI - \bV \bH)^{-1} \bV 
= \begin{bmatrix}
(\bI_{\boldsymbol{\alpha}}  - \bV_{\boldsymbol{\alpha}}  \bH_{\boldsymbol{\alpha}} )^{-1} & \mathbf{0} \\
\bV_{\bz\boldsymbol{\alpha}} \bH_{\boldsymbol{\alpha}}  (\bI_{\boldsymbol{\alpha}}  - \bV_{\boldsymbol{\alpha}}  \bH_{\boldsymbol{\alpha}} )^{-1} & \bI_\bz
\end{bmatrix}
\begin{bmatrix}
\bV_{\boldsymbol{\alpha}}  & \bV_{\boldsymbol{\alpha} \bz} \\
\bV_{\bz\boldsymbol{\alpha}} & \bV_{\bz}
\end{bmatrix}.
\end{align*}

Thus, the \((\boldsymbol{\alpha}, \boldsymbol{\alpha})\) block of the corrected covariance is:
$$
\widehat{\bSigma}_{\boldsymbol{\alpha}} 
= (\bI_{\boldsymbol{\alpha}} - \bV_{\boldsymbol{\alpha}}\bH_{\boldsymbol{\alpha}})^{-1} \bV_{\boldsymbol{\alpha}}.
$$

This significantly reduces computational complexity and avoids inverting the full Hessian across all moment parameters.

\clearpage
\section{Algorithms implementation}
\label{appendix:al_implement}
\paragraph{Learning rate.}
The learning rate is a scalar that determines how much the step size (weights) should be adjusted in response to the gradient during the optimization process. A high learning rate might result in excessively large step sizes, leading to an unstable training process. Conversely, a low learning rate could cause longer running time and increase the likelihood of getting stuck in local minima. The selection of an appropriate learning rate is crucial, as it directly influences the rate of convergence. A parameter-specific learning rate is often preferable and there are many different choices. For the parameters $\blambda$, we utilize the AdaDelta optimizer \citep{zeiler2012adadelta}, which adapts the learning rate across iterations. The AdaDelta algorithm, with an input rate $r$ and a smoothing term $\delta$ is defined as follows:
\begin{align}
    &\text{Compute original scale gradient of the full-batch ELBO $\mathcal{L}(\boldsymbol{\lambda})$ or the mini-batch ELBO} \nonumber\\
    & \calL_\calB(\blambda) \text{ with repsect to $\boldsymbol{\lambda}$ at iteration $t$: } g_{\boldsymbol{\lambda}}^{(t)} = \Delta_{\boldsymbol{\lambda}}  \mathcal{L}(\boldsymbol{\lambda}) \mbox{ or } g_{\boldsymbol{\lambda}}^{(t)} = \Delta_{\boldsymbol{\lambda}}  \mathcal{L}_\calB(\boldsymbol{\lambda}) \nonumber, \\
    &\text{Accumulate gradient at iteration $t$: } \mathbb{E}^{(t)}[g^2_{\boldsymbol{\lambda}}] = r\mathbb{E}^{(t-1)}[g^2_{\boldsymbol{\lambda}}] +(1-r)g^{(t)^2}_{\boldsymbol{\lambda}},\nonumber\\
     &\text{Compute the change at iteration $t$: } 
\Delta_{\boldsymbol{\lambda}}^{(t)}=\frac{\sqrt{\mathbb{E}^{(t-1)}[\Delta^2_{\boldsymbol{\lambda}}]+\delta}}{\sqrt{\mathbb{E}^{(t)}[g^2_{\boldsymbol{\lambda}}]+\delta}}g^{(t)}_{\boldsymbol{\lambda}},\nonumber\\
   &\text{Accumulate change at iteration $t$: }\mathbb{E}^{(t)}[\Delta^2_{\boldsymbol{\lambda}}] = r \mathbb{E}^{(t-1)}[\Delta^2_{\boldsymbol{\lambda}}] +(1-r)\Delta^{(t)^2}_{\boldsymbol{\lambda}}\nonumber,\\
   &\text{Use the adapted gradient at iteration $t$, $\Delta_{\boldsymbol{\lambda}}^{(t)}$, to update parameter}.  \label{eq:adadelta}
\end{align}

In the experiment, we set $r = 0.85$ and $\delta = 10^{-6}$.

\paragraph{Stopping rules.}
The stopping rules seen in the previous research include the following criteria: (1) the algorithm is stopped when the pre-fixed maximum iteration is reached, see \cite{wu2022variational,cao2023variational}; (2) the change in the evidence lower bound is below a threshold, see \cite{lee2024variational}; (3) the average of the lower bound does not improve after certain iterations, see \cite{tan2018gaussian,tran2020bayesian}. In our spVarBayes framework, the evidence lower bound can be computed to check the convergence and help to decide the stopping rule. Along with the fixed maximum iteration stopping rule, we also offer the option of stopping the algorithm decided by the ELBO. To reduce the variability in the estimated evidence lower bound depending on the number of Monte Carlo samples, we use the approach from \cite{tan2018gaussian}. Instead of using a single estimated lower bound, we calculate the average of the lower bound over the past $P$ iterations, denoted as $\bar{\mathcal{L}}$. We maintain a record of the maximum value of $\bar{\mathcal{L}}$ observed so far, referred to as $\bar{\mathcal{L}}_{\text{max}}$. The algorithm terminates either when the maximum number of iterations is reached or when the estimated evidence lower bound falls below $\bar{\mathcal{L}}_{\text{max}}$ for more than $K$ consecutive times.

\paragraph{Mini-Batch.} To further enhance computational efficiency, our methods can be readily extended to support mini-batch optimization. This integration leads to the development of a doubly stochastic gradient algorithm. The \emph{doubly stochastic} nature arises from two sources of randomness: firstly, the sub-sampling of the data through mini-batches, and secondly, the Monte Carlo simulation employed to estimate expectations $\mathbb{E}_q(\cdot)$ as in (\ref{eq:elbou}). The use of NNGP as prior for $\bw$ plays a crucial role in enabling this extension. Specifically, the NNGP log-likelihood decomposes into a sum of conditionally independent terms, each corresponding to a local neighborhood, which allows mini-batch evaluation of the joint likelihood without any matrix inversion. The ELBO corresponding to each mini-batch set $\mathcal{B}$ is expressed as follows:
\begin{align*}
    \mathcal{L}(q)_{\text{for $\mathcal{B}$}}
    =&\frac{n}{|\mathcal{B}|}\Bigg(\sum_{i \in \mathcal{B}} \mathbb{E}_q\Big[\frac{1}{2}\log\frac{1}{2\pi \tau^2} - \frac{1}{2\tau^2}(y_i- \bx_i^T\bbeta - w_i)^2 \Big] \\
    &+\sum_{i \in \mathcal{B}} \mathbb{E}_q\Big[ \frac{1}{2}\log \frac{1}{2\pi\sigma^2F_i} -\frac{1}{2\sigma^2}\frac{1}{F_i}(w_i - {\bb_{N[i]}\bw_{N[i]}})^2\Big]\\
    &+\mathbb{E}_q\Big[(a_{\tau}+1)\log \frac{1}{\tau^2} - \frac{b_{\tau}}{\tau^2} + (a_{\sigma}+1)\log \frac{1}{\sigma^2} - \frac{b_{\sigma}}{\sigma^2} + 
    \log(\frac{1}{\beta_{\phi}-\alpha_{\phi}})\mathbb{I}(\phi \in (\alpha_{\phi},\beta_{\phi}))\Big] \\
    &+\frac{|\mathcal{B}|}{2}+ \frac{1}{2}\sum_{i \in \mathcal{B}} \log{d_i}\Bigg)- \mathbb{E}_q\Big[q(\bbeta,\tau^2,\sigma^2,\phi)\Big].
\end{align*}

The closed-form gradient of $\mathcal{L}(\boldsymbol{\blambda})_{\text{for $\mathcal{B}$}}$ for the variational parameters $\boldsymbol{\lambda}$ can be computed similarly based on the reparameterization trick, the Monte Carlo simulations, and the vanishing gradient approximation. Gradient ascent algorithm can then be applied to find the values of parameters that maximize the minibatch ELBO. While we did not employ mini-batching in our simulation experiments to facilitate fair comparisons with existing alternatives, this capability is implemented in our software thereby enabling distributed computation for large datasets. 

\section{Details of spatial predictions}\label{sec:predapp}
Spatial predictions are necessary to estimate the response at new locations. Consider the set of locations $\calS_0 = \{\bs_{01},\bs_{02},...,\bs_{0r}\} \in \mathcal{D} \,\setminus\, \calS$ where we are interested in making predictions. The NNGP prior for spatial random effects can be constructed for locations outside the observations \citep{datta2016hierarchical},
where $N[\bs_{0j}]$ is the $m$-nearest neighbors of $\bs_{0j}$ in $\calS$,  $\bb_{\bs_{0j}} = C_{\bs_{0j},N[\bs_{0j}]}C^{-1}_{N[\bs_{0j}]}$ and $F_{\bs_{0j}} = C(\bs_{0j},\bs_{0j}) - C_{\bs_{0j},N[\bs_{0j}]}C^{-1}_{N[\bs_{0j}]}C_{N[\bs_{0j}],\bs_{0j}}$.  
We use the composition sampling approach for prediction.
We first generate $n_{\text{pred}}$ samples from the variational distributions $q(\bbeta)$, $q({\tau^2})$, $q({\sigma^2})$,  $q(\phi)$ and $q(\bw)$: $\bbeta^{(l)}, {\tau^2}^{(l)}, {\sigma^2}^{(l)}, \phi^{(l)}, w_{i}^{(l)}$ for $l = 1,2,...,n_{\text{pred}}$, and $i= 1,...,n $. Here, $l$ indicates the sample index while $i$ indicates the location index in the observed dataset. We update $\bb_{\bs_{0j}}^{(l)}$ and $F_{\bs_{0j}}^{(l)}$ for $j = 1,2,...,r$ given ${\sigma^2}^{(l)}$ and $ \phi^{(l)}$. The $l^{th}$ estimate for $\bw^{(l)}(\bs_{0j})$ given $\bw^{(l)}_{N[\bs_{0j}]}$ is obtained from the sample of a Gaussian distribution $N\Bigl(\bw(\bs_{0j})^{(l)}|\bb^{(l)}_{\bs_{0j}}\bw^{(l)}_{N[\bs_{0j}]},F^{(l)}_{\bs_{0j}}\Bigl)$.
The response variable ${y(\bs_{0j})}^{(l)}$ comes from a Gaussian distribution with mean ${\bx(\bs_{0j})}^T \bbeta^{(l)} + \bw(\bs_{0j})^{(l)}$ and variance ${\tau^2}^{(l)}$.

\clearpage
\section{Implementations of methods for simulation study
}\label{sec:impl}

The implementation of each method is as follows. For all methods, the prior we use for $\phi$ is set to be a Uniform prior $\text{Unif}(\phi_{min},\phi_{max})$. For $\tau^2$ and $\sigma^2$ we use $IG(1, 1)$ priors. In terms of NNGP prior for $\bw$, we choose the number of neighbors as $m = 15$. The default number of Monte Carlo samples is 30 for spVB-NNGP. The initial values for the mean vector of $q(\bw)$ in spVB-NNGP and spVB-MFA are set to be the residuals of the linear model without $\bw$. The initial covariance matrix for spVB-NNGP and spVB-MFA is the diagonal matrix with each element equal to $\frac{1}{1/{\sigma^2}^{(b)}+1/{\tau^2}^{(b)}}$. ${\sigma^2}^{(b)},{\tau^2}^{(b)}$ and the initial value of $\phi^{(0)}$ are the estimators from BRISC \citep{saha2018brisc}. $\phi_{min}$ and $\phi_{max}$ are based on the observed data. Given the maximum distance between locations, denoted as $d_{max}$, we aim for $\phi$ to control the decay of spatial correlation such that it is within the range of $(0.05, 0.5)$ when the distance between two locations is at its maximum. In other words, we want $\exp(-\phi d_{max})$ to fall within the range of $(0.05, 0.5)$. To achieve this goal, we can solve for the two boundaries, resulting in  $\phi_{min} = \frac{3}{d_{max}}$, $\phi_{max} = \frac{30}{d_{max}}$. The number of nearest neighbors in the variational distribution $m_q$ for spVB-NNGP is set to be 3. The influence of the number of nearest neighbors in variational distribution is investigated and can be found in Appendix \ref{appendix:num_nngp}. For spVB-MFA-LR, before fitting the spVB-MFA model, spatial points closer than a distance threshold of 0.015 units are removed to ensure that $\exp(-\phi d_{min})\geq 0.99$. This preprocessing step is necessary because closely positioned points cause numerical instability, leading to extreme values in the linear response correction. In the following comparison when calculating the KL divergence, we compare the corrected covariance of $\bw$ to the corresponding subset of the pseudo-reference posterior’s covariance and mean. In the simulation study, we use a stopping rule where we run the algorithm for a maximum number of epochs. We set the number of epochs for spVB-NNGP and spVB-MFA to 1500 and 1000, respectively. VNNGP runs for 500 epochs as suggested in \cite{wu2022variational}. The default number of epochs in DKLGP is 35 \citep{cao2023variational}. DKLGP also apply a fixed number of epochs as the stopping criteria, we determine the number of running epochs by monitoring the ELBO, stopping when it fails to improve for $K=10$ consecutive iterations. The number of epochs for DKLGP to achieve convergence is set to be $750$, $500$ and $500$ for $n = 1000, 5000, 10000$. Since DKLGP and VNNGP are not designed for capturing $\bX\bbeta$ in their model, before fitting the model, we regress out $\bX$ using linear models without spatial random effects and fit DKLGP and VNNGP using the residuals. For spNNGP,  the number of samples is adjusted according to the sample size: $5000, 7500, 10000$ for sample size $n = 1000, 5000, 10000$, respectively, to ensure convergence for each case. Inferences are made from the MCMC samples, with the first $2000$ samples used as burn-in. 

\section{Choices for the number of nearest neighbors for NNGP prior and variational family} \label{appendix:num_nngp}

The sizes of the neighbor set $m$ in the prior and $m_q$ in the variational distribution provide a trade-off between fast speed, achieved with a smaller $m$ or $m_q$, and high accuracy, attained with a larger $m$ or $m_q$. For the sizes of neighbor set $m$ in the prior, \cite{datta2016hierarchical} has shown that NNGP models with $m = 10 - 20$ provide an excellent approximation to the full Gaussian Process. We choose $m = 15$ in the prior to ensure similar results to the full Gaussian Process. For the $m_q$ in the variational distribution, empirically, we find that a smaller $m_q$ works for approximating the posterior, as the spatial correlation is weakened conditioning on the observed data. In our experiments, we explore this trade-off by setting a range of nearest neighbor counts, specifically from $1$ to $20$ to assess how the results vary with the number of nearest neighbors $m_q$. 

\begin{figure}[H]
\centering
\includegraphics[width=0.8\textwidth]{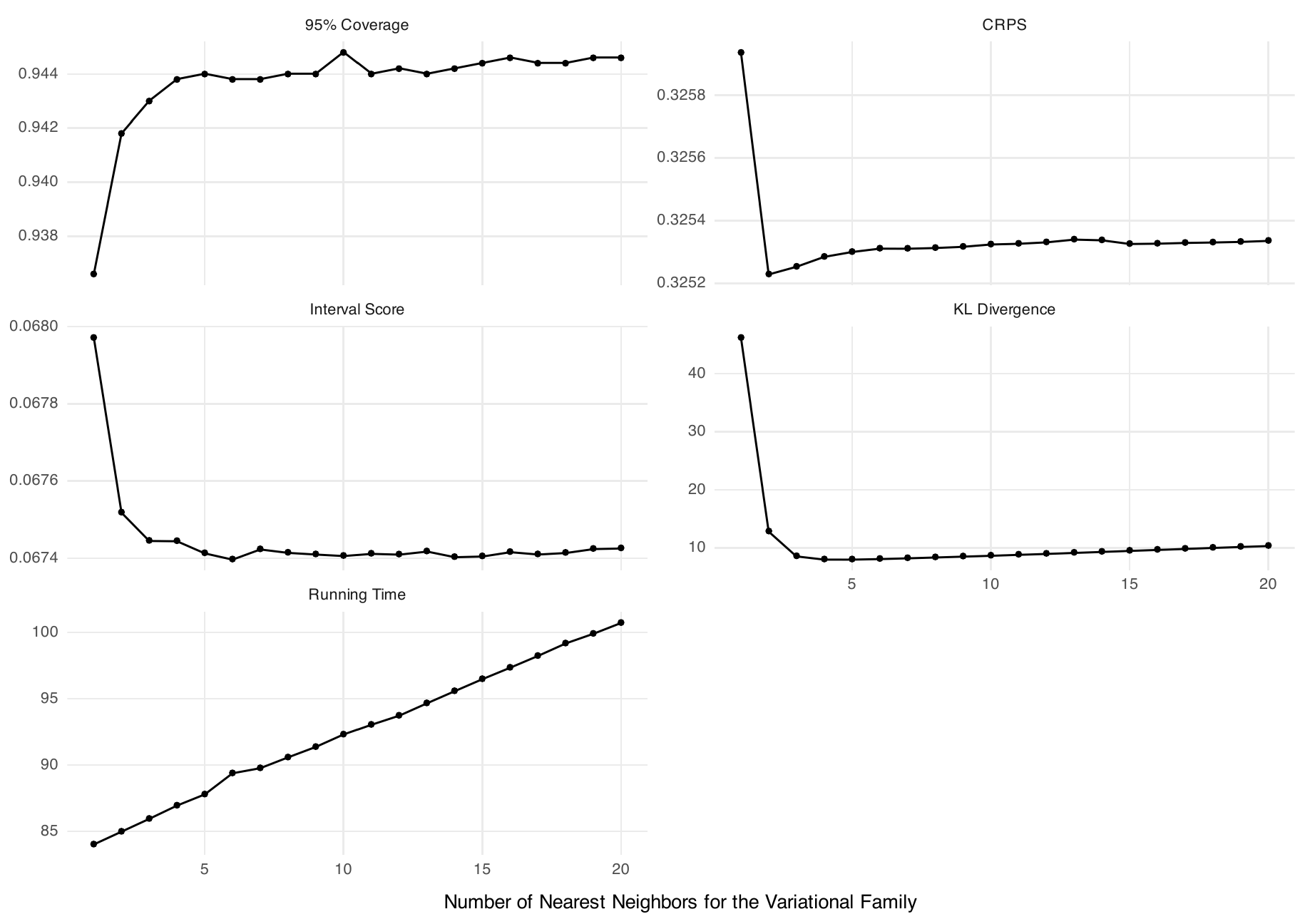}
\caption{Metrics for spVB-NNGP by different number of nearest neighbors in the variational distribution}
\label{fig:num_nn}
\end{figure}

\begin{figure}[H]
\centering
\includegraphics[width=0.8\textwidth]{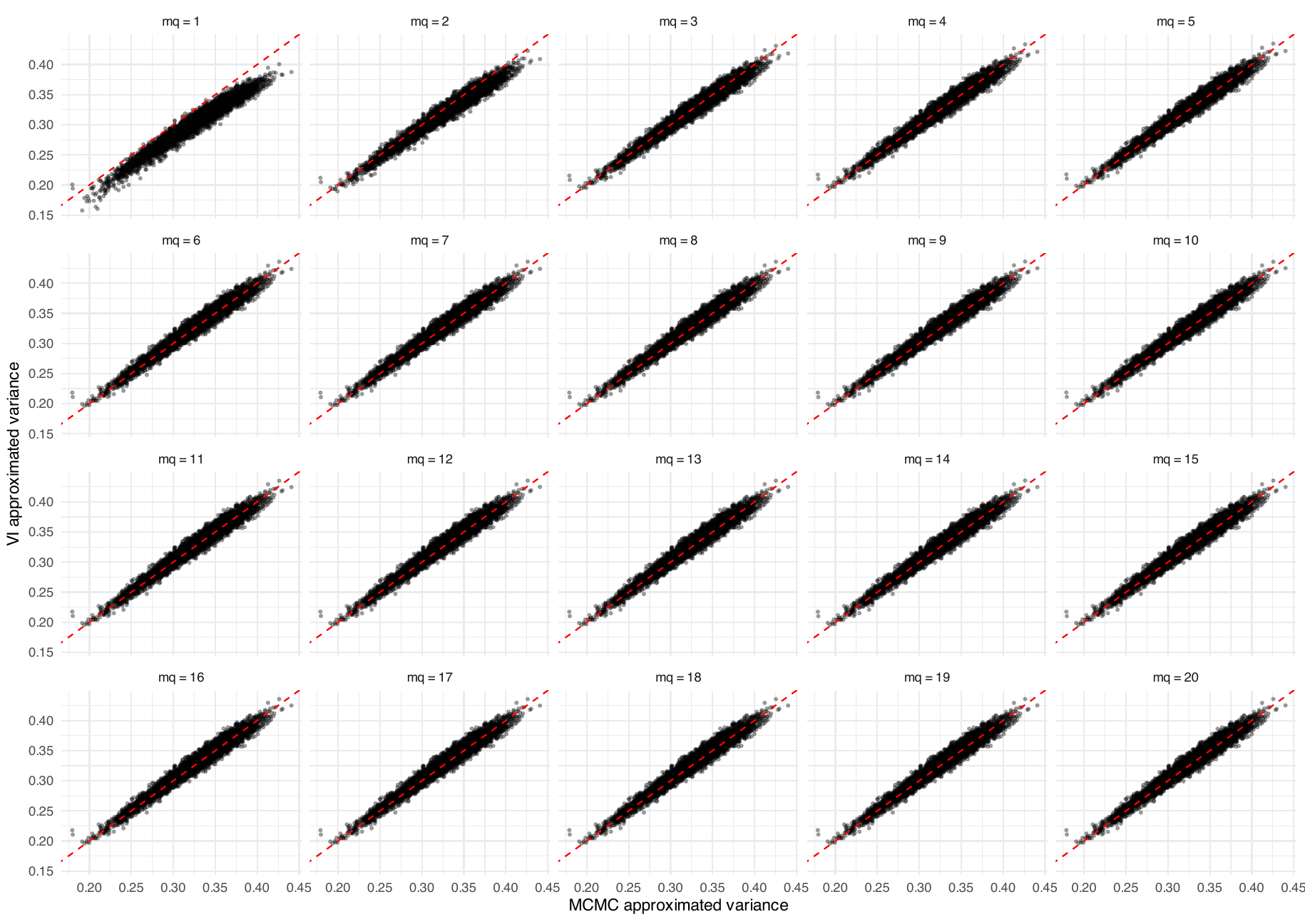}
\caption{Variance comparison for spVB-NNGP by different number of nearest neighbors in the variational distribution}
\label{fig:num_nn_var}
\end{figure}

To balance computational efficiency and statistical accuracy, we evaluate the performance of our variational method across a range of nearest neighbor sizes $m_q$ from $1$ to $20$. The experiment results are shown in Figure \ref{fig:num_nn}. We observe that key metrics, including posterior coverage, CRPS, 95\% weighted interval score, and KL divergence, stabilize quickly, with $m_q=3$ achieving near-optimal performance. The estimated posterior variances closely matched the spNNGP estimated variances from this point onward, as seen in Figure \ref{fig:num_nn_var}. Accordingly, we set $m_q=3$ for simulation experiments. For real-world applications, where spatial structures may be more irregular or noisy, we recommend using $m_q=3$ to $m_q=5$ as a default range to ensure both accuracy and computational efficiency.

\clearpage
\section{Inference on the spatial random effects} \label{appendix:sim_w}

\begin{figure}[H]
\centering
\includegraphics[width=0.9\textwidth]{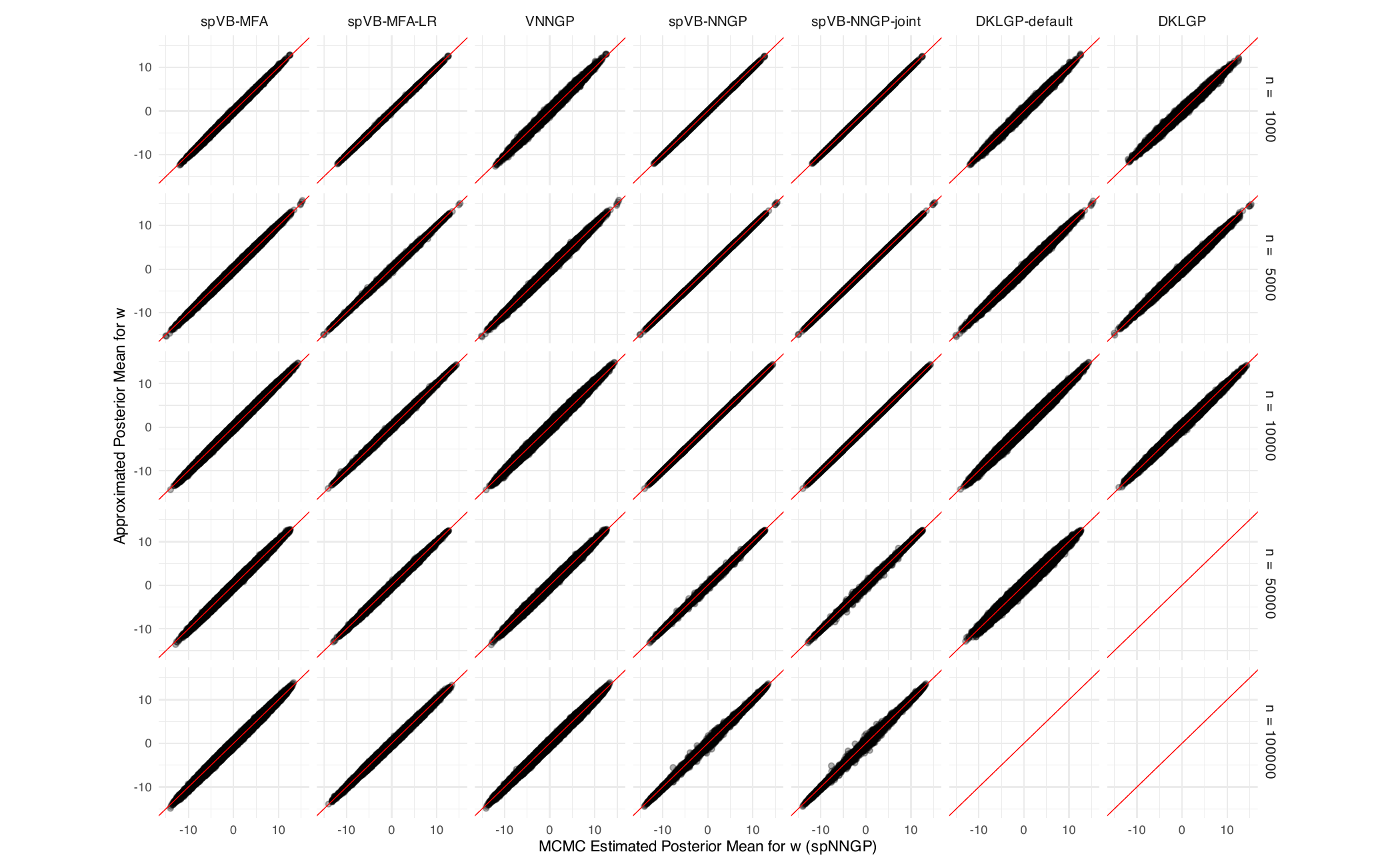}
\caption{VI approximated posterior mean vs MCMC estimated mean for $\bw$. Rows indicate sample sizes $n = 1000, 5000, 10000, 50000, 100000$. Columns correspond to spVB-MFA, spVB-MFA-LR, VNNGP, spVB-NNGP, spVB-NNGP-joint, DKLGP-default and DKLGP (optimized). The reference is MCMC-based method using spNNGP. DKLGP (optimized) is excluded at large $n$ due to failures and long running time. DKLGP-default is excluded at $n = 100000$ due to memory limits.}
\label{fig:app_w_mean_all}
\end{figure}

\begin{figure}[H]
\centering
\includegraphics[width=0.9\textwidth]{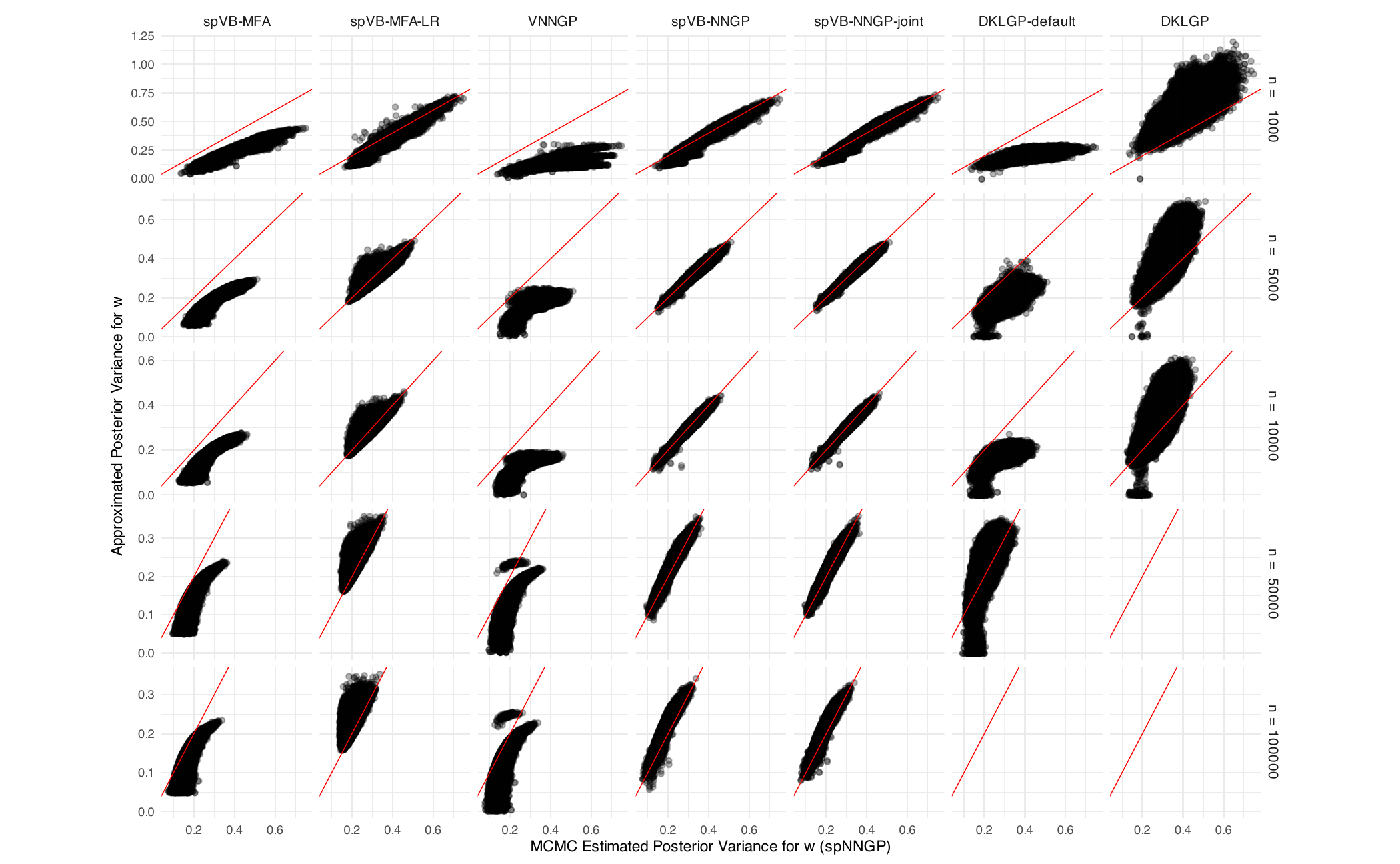}
\caption{VI approximated posterior variance vs MCMC estimated variance for $\bw$. Rows indicate sample sizes $n = 1000, 5000, 10000, 50000, 100000$. Columns correspond to spVB-MFA, spVB-MFA-LR, VNNGP, spVB-NNGP, spVB-NNGP-joint, DKLGP-default and DKLGP (optimized). The reference is MCMC-based method using spNNGP. DKLGP (optimized) is excluded at large $n$ due to failures and long running time. DKLGP-default is excluded at $n = 100000$ due to memory limits.}
\label{fig:app_w_var_all}
\end{figure}

We show the complete simulation results in this section. Across all VI methods, the approximated posterior means for the spatial random effects closely align with the MCMC estimated posterior means (Figure \ref{fig:app_w_mean_all}). spVB-NNGP and spVB-NNGP-joint provide variance approximations that lie close to the diagonal line, as seen in Figure \ref{fig:app_w_var_all}. In contrast, spVB-MFA and VNNGP underestimate the variance across all settings. The enhanced version of spVB-MFA, spVB-MFA-LR, effectively corrects the variance estimates. The DKLGP-default results show substantial deviations from the diagonal, likely due to convergence issues. When running longer, DKLGP shows improvement, though most variance estimates are still overestimated.

\begin{figure}[H]
\centering
\includegraphics[width=\textwidth]{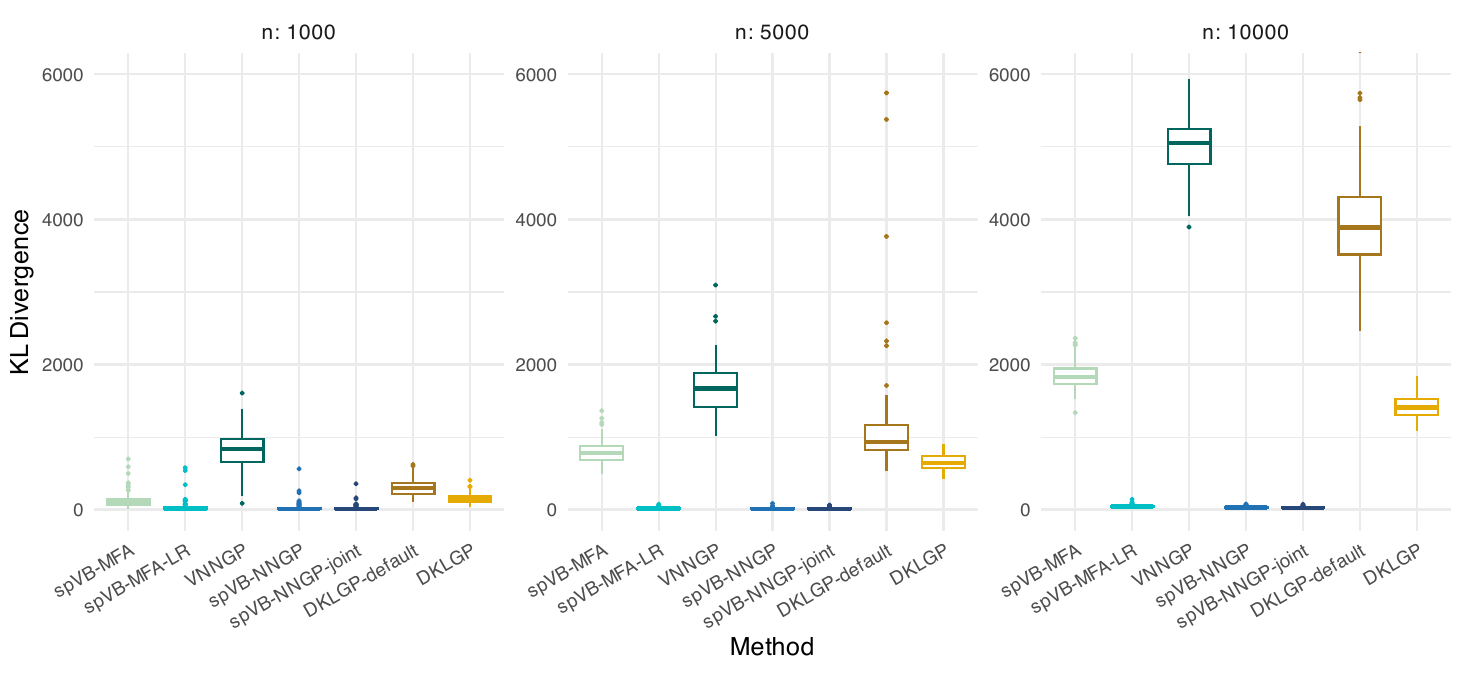}
\caption{Kullback–Leibler (KL) divergence from VI approximated posterior distribution to pseudo-reference posterior distribution under different sample size settings with $100$ simulated replicates. }\label{fig:KL_all}
\end{figure}

\begin{figure}[H]
\centering
\includegraphics[width=\textwidth]{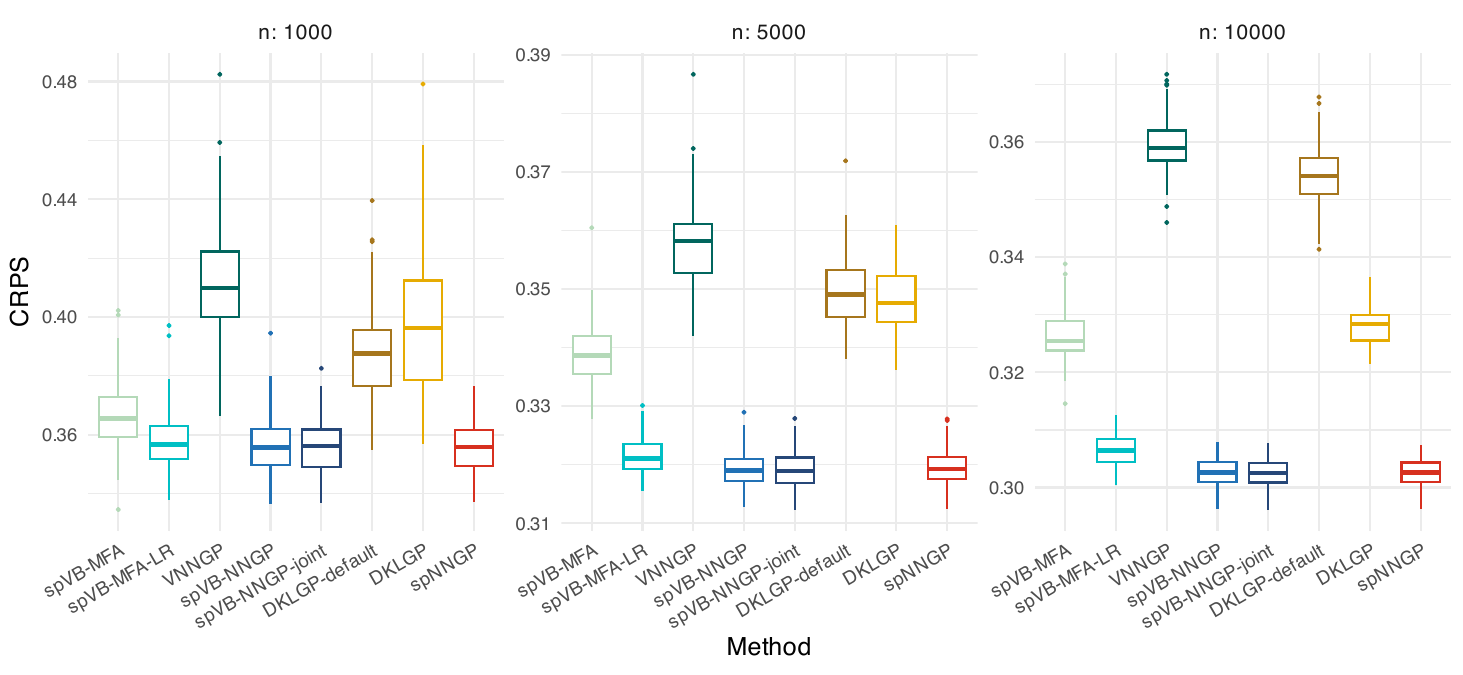}
\caption{CRPS under different sample size settings with $100$ simulated replicates. 
}\label{fig:crps_all}
\end{figure}

\begin{figure}[H]
\centering
\includegraphics[width=\textwidth]{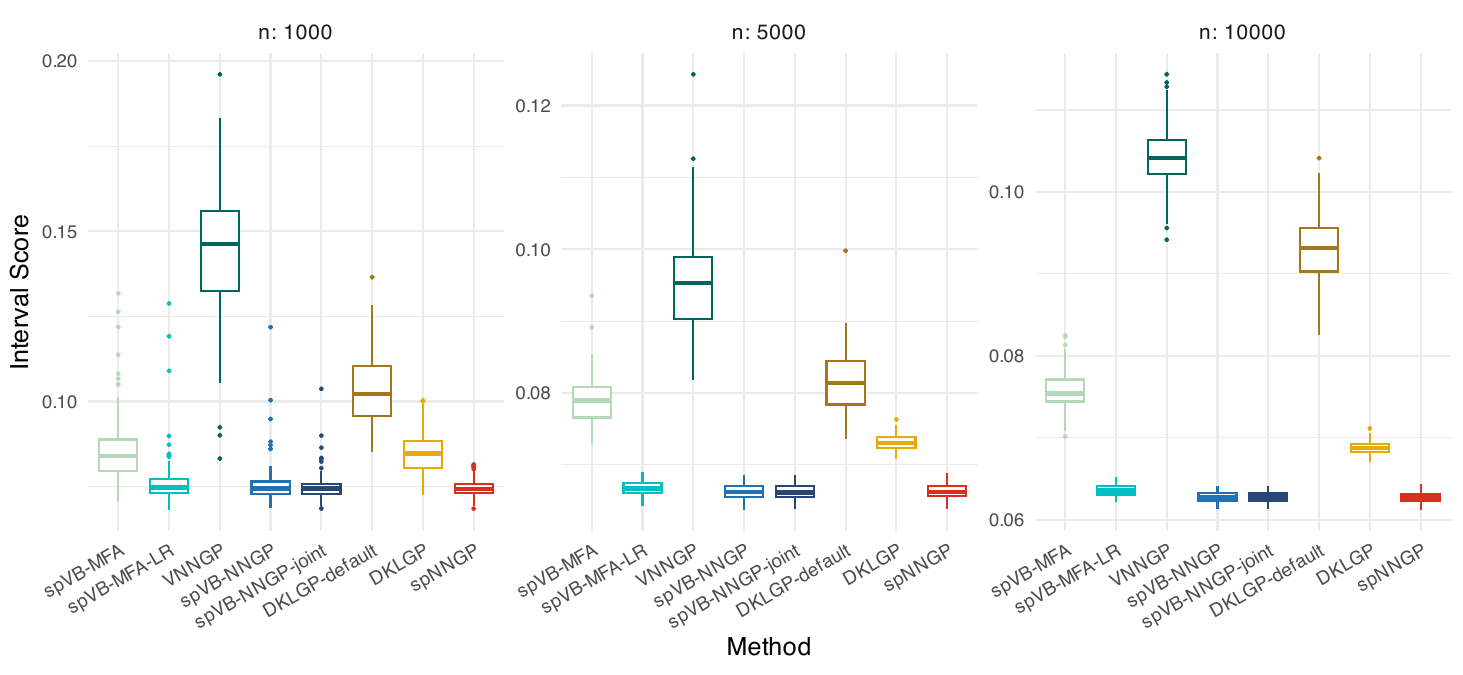}
\caption{95\% weighted interval score under different sample size settings with $100$ simulated replicates. 
}\label{fig:is_all}
\end{figure}

\begin{figure}[H]
\centering
\includegraphics[width=\textwidth]{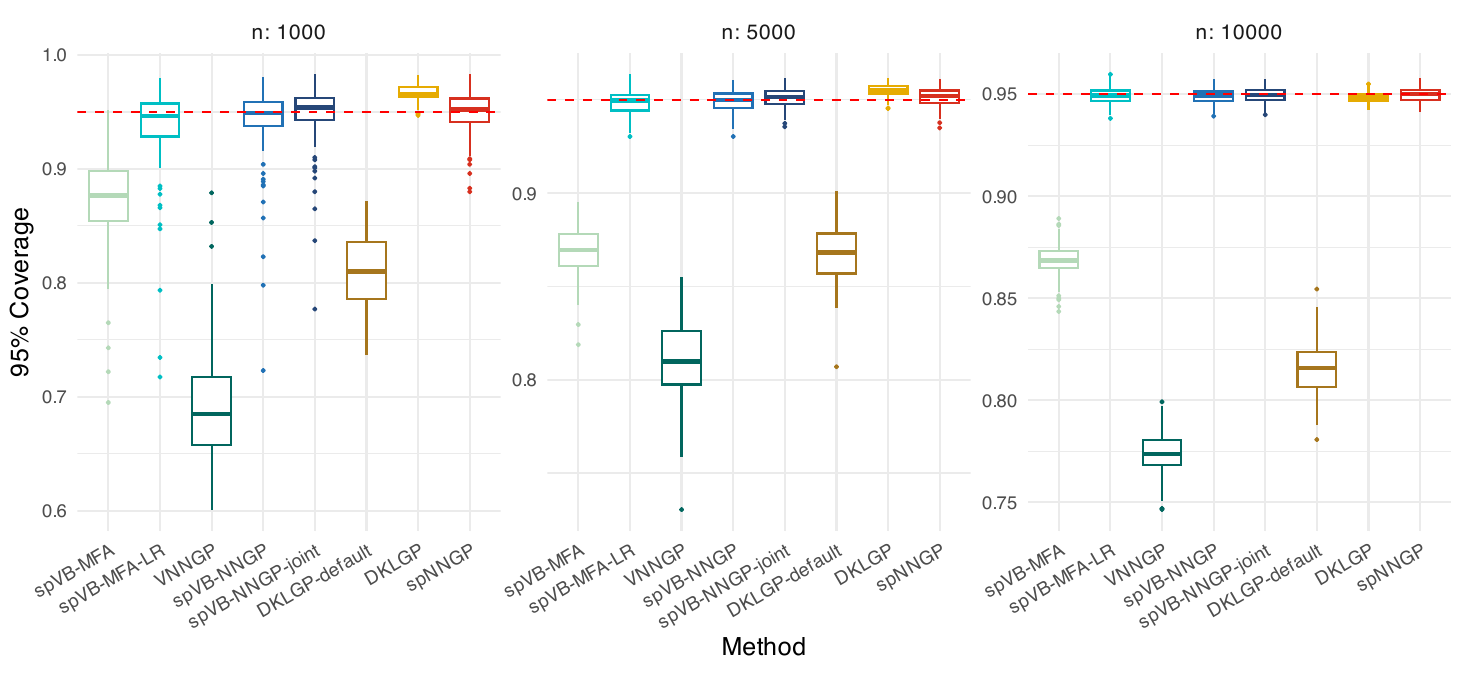}
\caption{95\% coverage under different sample size setting with $100$ simulated replicates. 
}\label{fig:coverage_all}
\end{figure}

Since we can compute the pseudo-reference posterior distribution and know the true $\bw$ in the simulations, we compare the Kullback–Leibler (KL) divergence between the VI approximated posterior distribution for $\bw$ and the pseudo-reference posterior in the training dataset (Figure \ref{fig:KL_all}). Accuracy comparisons using metrics such as the continuous ranked probability score (CRPS), the 95\% weighted interval score, and 95\% coverage for posterior samples versus the true generated values of $\bw$ are presented in Figures \ref{fig:crps_all}, \ref{fig:is_all}, and \ref{fig:coverage_all}, respectively.

For KL divergence, across all sample size settings, spVB-NNGP and spVB-NNGP-joint achieve the lowest values. The spVB-MFA-LR benefits from variance correction and shows improved performance over spVB-MFA, while VNNGP, DKLGP-default, and DKLGP report higher divergences. Regarding 95\% coverage, spVB-NNGP, spVB-NNGP-joint, spVB-MFA-LR, and the MCMC-based spNNGP method, as well as DKLGP, all achieve coverage close to 95\%. For CRPS and the 95\% weighted interval score, where lower values indicate better performance, spVB-NNGP, spVB-NNGP-joint, spVB-MFA-LR, and spNNGP exhibit similar and lowest values. These results suggest that our proposed variational inference methods not only outperform existing VI methods but also achieve accuracy comparable to the MCMC-based spNNGP.

\clearpage
\section{Inference on the spatial variance and random error variance parameters}

We show box plots of point estimates for $\sigma^2$ and $\tau^2$ across 100 replicates in the simulation study with small sample sizes, $n = 1000, 5000, 10000$. For spatial variance, spVB-NNGP, spVB-NNGP-joint, DKLGP with converged option, and spNNGP provide estimates that are approximately close to the true values. For the random error variance, only spVB-NNGP, spVB-NNGP-joint, and spNNGP yield estimates near the true values. In contrast, spVB-MFA and VNNGP consistently underestimate $\tau^2$, while DKLGP (with the converged option) tends to overestimate it.

\label{appendix:other_var}
\begin{figure}[H]
\centering
\includegraphics[width=\textwidth]{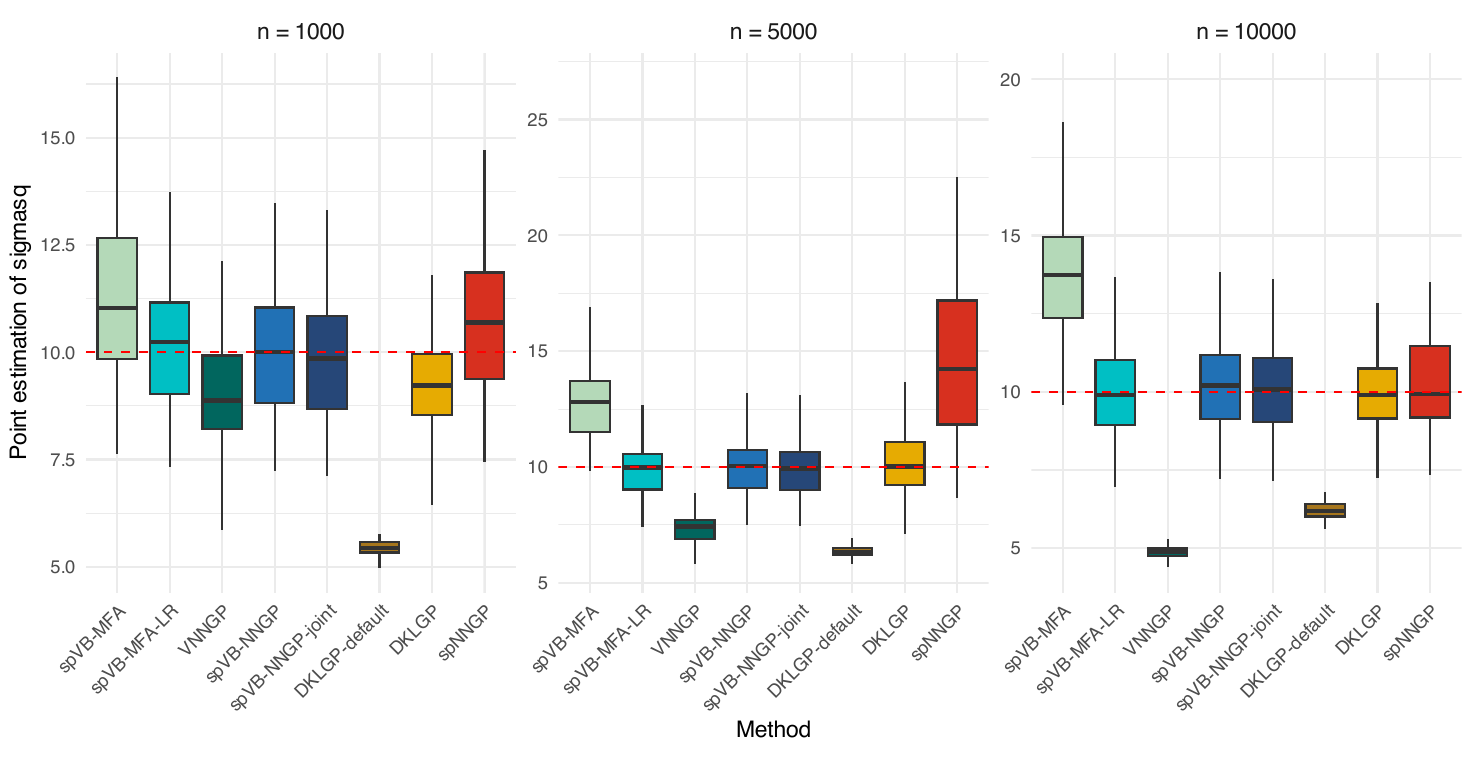}
\caption{Point estimation on the spatial variance across 100 simulated replicates}\label{fig:sigmasq-est}
\end{figure}

\begin{figure}[H]
\centering
\includegraphics[width=\textwidth]{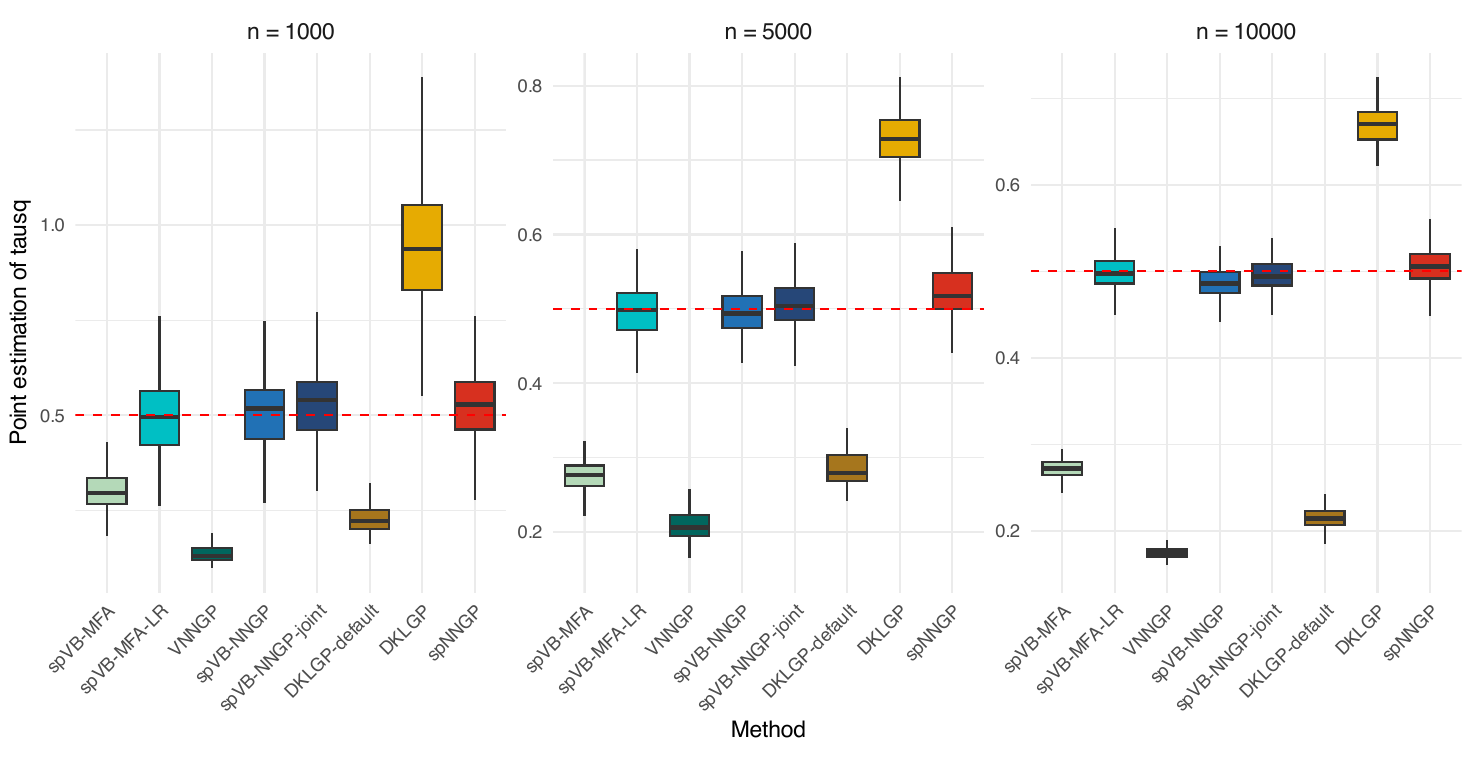}
\caption{Point estimation on the random error variance across 100 simulated replicates}\label{fig:tausq-est}
\end{figure}

\clearpage
\section{Prediction}\label{appendix:results_pred}

\begin{figure}[H]
\centering
\includegraphics[width=\textwidth]{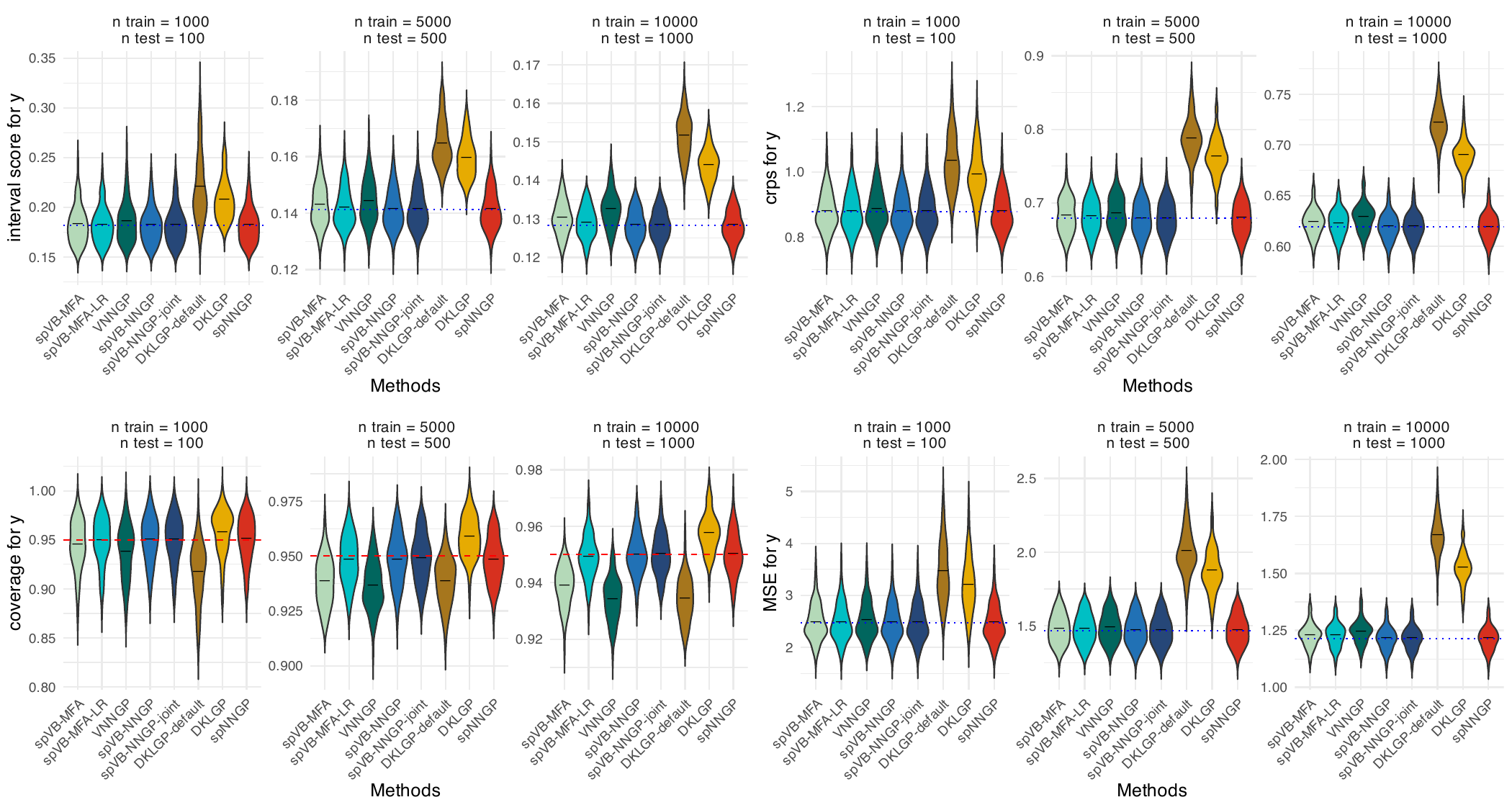}
\caption{Prediction comparison on the outcome across different methods. The blue dashed line indicates the lowest averaged scores across all competing methods by sample size. The red dashed line indicates the line of 0.95.}\label{fig:pred_y}
\end{figure}

\begin{figure}[H]
\centering
\includegraphics[width=\textwidth]{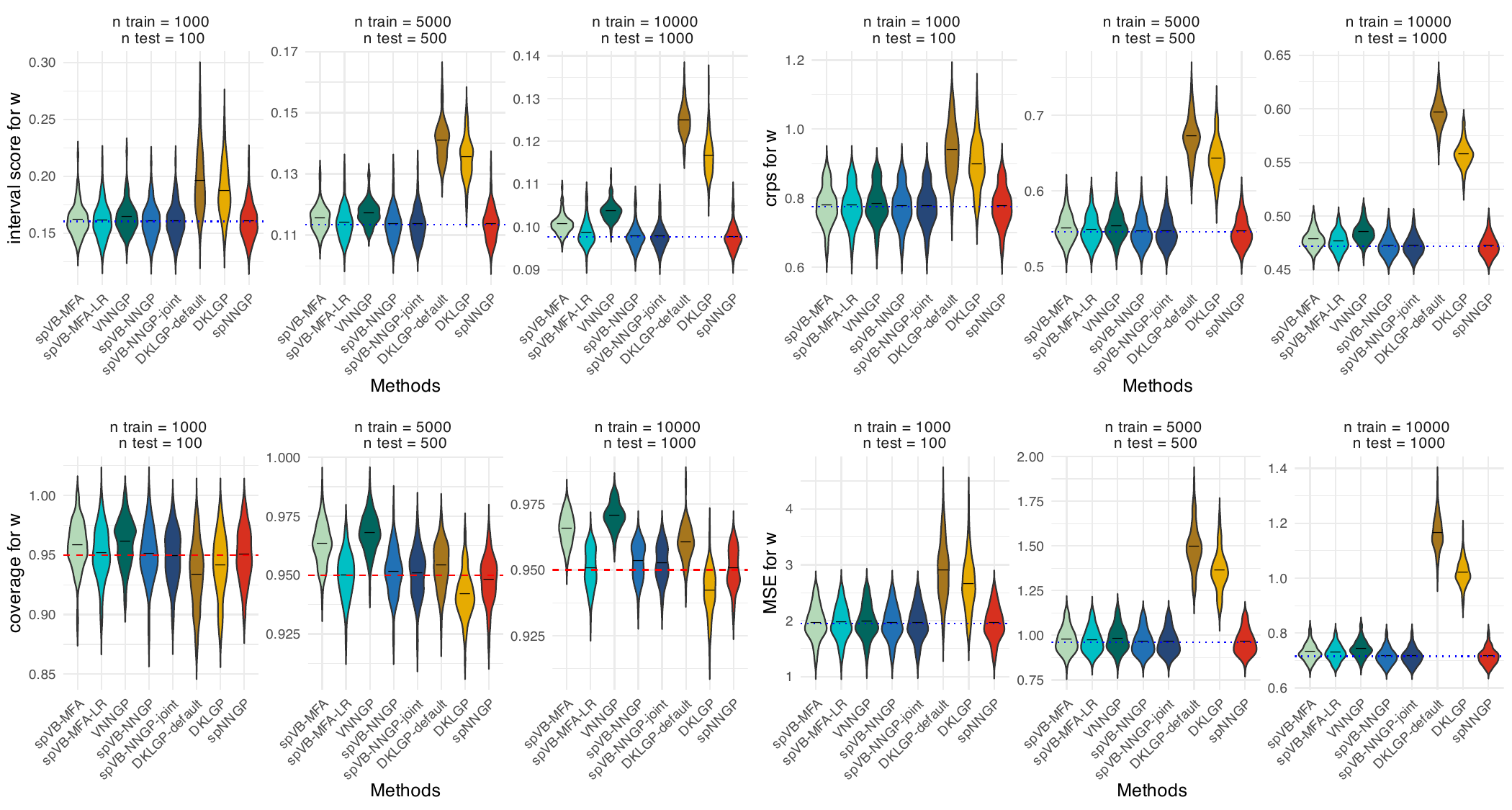}
\caption{Prediction comparison on the spatial random effects across different methods. The blue dashed line indicates the lowest averaged score across all competing methods by sample size. The red dashed line indicates the line of 0.95.}\label{fig:pred_w}
\end{figure}

We further assess predictive performance by conducting simulations focused on the response $\by$ and the spatial random effects $\bw$. For each test replicate, we compute the average 95\% weighted interval score, continuous ranked probability score (CRPS), mean squared error (MSE), and 95\% coverage across all test locations. Figure \ref{fig:pred_y} and \ref{fig:pred_w} displays violin plots of these average metrics, aggregated over 100 replicated simulation datasets.

For both the interval score and CRPS, where lower values indicate better predictive distributions, spVB-NNGPs, spVB-MFA-LR, and spNNGP consistently demonstrate superior performance across all sample sizes. The performance of spVB-MFA and VNNGP is comparable but generally less favorable. DKLGP-default with 35 training epochs performs worse, while its converged version shows improvement but still lags behind our proposed methods. The MSE results are broadly consistent across methods, with relatively small differences. However, spVB-NNGPs, spVB-MFA-LR, and spNNGP achieve the lowest prediction errors, indicating high point-prediction accuracy. The interval score, CRPS, and MSE metrics exhibit similar patterns for predictions of $\by$ and spatial random effects $\bw$.

In terms of 95\% coverage, spVB-NNGPs, spVB-MFA-LR, and spNNGP are closely achieving a 95\% level. In contrast, spVB-MFA, VNNGP, and DKLGP (default) tend to overcover the spatial random effects but undercover the response. This may result from these methods underestimating the random error variance. While the converged version of DKLGP slightly improves, reflecting broader but less efficient uncertainty intervals.

In summary, our results demonstrate that spVB-NNGPs and spVB-MFA-LR not only improve upon existing variational inference approaches such as VNNGP and DKLGP in predictive accuracy but are also comparable to the MCMC-based spNNGP. 

\label{appendix:pred_w}

\end{document}